\documentclass[11pt]{article}
\pdfoutput=1 
\usepackage[numbers,sort&compress]{natbib}
\usepackage{shorthand}

\usepackage{mathtools}
\usepackage{booktabs}
\usepackage[english]{babel}
\usepackage{amsmath,amssymb,amsbsy,amstext, amsthm, simplewick, amsfonts}
\usepackage{hyperref}
\usepackage[capitalise]{cleveref}
\usepackage{comment}
\usepackage{graphicx}
\usepackage[small,compatibility=false]{caption}
\usepackage[svgnames,dvipsnames,x11names]{xcolor}
\usepackage[utf8x]{inputenc}
\usepackage{float}
\usepackage{anyfontsize}
\usepackage{enumerate}

 \usepackage[T1]{fontenc} 
 \usepackage{braket}

\definecolor{myblue}{RGB}{0, 100, 200}
\definecolor{myred}{RGB}{214, 39, 40}
\definecolor{mygreen}{RGB}{44, 160, 44}
\definecolor{mybrown}{RGB}{123, 64, 26}
\definecolor{mydarkblue}{RGB}{44, 77, 118}
\definecolor{mycyan}{RGB}{0, 255, 255}
\definecolor{revisionpurple}{RGB}{128, 0, 160}
\definecolor{clarifyorange}{RGB}{230, 110, 0}
\definecolor{clarifyred}{RGB}{190, 35, 35}

\long\def\clarifyold#1{{\color{clarifyred}#1}}
\long\def\clarifytext#1{{\color{clarifyorange}#1}}

\hypersetup{
	colorlinks=true,
	linkcolor=blue,
	filecolor=blue,      
	urlcolor=blue,
	citecolor=myred,
}
\crefname{equation}{Eq.}{Eqs.}
\Crefname{equation}{Eq.}{Eqs.}
\crefname{figure}{Fig.}{Figs.}
\Crefname{figure}{Fig.}{Figs.}
\crefname{table}{Table}{Tables}
\Crefname{table}{Table}{Tables}
\crefname{section}{Sec.}{Secs.}
\Crefname{section}{Sec.}{Secs.}

\allowdisplaybreaks[4]

\begin{document}

\hypersetup{pageanchor=false}
\begin{titlepage}
\renewcommand{\thefootnote}{\fnsymbol{footnote}}
\setcounter{page}{1} \baselineskip=15.5pt 
\thispagestyle{empty}

\begin{center}
{\fontsize{15}{18} \bf {On Cosmological Correlators with Boundary Contributions}} \\
\end{center}

\vskip 25pt

\begin{center}
\noindent
{\fontsize{10}{18}\selectfont Yanjiao Ma$^{a,\text{ }b,}$ \footnote{\href{mailto:ymaby@connect.ust.hk}{ymaby@connect.ust.hk}}, Dong-Gang Wang$^{a,\text{ }b,\text{ }c,}$ \footnote{\href{mailto:wdgang@ust.hk}{wdgang@ust.hk}}, Xiangwei Wang$^{a,\text{ }b,}$ \footnote{\href{mailto:xwanglh@connect.ust.hk}{xwanglh@connect.ust.hk}}, Yi Wang$^{a,\text{ }b,}$ \footnote{\href{mailto:phyw@ust.hk}{phyw@ust.hk}}, Wenqi Yu$^{a,\text{ }b,}$ \footnote{\href{mailto:wyuaz@connect.ust.hk}{wyuaz@connect.ust.hk}}}
\end{center}

\vskip 15pt

\begin{center}
  \vskip 8pt
   {$^a$ \fontsize{12}{18}\it Department of Physics, The Hong Kong University of Science and Technology, Clear Water Bay, Kowloon, Hong Kong S.A.R., P.R. China
}

   {$^b$ \fontsize{12}{18}\it Jockey Club Institute for Advanced Study, The Hong Kong University of Science and Technology, Hong Kong S.A.R., P.R. China
} 

{$^c$ \fontsize{12}{18}\it Department of Applied Mathematics and Theoretical Physics,\\ University of Cambridge,
Wilberforce Road, Cambridge, CB3 0WA, UK}
 
\end{center}

\vspace{0.4cm}
 \begin{center}{\bf Abstract} 
 \end{center}
 \noindent
Cosmological correlators receive contributions from both field interactions in the bulk of quasi-de Sitter (dS) spacetime and boundary terms at the end of inflation. While most of the research efforts focus on the former, boundary contributions are normally believed to be negligible or related to field redefinitions that are associated with redundancies of the bulk Lagrangian.
In this paper, we revisit this topic in the light of the cosmological bootstrap. We first establish the correspondence between boundary terms and field redefinitions for cosmological correlators. This result provides a set of criteria for determining when boundary terms lead to non-vanishing contributions to cosmological observables.
Next, we apply this general understanding to concrete examples of correlators from massive-exchange diagrams, in both dS-invariant and boost-breaking scenarios. 
For theories with dS isometries, both IBP and field redefinitions are used to relate derivative and non-derivative exchange diagrams, from which the leading boundary contributions and IR-divergent pieces are extracted; we also derive recursion relations among higher-derivative exchange correlators.
For theories with broken dS boosts, we use IBP and EoM to reduce exchange diagrams to a basis of independent templates, and then present a classification of the boundary contributions in the general effective field theory framework.
These results pave the way for a more systematic investigation on the boundary physics of the inflationary spacetime.
\noindent

\end{titlepage}

\tableofcontents

\newpage
\hypersetup{pageanchor=true}

\section{Introduction} 


Cosmological correlators provide a unique probe for both the beginning of our Universe and also new physics at extremely high-energy scales \cite{Meerburg:2019qqi,Achucarro:2022qrl}. Recently, significant progress has been made on the analytic understanding of quantum field theory (QFT) in de Sitter spacetime. In particular, new approaches, such as the cosmological bootstrap, inspired by modern scattering amplitudes and conformal field theory studies, lead to a boundary perspective towards the equal-time correlation functions at the end of inflation \cite{Arkani-Hamed:2018kmz,Baumann:2019oyu,Baumann:2020dch, Arkani-Hamed:2017fdk, Benincasa:2018ssx, Sleight:2019mgd, Sleight:2019hfp, Goodhew:2020hob, Cespedes:2020xqq, Pajer:2020wxk, Jazayeri:2021fvk, Bonifacio:2021azc, Melville:2021lst, Goodhew:2021oqg,Baumann:2022jpr}.
Starting with basic principles, such as locality, unitarity and symmetries, these inflationary observables are systematically classified in a model-independent fashion, largely removing the redundancies in the bulk theory.


In the study of QFT in Minkowski spacetime, it is well-known that there are two types of redundancies: the interactions can be related by integration by parts (IBP), and terms in the Lagrangian proportional to the equations of motion (EoM).
The IBP terms can be removed because for S-matrix in Minkowski spacetime, we assume that fields approach the vacuum state at future and past infinity. The EoM terms are not relevant either because the field redefinitions performed to remove them can only generate first-order poles in the correlators and thus do not survive the LSZ reduction formula, which requires at least second-order poles to realise a non-trivial S-matrix element.

The same reasoning is not directly applicable to the study of cosmological correlators. 
In de Sitter spacetime, there exists a future boundary, namely the reheating surface at the end of inflation, where the equal-time correlators are defined. Unlike scattering amplitudes in flat spacetime, these observables contain interesting features due to the presence of the spacelike boundary, which evades the Minkowski arguments above. 
Consequently, IBP and EoM manipulations in the bulk can leave remnants on boundary correlators. Meanwhile, these late-time observables are also affected by field redefinitions.
From the boundary perspective of the cosmological bootstrap, one natural question arises: what are the roles of the boundary terms and field redefinitions? To be more concrete, when we remove the redundancies in the bulk interactions in the bootstrap analysis, can we safely neglect these boundary contributions, or may they become important in certain circumstances?

For single-field inflation scenarios, the above question has been systematically addressed in the context of the boostless bootstrap \cite{Pajer:2020wxk}. There, these boundary contributions are highly constrained by spacetime symmetries and locality. In multi-field models, there have been extensive studies previously using the $\delta N$ formalism under the separate universe approximation \cite{Starobinsky:1985ibc,Sasaki:1995aw,Lyth:2005fi}. In these previous studies, the boundary contributions often lead to local non-Gaussianities, as only massless fields and local field redefinitions are concerned.


However, a similar level of understanding has not been reached for the cosmological collider scenario. 
There, massive intermediate particles during inflation leave characteristic imprints in the cosmological correlators through tree-level exchange or loop diagrams \cite{Chen:2009zp, Baumann:2011nk, Noumi:2012vr, Arkani-Hamed:2015bza,Chen:2009we, Assassi:2012zq, Chen:2012ge, Pi:2012gf,  Chen:2015lza, Lee:2016vti, Chen:2016uwp, Chen:2016hrz, Chen:2017ryl, Kehagias:2017cym, Kumar:2017ecc, An:2017hlx,  Chen:2018xck, Bordin:2018pca,  Kim:2019wjo, Alexander:2019vtb,  Wang:2019gbi, Wang:2019gok, Wang:2021qez, Tong:2021wai, Cui:2021iie, Tong:2022cdz, Chen:2022vzh, Aoki:2024uyi, McCulloch:2024hiz,Pajer:2024ckd,Jiang:2025mlm,Green:2026yev}. 
On the theory frontier, the recent advances focus on the analytic computations of these correlators from bulk interactions \cite{Pimentel:2022fsc, Jazayeri:2022kjy, Qin:2022fbv, Xianyu:2022jwk, Wang:2022eop,Qin:2023ejc, Jazayeri:2023xcj,  Liu:2024xyi,  Cespedes:2025dnq, Pimentel:2025rds, Wang:2025qww, wang_interact_2025, Qin:2025xct, deRham:2025mjh, Kumar:2025anx,Jazayeri:2025vlv,Cheung:2025dmc,Xianyu:2025lbk, Arundine:2026myr,Aoki:2026vbc}.

Meanwhile, certain examples exist in the literature that highlight the effects of boundary contributions in cosmological correlators when additional degrees of freedom are present.
Recently, it has been shown in \cite{wang_interact_2025} that, for some mixings
between 
massless and massive scalars, field redefinitions and boundary terms can generate a non-trivial analytic structure at the loop level with a total energy singularity, which was widely believed to be a particular feature of generic bulk interactions. 
At the tree level, boundary terms can also become non-negligible when we encounter correlators with infrared (IR) divergences \cite{Cespedes:2025ple}. 
This motivates a more systematic investigation of boundary contributions in cosmological correlators that have been overlooked in previous works.

In this work, we develop such an improved understanding by linking different bulk interactions and classifying the accompanying boundary contributions. We first use Schwinger--Dyson equations to formulate how local field redefinitions act in the Schwinger--Keldysh path integral: they change the late-time operator, generate bulk EoM terms, and induce boundary terms on the final slice. This gives a diagrammatic approach for tracking their effect on perturbative correlators, and leads to practical criteria for deciding which boundary contributions survive the late-time limit.

We then apply this general understanding to massive-exchange diagrams. For dS-invariant theories, field redefinitions together with IBP relations connect derivative and non-derivative exchange diagrams. By comparing the resulting diagrams, we extract the leading effects of the surviving boundary contributions and identify the IR-divergent pieces generated by the reduction. We also extend our analysis to higher-derivative interactions and derive recursion relations among the resulting exchange correlators.
For boost-breaking scenarios, we present the independent exchange trispectra and bispectra after performing IBP and EoM reduction, and also provide a complete classification of non-vanishing boundary contributions.

The structure of the paper is organized as follows.

\subsection{Outline}
In \cref{CH2}, we first establish a perturbative correspondence between boundary terms and field redefinitions. This correspondence can be understood as a series of reductions on Feynman diagrams, in which boundary terms involving a single time derivative --- accompanied by on-shell vanishing terms (which we refer to as EoM terms) --- act as precise scalpels. We then present several criteria for identifying when these boundary terms --- those involving a single time derivative --- yield non-vanishing physical effects in \cref{criteria}. 
In \cref{path integral issues}, we demonstrate the complete invariance of the path integral under field redefinitions. After establishing the invariance of both the path integral measure and the initial conditions that define the quantum state in \cref{equivalence}, we derive the corresponding Schwinger--Dyson equations in \cref{SD-eqn}.

In \cref{CH3}, we investigate the roles of boundary contributions in dS-invariant correlators, 
with a focus on Feynman diagrams involving the exchange of massive particles. In the principal series ($m>\frac{3H}{2}$), we employ two complementary approaches---IBP in \cref{IBP} and field redefinition in \cref{field-redef}---to classify the connections among various dS-invariant interactions, including the non-derivative ones with infrared divergences and the higher-derivative couplings.
We further illustrate the cosmological bootstrap computation for these diagrams in \cref{Bootstrap details}, where late-time divergences are interpreted as manifestations of boundary conformal anomalies.
In the light-mass regime ($m<\frac{3H}{2}$), where late-time divergences cannot be fully resolved by IBP or field redefinition, we show in \cref{Light mass} that the bootstrap equations receive late-time decaying corrections, and that the massless limit can be consistently achieved by appropriately fixing the coefficients of the homogeneous solutions.

In \cref{CH4}, we extend the analysis to the boost-breaking theories in the framework of the EFT of inflation. Through IBP and EoM, we first reduce the exchange correlators from boost-breaking interactions to a complete basis of shape functions. Then we classify all possible boundary contributions to the late-time bispectra and trispectra.

We conclude in \cref{Conclusion} with an outlook for future directions.

\subsection{Notation and Conventions}
The spacetime background is approximated by the flat slicing of de Sitter universe with a late-time boundary at $\eta_0\rightarrow0^-$. The metric is given by
\begin{equation}
\label{eqn001}
    ds^2=a^2(\eta)(-d\eta^2+d\vec{x}^2)=\frac{1}{H^2\eta^2}(-d\eta^2+d\vec{x}^2),
\end{equation}
where $H$ is the Hubble constant. We use a prime ($  '  $) and a dot ($  \dot{~}$) to denote derivatives with respect to conformal time $  \eta  $ and physical time $  t  $, respectively. They are related by
$  X' = a \dot{X}  $.
We study theories of $N$ real scalar fields $\Phi_i$ with a general interaction Lagrangian. When deriving the field-redefinition identities in \cref{CH2,path integral issues}, we assume local interactions involving at most first time derivatives. The corresponding action takes the form
\begin{equation}
\label{eqn002}
    S[\Phi_i]=\int d^4 x \hspace{2pt}\sqrt{-g}\left[\sum_{i}\left(-\frac{1}{2}(\partial{\Phi_i})^2-\frac{1}{2}m_i^2\Phi_i^2\right) + \mathcal L_{\rm int}[\Phi_i,\nabla_\mu\Phi_i,\ldots]\right].
\end{equation}
We introduce the following dimensionless parameters:
\begin{equation}
\label{eqn003}
\nu_i \equiv \sqrt{\frac{9}{4} - \frac{m_i^2}{H^2}}, \quad \Delta_i \equiv \frac{3}{2} - \nu_i.
\end{equation}
In \cref{CH3,CH4}, as well as \cref{Bootstrap details,Light mass}, we focus on two-field models involving a massless field $\phi$ and a massive field $\sigma$. We consider two representative mass regimes:
\begin{itemize}
    \item $m_{\phi}=0, m_{\sigma}=m>\frac{3H}{2}$: Here, $\phi$ plays the role of the inflaton, while $\sigma$ lies in the principal series. The heavy field $\sigma$ has negligible backreaction on the inflationary background but can produce distinctive cosmological collider signals. 
    \item $m_{\phi}=0, m_{\sigma}=m<\frac{3H}{2}$: In this case, $  \sigma  $ belongs to the complementary series. 
    A particularly interesting limit occurs when $  m \to 0  $, where additional late-time decaying modes emerge in the correlators, revealing rich and nontrivial analytic structures.
\end{itemize}
We employ the Schwinger--Keldysh formalism to compute cosmological correlators. The relevant bulk-to-boundary propagators, expressed in terms of the mode function $u_k(\eta)$, are defined as:
\begin{equation}
\label{eqn004}
\begin{split}
    & K_{+}(k,\hspace{2pt}\eta)=u^\ast(k,\hspace{2pt}\eta)u(k,\hspace{2pt}\eta_0),\\
    & K_{-}(k,\hspace{2pt}\eta)=u(k,\hspace{2pt}\eta)u^\ast(k,\hspace{2pt}\eta_0),
\end{split}
\end{equation}
while the bulk-to-bulk propagators are as follows:
\begin{equation}
\label{eqn005}
\begin{split}
    & D_{++}(k,\hspace{2pt}\eta_1,\hspace{2pt}\eta_2)=u(k,\hspace{2pt}\eta_1)u^\ast(k,\hspace{2pt}\eta_2)\theta(\eta_1-\eta_2)+u^\ast(k,\hspace{2pt}\eta_1)u(k,\hspace{2pt}\eta_2)\theta(\eta_2-\eta_1),\\
    & D_{+-}(k,\hspace{2pt}\eta_1,\hspace{2pt}\eta_2)=u^\ast(k,\hspace{2pt}\eta_1)u(k,\hspace{2pt}\eta_2),\\
    & D_{-+}(k,\hspace{2pt}\eta_1,\hspace{2pt}\eta_2)=u(k,\hspace{2pt}\eta_1)u^\ast(k,\hspace{2pt}\eta_2),\\
    & D_{--}(k,\hspace{2pt}\eta_1,\hspace{2pt}\eta_2)=u^\ast(k,\hspace{2pt}\eta_1)u(k,\hspace{2pt}\eta_2)\theta(\eta_1-\eta_2)+u(k,\hspace{2pt}\eta_1)u^\ast(k,\hspace{2pt}\eta_2)\theta(\eta_2-\eta_1),\\
\end{split}
\end{equation}
where the step function is defined as
\begin{equation}
\label{eqn006}
    \theta(\eta)=
    \begin{cases}
        0,&\eta<0\\
        1,&\eta>0 
    \end{cases},
\end{equation}
with $\theta(0)=1/2$.
After applying IBP to move derivatives away from interaction vertices, we adopt the following consistent prescription:

The time integral associated with each bulk vertex is taken from the infinite past ($  \eta = -\infty  $) up to the late-time cutoff $  \eta_0 $ (approaching 0 from below), capturing the bulk contribution.
The endpoint contributions generated by IBP are not included in this bulk integral; they are collected and evaluated separately on the future hypersurface $  \eta = \eta_0  $.

In the Schwinger--Keldysh diagrammatic expansion, each original interaction vertex must be replaced by a sum over the two types of vertices: the time-ordered ($  +  $) vertex and the anti-time-ordered ($  -  $) vertex.

The Bunch–Davies mode function for a scalar field with mass parameter $\nu$ is given by
\begin{equation}
\label{eqn007}
    u(k,\hspace{2pt}\eta)=-i \frac{\sqrt{\pi}}{2} e^{i\pi(\nu/2 + 1/4)} H (-\eta)^{3/2} \mathrm{H}^{(1)}_\nu(-k\eta), 
\end{equation}
where $  \mathrm{H}^{(1)}_\nu  $ is the Hankel function of the first kind. For the massless case ($  \nu = 3/2  $), this reduces to the familiar form
\begin{equation}
\label{eqn008}
u(k,\hspace{2pt}\eta) \xrightarrow{\nu\rightarrow 3/2} \frac{H}{\sqrt{2k^3}}(1+ik\eta)e^{-ik\eta}.
\end{equation}
We collect below several useful identities satisfied by the Schwinger--Keldysh propagators (with $  b,c = \pm  $ labelling the contour branches):
\begin{equation}\label{properties-eq009}
    \begin{split}
        &K_{\pm}(k,\eta_0)=|u(k,\eta_0)|^2,\\
      &D_{b\pm}(k,\eta,\eta_0)=K_{b}(k,\eta),\\
    &\partial_{\eta}(K_{+}-K_{-})|_{\eta=\eta_0}=iH^2\eta_0^2,\\
      &\partial_{\eta_2}(D_{b+}(k,\eta_1,\eta_2)-D_{b-}(k,\eta_1,\eta_2))|_{\eta_1=\eta_2=\eta_0}=iH^2\eta_0^2 ,\\
      &\partial_{\eta_1}(D_{b+}(k,\eta_1,\eta_2)-D_{b-}(k,\eta_1,\eta_2))|_{\eta_1=\eta_2=\eta_0}=-iH^2\eta_0^2, \\
      &\partial_{\eta_1}\partial_{\eta_2}(D_{+b}(k,\eta_1,\eta_2)-D_{-b}(k,\eta_1,\eta_2))|_{\eta_1=\eta_0}=iH^2\eta_0^2\delta(\eta_2-\eta_0),\\  
      &(\square_{\eta}-m^2) K_{\pm}(k,\eta)=0,\\
      &(\square_{\eta}-m^2)D_{bc}(k,\eta,\eta')=i b H^4\eta^4\delta(\eta-\eta^\prime)\delta_{bc},
    \end{split}
\end{equation}
where $  \square_\eta  $ is the d'Alembertian with respect to conformal time $\eta$; and in the RHS of the last equation $b$ takes values of $\pm 1$ depending on the contour branches.  

\section{Correspondence between Boundary Terms and Field Redefinitions}\label{CH2}
In this section, we explore the connection between boundary terms and local field redefinitions, which we summarize as several diagrammatic relations. As a straightforward application, we obtain precise criteria for when boundary terms yield non-vanishing contributions to cosmological correlators.

Our approach is to start with the invariance of the Schwinger--Keldysh generating functional under field redefinitions, and derive the diagrammatic relations between boundary terms and field redefinitions using the Schwinger--Dyson equation.
This shares some similarities with \cite{Cohen:2024fak}, which focused on field redefinitions in flat spacetime and established the diagrammatic relations for scattering amplitudes.
For cosmology, as the objects of interest are equal-time correlators in the Schwinger--Keldysh formalism, we shall show that the boundary terms from IBP will give nonzero contributions, yielding new diagrammatic relations.

In particular, we focus on the redefinition of a single real scalar field $  \Phi  $ with mass $  m  $ and suppress all other fields in the expressions below. We consider general late-time correlators $  \braket{\mathcal{O}[\Phi(\eta_0)]}  $, where $  \mathcal{O} $ is a local polynomial of the field operator on the late-time boundary $  \eta = \eta_0  $. 
Our analysis relies on a simple fact: under a local invertible field redefinition of the form $  \Phi = \tilde{\Phi} + f[\tilde{\Phi}]  $, the Schwinger--Keldysh path integrals for arbitrary late-time correlators 
\begin{equation}
\label{eqn010}
\bra{\Omega}\mathcal{O} [\Phi]\ket{\Omega}=\mathcal{O}\left[\frac{\delta}{i\delta J^+}\right]Z[J^+,J^-]|_{J^+=J^-=0},
\end{equation}
remain invariant, where the generating functional is defined as
\begin{equation}
\label{eqn011}
     Z[J^+,J^-] = \int ^{\Phi^+(\eta_0)=\Phi^-(\eta_0)}_{\text{initial conditions}}\mathcal D\Phi^+\mathcal D\Phi^- \exp \left[i\sum_{b=\pm}b\left(S[\Phi^b]+\int d^4 x \sqrt{-g} J^b\Phi^b\right)\right].
\end{equation}
To demonstrate how boundary terms are related to field redefinitions, we treat the field redefinition as a field variation $  \Delta\Phi = f[\Phi]  $. 
After performing IBP, the variation of the action splits into a boundary term and an on-shell vanishing term, which we refer to as the EoM term:
\begin{equation} \label{IBP2.1-eq012}
\begin{split}
    \Delta S=&  \int d^4x\hspace{2pt}a^4(\eta)\left[\nabla_\mu\left(\frac{\partial\mathcal{L}}{\partial (\nabla_\mu\Phi)}\Delta \Phi\right)+\left(\frac{\partial\mathcal L }{\partial\Phi}-\nabla_\mu \left(\frac{\partial \mathcal L}{\partial(\nabla_\mu \Phi)}\right)\right)\Delta\Phi\right]\\
    =&\left. \int d^3x\hspace{2pt}a^4(\eta)\Pi[\Phi]f[\Phi] \right|_{\eta=\eta_0}+\int d^4x \hspace{2pt}a^4(\eta)E[\Phi]f[\Phi] ,
\end{split}
\end{equation}
where $  \Pi  $ denotes the field momentum. For the free theory, $  \Pi(\Phi) =\Pi_0= a^{-2} \partial_\eta\Phi  $ and $  E(\Phi) =E_0= (\square - m^2)\Phi  $.

\subsection{Diagrammatic Reduction Rules}\label{reduction}

Exploiting the invariance of path integrals under field variations, we can systematically derive perturbative Schwinger--Dyson equations. These equations provide the desired correspondence between boundary terms and the variation of late-time correlators under field redefinitions. 

The basic idea of the derivation is as follows. A field redefinition $\Phi\to\Phi+f[\Phi]$ is a change of variables in the Schwinger--Keldysh path integral. Since the path integral is invariant, the variation of the late-time operator $\mathcal {O}$ must be compensated by the variation of the action. After integrating by parts, this variation splits into a boundary piece, generated by $\Pi f$ on the final time slice, and a bulk piece proportional to the equation of motion, $E f$.
Here we present the Schwinger--Dyson equations directly, leaving their derivation to \cref{path integral issues}, and then explain their diagrammatic implications.

The first-order Schwinger--Dyson equations under field redefinitions are given by:
\begin{equation}\label{2.2-eq013}
    \begin{split}
     \left(\Delta \mathcal{O}\left[\frac{\delta}{i\delta J^+}\right]+\mathcal{O}\left[\frac{\delta}{i\delta J^+}\right]\left(\sum_{b=\pm}ib\int d^3x \hspace{2pt}a^4(\eta_0) \Pi\left[\frac{\delta}{ib\delta J^b}\right]f\left[\frac{\delta}{ib\delta J^b}\right]\right)\right)Z[J^\pm]|_{J^\pm=0} =&0,\\
    \mathcal{O}\left[\frac{\delta}{i\delta J^+}\right]\left(\sum_{b=\pm}ib\int d^4x \hspace{2pt}a^4(\eta)E\left[\frac{\delta}{ib\delta J^b}\right]f\left[\frac{\delta}{ib\delta J^b}\right]\right)
    Z[J^\pm]|_{J^\pm=0} =&0.
\end{split}
\end{equation}
The first one is the boundary identity: it relates the variation of the late-time operator to the boundary vertex $\Pi f$. The second is the bulk EoM identity: it states that the bulk term $Ef$ does not produce an independent contribution. To obtain diagrammatic rules, we expand perturbatively in the interactions and decompose
\begin{equation}
\label{eqn014}
\Pi=\Pi_0+\Pi_{\rm int},\qquad E=E_0+E_{\rm int}.
\end{equation}
The free pieces $\Pi_0$ and $E_0$ act as inverse propagators. They either remove an external line, or collapse an internal line ending on an interaction vertex, reducing the diagrams to simpler ones, which reproduce or cancel the variations of boundary operators $\mathcal O$ or interactions $\Pi_{\rm int}f$ and $E_{\rm int}f$.

Below, we shall apply the above two identities to derive the reduction rules on Feynman diagrams.
In the following, we use $\mathcal P_b^f$ to represent the product of propagators attached to the fields inside $f[\Phi]$, while $\mathcal P_b^{\rm int}$ denotes the product of spectator propagators attached to the remaining fields in the interaction vertex after the line acted on by $\Pi_0$ or $E_0$ has been singled out.
For example, consider a boundary vertex 
$f\Pi_0=\lambda a^2(\eta_0)\phi^2\phi'$ on the $b$ branch. If the two $\phi$ fields inside $f=\lambda\phi^2$ are connected to two other vertices on branches $c_1$ and $c_2$, at times $\eta_1$ and $\eta_2$, by internal lines with momenta $\vec k_1$ and $\vec k_2$, then
\begin{equation}
\label{eqn015}
\mathcal P_b^f
=
\lambda D_{bc_1}^{\phi}(k_1,\eta_0,\eta_1)
D_{bc_2}^{\phi}(k_2,\eta_0,\eta_2).
\end{equation}
For an interaction vertex 
$g a^3(\eta)\phi'(\nabla\phi)^2$ on the $b$ branch, suppose that the $\phi'$ leg is the distinguished line acted on by an $E_0$ or $\Pi_0$ operator. If the two $\nabla\phi$ legs are connected to two other vertices on branches $c_1$ and $c_2$, at times $\eta_1$ and $\eta_2$, by internal lines with momenta $\vec k_1$ and $\vec k_2$, then
\begin{equation}
\label{eqn016}
\mathcal P_b^{\rm int}
=
g a^3(\eta)\left(\vec k_1 \cdot \vec k_2\right)
D_{bc_1}^{\phi}(k_1,\eta,\eta_1)
D_{bc_2}^{\phi}(k_2,\eta,\eta_2).
\end{equation}

We begin with the first Schwinger--Dyson equation.
As a simple illustration, consider a field redefinition $\Phi\to\Phi+\lambda\chi^2$. If the late-time operator contains one external $\Phi$, say $\mathcal O=\Phi(\mathbf k)X$, then $\Delta\mathcal O=\lambda\chi^2(\mathbf k)X$. Then the first identity in \cref{2.2-eq013} says that this variation is equivalently produced by attaching the boundary vertex $\lambda\Pi f=\lambda\Pi\chi^2$ to the external $\Phi$ line. Its free part $\lambda\Pi_0\chi^2$ cuts the external line by the equal-time canonical commutator and leaves $\lambda\chi^2$ on the boundary. If the theory also contains a derivative interaction such as $\mathcal L_{\rm int}=g\,\Phi'\sigma^2$, the same $\Pi_0$ operator may act on the $\Phi'$ inside the interaction vertex. It then cuts the internal line at the final boundary and reproduces the interaction-dependent boundary variation $\Pi_{\rm int}f=g\,\sigma^2\,\lambda\chi^2$.
Thus, we reach the following two reduction rules of Feynman diagrams.

\begin{enumerate}[(1)]
    \item \textbf{$\Pi_0$ on external-line}: When a $  \Pi_0 f  $ boundary vertex is connected to a $  \Phi(\vec{k})  $ operator in $  \mathcal{O}  $ through an external line, the $  \Pi_0  $ operator cuts the external line, and glues  $  f  $ onto the boundary, thus cancels the contribution from the variation $  \Delta \mathcal{O}  $:
    \begin{equation}
\label{eqn017}
        \sum_{b=\pm}ib\,a^2(\eta_0)\,\partial_{\eta_0} K_{b}(k,\eta_0)\,\mathcal P_b^f(\eta_0)=-\mathcal P^f(\eta_0).
    \end{equation}
    
    \item \textbf{$\Pi_0$ on internal-line}: For a $  \Pi_0 f  $ vertex connected to a $  \Phi'  $ operator in $  S_{\rm int}  $ through an internal line, the $  \Pi_0  $ operator annihilates with the $  \Phi'  $ operator, cuts the internal line and glues $  \frac{\partial\mathcal{L}_{\rm int}}{\partial\Phi'}  $ together with $  f  $ on the boundary, thus cancels the contribution from $  \Pi_{\rm int} f $ -- the boundary part of the variation $\Delta S_{int}$ (with $  \Pi_{\rm int} = \Pi - \Pi_0  $): 
    \begin{equation}
\label{eqn018}
    \begin{split}
         &\sum_{c=\pm} ic \int d\eta \, a^4(\eta) \, \mathcal P_c^{\rm int}(\eta) \Bigl[ \sum_{b=\pm} ib \, a^2(\eta_0) \, \partial_{\eta_0}\partial_\eta D_{bc}(k,\eta_0,\eta) \, \mathcal P_b^f(\eta_0) \Bigr] \\
         =& -\sum_{c=\pm} ic \, a^4(\eta_0) \, \mathcal P_c^{\rm int}(\eta_0) \, \mathcal P^f(\eta_0).
    \end{split}
    \end{equation}
\end{enumerate}
    
The second identity can be illustrated by a non-derivative interaction, for instance $\mathcal L_{\rm int}=h\,\Phi\sigma^2/2$. The bulk operator $E_0$ acting on an external $\Phi$ leg gives a vanishing result, because the bulk-to-boundary propagator obeys the free equation of motion. When attached to an internal $\Phi$ propagator ending on the interaction vertex, however, the Green-function identity collapses that propagator and produces the interaction-dependent EoM variation $E_{\rm int}f=(h/2)\sigma^2\,\lambda\chi^2$, up to the sign fixed by the convention for $E$. Therefore, the second identity generates two more diagrammatic reduction rules below.

\begin{figure}
    \centering
    \includegraphics[width=1\linewidth]{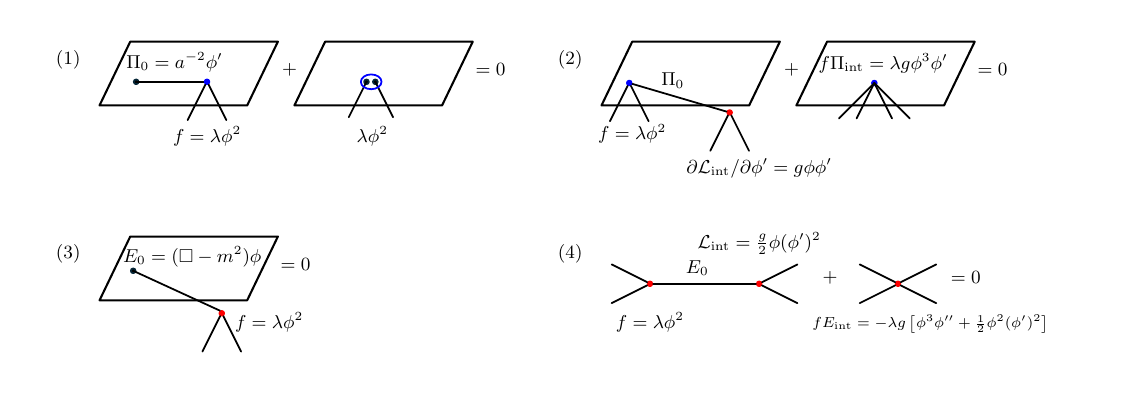}
    \caption{Diagrammatic reduction rules, illustrated with the single-field example $f=\lambda\phi^2$ and $\mathcal L_{\rm int}=g\phi(\phi')^2/2$. In this example $\Pi_0=a^{-2}\phi'$, $E_0=(\Box-m^2)\phi$, $\partial \mathcal L_{\rm int}/\partial\phi'=g\phi\phi'$, $\Pi_{\rm int}=g\phi\phi'$, and $fE_{\rm int}=-\lambda g\left(\phi^3\phi''+\frac12\phi^2(\phi')^2\right)$. Blue dots denote boundary vertices, while red dots denote bulk vertices. Parallelograms are the future boundary hypersurface at $\eta_0$, on which all late-time correlators are evaluated. Black dots denote boundary operators. 
The four panels respectively correspond to reduction rule (1)--(4). }
    \label{reductions}
\end{figure}

\begin{enumerate}
    \item [(3)]\textbf{$E_0$ on external-line}: An $  E_0 f  $ vertex connected to a boundary $  \Phi  $ operator through an external line in $  \mathcal{O} $ gives vanishing result, as expected from the free equation of motion:
    \begin{equation}
\label{eqn019}
        \sum_{b=\pm} ib \int d\eta \, a^4(\eta) \, (\square - m^2) K_b(k,\eta) \, \mathcal P_b^f(\eta) = 0.
    \end{equation}
    
    \item [(4)]\textbf{$E_0$ on internal-line}: When an $  E_0 f  $ vertex is connected to $  S_{\rm int}  $ through an internal line, it cancels the contribution from $  E_{\rm int} f  $ -- the EoM part  of the variation $\Delta S_{int}$ (with $  E_{\rm int} = E - E_0  $). For instance, in the case of connecting to a $  \Phi'  $ operator:
    \begin{equation}
\label{eqn020}
        \begin{aligned}
&\sum_{c=\pm} ic \int d\eta_1 \, a^4(\eta_1) \, \mathcal P_c^{\rm int}(\eta_1) \sum_{b=\pm} ib \int d\eta_2 \, a^4(\eta_2) \, \mathcal P_b^f(\eta_2) \, \partial_{\eta_1} (\square_{\eta_2} - m^2) D_{bc}(k,\eta_2,\eta_1) \\
=& -\sum_{b=\pm} ib \int d\eta \, a^4(\eta) \, \mathcal P_b^f(\eta) \, \partial_\eta \mathcal P_b^{\rm int}(\eta). \end{aligned}
    \end{equation}
Similar cancellations occur for connecting to $  \Phi  $ or its spatial derivatives. Summing over all these operations recovers exactly the contribution from $E_{\rm int} f$.
\end{enumerate} 
{In \cref{reductions}, we illustrate the four reduction rules using the concrete single-field example $f=\lambda\phi^2$ and $\mathcal L_{\rm int}=g\phi(\phi')^2/2$, making explicit how the general functions of $\Pi_0 f$, $\Pi_{\rm int}f$, $E_0f$, and $E_{\rm int}f$ are realised diagrammatically.}

Next, we examine the second-order Schwinger--Dyson equation under field variations:
\begin{equation} \label{2.3-eq021}
    \begin{split}
&\mathcal{O}\left[\frac{\delta}{i\delta J^+}\right]\Biggl(\sum_{b=\pm}ib\,\Delta^2S\left[\frac{\delta}{ib\delta J^b}\right] + \Biggl(\sum_{b=\pm}ib\int d^4x\,a^4(\eta)\,E\left[\frac{\delta}{ib\delta J^b}\right]f\left[\frac{\delta}{ib\delta J^b}\right]\Biggr)^2\Biggr|_{c}\Biggr) \\
\times&Z[J^\pm]\Big|_{J^\pm=0}=0,
\end{split}
\end{equation}
where the subscript $c$ retains only diagrams where the two EoM operators are connected through an internal line, and $  \Delta^2S  $ takes the form
\begin{equation}
\label{eqn022}
    \Delta^2S = \int d\eta \int d^3x \, a^4(\eta) \left[ \frac{\partial^2\mathcal{L}}{\partial\Phi^2} f^2 + 2\frac{\partial^2\mathcal{L}}{\partial\Phi\,\partial_\eta\Phi} f f' + \frac{\partial^2\mathcal{L}}{\partial(\partial_\eta\Phi)^2} (f')^2 \right][\Phi].
\end{equation}
This equation is derived by subtracting the perturbative corrections generated by the first-order Schwinger--Dyson identity from the original second-order variation, thereby isolating a nontrivial reduction relation among Feynman diagrams. In the free theory case, this relation manifests most clearly as follows:
\begin{itemize}
    \item Two $  E_0  $ vertices linked by an internal propagator reduce to a single effective vertex $  -\Delta^2S_0  $, where
    \begin{equation}
\label{eqn023}
        -\Delta^2S_0 = \int d\eta \int d^3x \left[ (\partial f[\Phi])^2 + m^2 f[\Phi]^2 \right].
    \end{equation}
    The explicit identity comes from
\end{itemize}
    \begin{equation}
\label{eqn024}
        \begin{split}
&\sum_{b,c=\pm} (-bc) \int d\eta_1 \, a^4(\eta_1) \int d\eta_2 \, a^4(\eta_2) \, \mathcal P_b^f(\eta_1) \mathcal P_c^f(\eta_2) (\square_{\eta_1}-m^2)(\square_{\eta_2}-m^2) D_{bc}(k,\eta_1,\eta_2) \\
=& \sum_{b=\pm} ib \int d\eta \, a^4(\eta) \Bigl[ a^{-2}(\eta) \bigl( -(\partial_\eta \mathcal P_b^f(\eta))^2 + k^2 \mathcal P_b^f(\eta)^2 \bigr) + m^2 \mathcal P_b^f(\eta)^2 \Bigr].
\end{split}
    \end{equation}
This identity shows that higher-order field-redefinition effects do not introduce new independent diagrammatic operations: after the first-order identities are subtracted, the diagram with two EoM operators connected by an internal propagator is reduced to a simpler diagram with a joint vertex. In the applications in \cref{CH3}, we will use this relation together with reduction rule (3) to connect off-shell EoM contributions after IBP with the diagrams generated by the second-order variation.

\subsection{Criteria for Non-vanishing Boundary Effects}\label{criteria}

In this subsection, we derive practical criteria for whether boundary terms can contribute to late-time correlators. The basic operation behind these criteria is the equal-time contraction between a boundary field and its conjugate momentum. For the $\Pi_0 f$ boundary terms considered in \cref{CH3}, this is precisely reduction rule (1): the $\Pi_0$ operator cuts the external line attached to a boundary $\Phi$ operator and leaves the remaining factor $f$ on the future boundary. For more general boundary vertices considered in \cref{CH4}, the same commutator logic can still be applied even when a direct field-redefinition interpretation is not manifest. The criteria below are therefore late-time survival criteria after the relevant boundary momentum-field contractions have been performed.\\

\textbf{Criterion 1: Strict boundary contraction}\\
A Feynman diagram can survive only if every boundary conjugate-momentum operator needed to avoid the branch-sum cancellation is contracted with a boundary field operator through the equal-time commutator. For the $\Pi_0 f$ vertices considered in \cref{CH3}, this reduces to the statement that every $\Pi_0$ operator must be directly connected to one of the boundary $\Phi$ operators in the correlator $\mathcal O$ through an external line.

\textbf{Criterion $1'$: No remaining field momentum of the massless scalar}\\
After all such boundary contractions have been performed, the remaining operators in the boundary vertex are evaluated at $\eta_0$. A remaining time derivative acting on an external massless field gives an extra positive power of $\eta_0$ and therefore decays in the late-time limit. Hence a non-vanishing boundary contribution requires that no such $\phi'$ remains after the boundary contractions. For the single-momentum vertices $\Pi_0 f$ used in \cref{CH3}, this condition is automatically satisfied. We shall see that this criterion rules out certain boundary contributions in boost-breaking theories. \\

After the boundary momentum contractions, the remaining structure consists of boundary-to-boundary contractions and lines connecting the boundary operators to bulk vertices. Each boundary-to-boundary contraction between two $\Phi_i$ fields leaves a factor $|u_{\nu_i}(k,\eta_0)|^2\sim O((-\eta_0)^{2\Delta_i})$. This motivates:\\ 

\textbf{Criterion 2: Absence of massive contraction}\\
For the diagram to remain finite in the $\eta_0\to0^-$ limit, all contractions after the boundary reductions must only involve operators with $\Delta\simeq0$, i.e. massless or light fields with negligible late-time suppression.\\ 

Furthermore, when the boundary operators left after the reduction include spatial derivatives, rotation symmetry requires them to appear in inner-product form, such as $(-H\eta_0)^2\nabla\Phi_1\cdot\nabla\Phi_2$. This leaves an explicit factor $(-H\eta_0)^2(\vec k_1\cdot\vec k_2)$ multiplying the reduced correlator. Since the undressed correlator cannot exhibit a compensating power-law divergence without signalling a tachyonic instability, the full contribution is suppressed at least as $O(\eta_0^2)$. We therefore conclude:\\ 

\textbf{Criterion 3: Absence of spatial derivatives}\\
Boundary terms contribute non-negligibly only if the operators left on the boundary after the reduction contain no spatial derivatives.\\


We will use these criteria in \cref{CH3} for the boundary terms generated by dS-invariant IBP relations, and in \cref{CH4} for the classification of more general boost-breaking boundary contributions.

~

\section{De Sitter Invariant Correlators with Boundary Contributions}\label{CH3}

In this section, we shall apply the connection between field redefinitions and boundary terms in the cosmological correlators with de Sitter (dS) isometries.
Our analysis closely follows the development of the dS bootstrap~\cite{Arkani-Hamed:2018kmz,Baumann:2019oyu,Baumann:2020dch,Baumann:2022jpr}, which computes cosmological correlators by translating dS isometries into momentum-space differential operators. In particular, here we study $\phi\phi\sigma$-type interactions through IBP and field redefinitions. 
We show that IBP relates derivative and non-derivative couplings by generating both boundary terms and EoM terms; when an EoM operator acts on an internal massive propagator, it collapses the exchange line and produces a contact diagram that may lead to late-time divergences. The same relations are reproduced by field redefinitions, with a diagrammatic dictionary \cref{fig:full_dictionary} linking to \cref{reduction}. 

Extending to higher-derivative couplings yields recursion relations that reduce exchange diagrams to a basis plus contact contributions---the bulk counterpart of the bootstrap recursion relations of~\cite{Arkani-Hamed:2018kmz}. The full bootstrap computation of all results is provided in \cref{Bootstrap details}, where seed functions are related to the contact sources $\hat C_n$ and exchange solutions $\hat F_n$; the light-mass regime, where neither IBP nor field redefinitions can fully resolve late-time divergences, is treated in \cref{Light mass}.

\subsection{IBP Approach}\label{IBP}
We start with the simple example of two dS-invariant interactions with lowest derivatives: $(\nabla \phi)^2 \,\sigma$ and $\phi^2 \sigma$. 
This example has been used in the dS bootstrap \cite{Arkani-Hamed:2015bza,Arkani-Hamed:2018kmz} to demonstrate that cosmological correlators are fixed by conformal Ward identities as long as the dS isometries are respected. Here, more explicitly, we apply IBP to build the precise connection and highlight the roles of terms with late-time divergences.
We can apply the IBP twice so that the two-derivative coupling $(\nabla\phi)^2\sigma$ is related to the non-derivative coupling $\phi^2\sigma$:
\begin{equation}
    \label{eq:IBP_cubic_general-eq025}
    \begin{split}
         \int d^4x \,a^4(\eta)\,(\nabla \phi)^2 \,\sigma 
    =& \left.\frac{1}{2} \int d^3x\, a^2(\eta)\, \left[ (\partial_\eta \phi^2) \,\sigma - \phi^2 \,\partial_\eta \sigma \right]\, \right|_{\eta=\eta_0}
    - \int d^4x \,a^4(\eta) \,\phi\square \phi \,\sigma 
    \\ +&\frac{1}{2} \,\int d^4x \,a^4(\eta)\,\phi^2 \,(\square-m^2) \sigma + \frac{m^2}{2} \,\int d^4 x \,a^4(\eta)\,\phi^2 \, \sigma .
    \end{split}
\end{equation}
For convenience, we introduce the following shorthand notations:
\begin{equation}
\label{eqn026}
    \mathcal{B}_1=\left.\int d^3x\, a^2(\eta)\,\phi^2 \,\partial_\eta \sigma \right|_{\eta=\eta_0} , \quad \mathcal{B}_2=\left.\int d^3x\, a^2(\eta)\, (\partial_\eta \phi^2) \,\sigma \right|_{\eta=\eta_0}.
\end{equation}
The difference between the two couplings $\phi^2\sigma$ and $(\nabla\phi)^2\sigma$ is captured by the non-trivial boundary terms evaluated on the late-time $\eta_0$ cut-off accompanied by EoM terms which may give rise to off-shell contributions when the vertices are connected to internal lines. This observation motivates the computational strategy adopted in this section. Instead of solving the boundary differential equations directly---a procedure we rigorously perform in \cref{Bootstrap details} for validation---we employ the IBP relations to compute the correlators with $\phi^2\sigma$ interactions. 
This approach offers two advantages. First, it simplifies the computation by expressing correlators from non-derivative couplings in terms of those from derivative couplings, whose results are already established, together with the associated boundary contributions and contact contributions. Second, it attributes the late-time divergences to the contact diagrams emerging through the IBP procedure.

\begin{figure}
    \centering
    \includegraphics[width=0.95\textwidth]{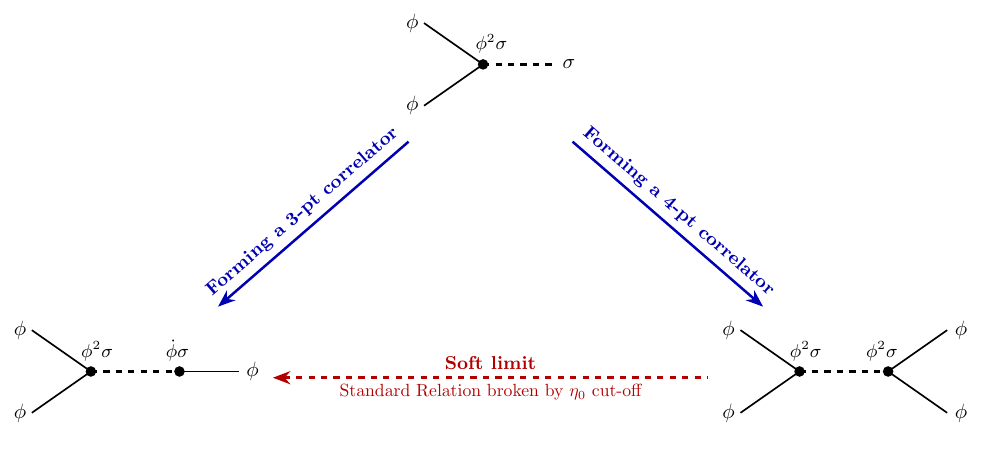}
    \caption{The diagrammatic triangle relating the three cases. \textbf{Top:} The correlator $\langle \phi\phi\sigma \rangle'$ generated by interaction $\phi^2\sigma$. \textbf{Left:} The 3-pt correlator $\langle \phi\phi\phi \rangle'$ formed by connecting the $\sigma$ leg to $\phi$ via a two-point mixing $\dot{\phi}\sigma$. \textbf{Right:} The 4-pt correlator $\langle \phi\phi\phi\phi \rangle'$ formed by connecting two cubic vertices with a massive $\sigma$-propagator. Bottom: The standard relation between the soft limit of the 4-pt correlator and the 3-pt correlator is modified due to the late-time cutoff.}
    \label{fig:triangle_relation}
\end{figure}

To demonstrate this method explicitly, we analyse three specific cases, whose interconnections are visualised in \cref{fig:triangle_relation}. In \cref{sec:building_block}, we study a simple case $\langle \phi\phi\sigma \rangle'$. Although this object is not directly observable in a quasi-single-field scenario, it serves as the simplest example. In \cref{sec:3pt_observable}, we extend this case to the observable sector by attaching a mixing vertex, yielding the standard bispectrum $\langle \phi\phi\phi \rangle'$. In \cref{sec:4pt_exchange}, we construct the 4-pt exchange diagram by gluing two cubic vertices via a massive propagator and then discuss the modified consistency relation in the soft limit in \cref{sec:4pt_to_3pt}. 

\subsubsection{\texorpdfstring{A Simple Case: 3-pt Correlator $\langle \phi\phi\sigma \rangle'$ from Contact Diagram}{A Simple Case: 3-pt Correlator phi phi sigma from Contact Diagram}}
\label{sec:building_block}

We begin with a simple case: the mixed 3-pt correlator $\braket{\phi\phi\sigma}^\prime$ with the interaction $\frac{1}{2}g_3 \phi^2\sigma$. Since both $\phi$ and $\sigma$ propagators serve as external lines in this contact diagram, the on-shell terms have no contribution. Utilising the IBP relation~\cref{eq:IBP_cubic_general-eq025}, we can translate the computation of the $\frac{1}{2}g_3 \phi^2\sigma$ interaction into that of a derivative interaction $\frac{1}{2}\lambda_3 (\nabla\phi)^2\sigma$ and boundary terms, with the coupling mapping $g_3 = \lambda_3 m^2$. The full relation is given by:
\begin{equation}
	\label{eq:phiphisigma_decomp-eq027}
	\langle\phi\phi\sigma\rangle'_{g_3\phi^2\sigma} = 2\langle\phi\phi\sigma\rangle'_{\lambda_3(\nabla\phi)^2\sigma} + \langle\phi\phi\sigma\rangle'_{\mathcal{B}_1} - \langle\phi\phi\sigma\rangle'_{\mathcal{B}_2}\,,
\end{equation}
where the terms with subscripts $\mathcal{B}_1$ and $\mathcal{B}_2$ represent the boundary contributions arising from $\phi^2 \partial_\eta \sigma$ and $(\partial_\eta \phi^2) \sigma$, respectively. On the RHS in \cref{eq:phiphisigma_decomp-eq027}, the first term $\langle\phi\phi\sigma\rangle'_{\lambda_3}$ decays as $\eta_0 \to 0$, as does the contribution from the second boundary term $\langle\phi\phi\sigma\rangle'_{\mathcal{B}_2}$. Consequently, the total result is entirely dominated by the remaining boundary term $\langle\phi\phi\sigma\rangle'_{\mathcal{B}_1}$. Evaluating this term yields the final expression for the correlator
\begin{equation}
\label{eqn028}
	\langle\phi\phi\sigma\rangle'_{g_3\phi^2\sigma} \simeq \langle\phi\phi\sigma\rangle'_{\mathcal{B}_1} = \frac{g_3 H^4}{4m^2 k_1^3 k_2^3}\,.
\end{equation}
Therefore, this mixed 3-pt function takes a local form.
 One may wonder why this correlator survives in the $\eta_0\rightarrow 0$ limit even though it contains a massive external leg. In the direct bulk integration, we see that the late-time divergence cancels the decaying behaviour of a massive external leg.

\subsubsection{\texorpdfstring{Inflationary Bispectrum $\langle \phi\phi\phi \rangle'$ from Exchange Diagram}{Inflationary Bispectrum phi phi phi from Exchange Diagram}}
\label{sec:3pt_observable}

We now turn to a related observable: the 3-pt correlator of the massless inflaton. As illustrated in the left panel of \cref{fig:triangle_relation}, this correlator is generated by the cubic interaction $\frac{1}{2}g_3 \phi^2\sigma$, and the massive field $\sigma$ is converted into an inflaton $\phi$ via the quadratic coupling $\lambda_2 \dot{\phi}\sigma$.

Applying the IBP relation in \cref{eq:IBP_cubic_general-eq025} to the cubic interaction vertex while keeping the other vertex unchanged, we can simplify the calculation in a similar way. The crucial difference from the previous subsection is that $\sigma$ now propagates as an internal line, and the corresponding propagator satisfies the inhomogeneous equation $(\square_{\eta}-m^2)D_{bc}(k,\eta,\eta')=ib(H\eta)^4\delta(\eta-\eta^\prime)\delta_{bc}$. Consequently, when the $\Box-m^2$ operator from the IBP relation acts on the internal line, a new contact diagram is generated, which is absent in the previous calculation for $\langle \phi\phi\sigma \rangle'$. In summary, the full decomposition involves a bulk exchange part, a contact part, and two boundary terms:
\begin{equation}
	\label{eq:phiphiphi_decomp-eq029}
	\langle\phi\phi\phi\rangle'_{g_3\phi^2\sigma \times \lambda_2\dot{\phi}\sigma} = 2\langle\phi\phi\phi\rangle'_{\lambda_3(\nabla\phi)^2\sigma \times \lambda_2\dot{\phi}\sigma} + \langle\phi\phi\phi\rangle'_{\lambda_2\lambda_3\phi^2\dot{\phi}} + \langle\phi\phi\phi\rangle'_{\mathcal{B}_1 \times \lambda_2\dot{\phi}\sigma} - \langle\phi\phi\phi\rangle'_{\mathcal{B}_2 \times \lambda_2\dot{\phi}\sigma}\,.
\end{equation}
In the late-time limit $\eta_0 \to 0$, the last two terms on the RHS of \cref{eq:phiphiphi_decomp-eq029} can be discarded. Specifically, the third term $\langle\phi\phi\phi\rangle'_{\mathcal{B}_1 \times \lambda_2\dot{\phi}\sigma}$, arising from the boundary operator $\phi^2 \partial_\eta \sigma$, vanishes strictly by the branch sum. The fourth term $\langle\phi\phi\phi\rangle'_{\mathcal{B}_2 \times \lambda_2\dot{\phi}\sigma}$, originating from the boundary term $(\partial_\eta \phi^2) \sigma$, decays as $\eta_0 \to 0$. Consequently, the physical result is determined solely by the first two terms: the exchange diagram with derivative coupling and the contact diagram.

The first term, $\langle \phi\phi\phi \rangle'_{\lambda_3(\nabla\phi)^2\sigma\times \lambda_2\dot{\phi}\sigma}$, corresponds to the exchange diagram with derivative interactions $\frac{1}{2}\lambda_3 (\nabla\phi)^2\sigma$. This diagram is free of late-time divergence, and its result has been well-studied using bootstrap methods, specifically via the weight-shifting operator acting on a conformal seed integral~\cite{Pimentel:2022fsc}. We provide the explicit expression for this term in \cref{Bootstrap details} (see~\cref{eq:3pthomsol-eq128} and \cref{eq:3ptinhsollamda3-eq130}).

The second term, $\langle \phi\phi\phi \rangle'_{\lambda_2\lambda_3\phi^2\dot{\phi}}$, as mentioned, comes from the on-shell equation of motion
\begin{equation}
\label{eqn030}
(\square_{\eta}-m^2)D_{bc}^{\sigma}(k,\eta,\eta')=ib(H\eta)^4\delta(\eta-\eta^\prime)\delta_{bc}.
\end{equation} 
The Dirac-$\delta$ function effectively ``erases'' the internal $\sigma$-propagator and ``glues'' the two cubic vertices together, collapsing the exchange diagram into a contact diagram with coupling $\lambda_2\lambda_3\phi^2\dot{\phi}$. Explicit evaluation yields
\begin{equation}
\label{eq:3pt_contact_res-eq031}
\begin{aligned}
    \langle \phi\phi\phi \rangle'_{\lambda_2\lambda_3\phi^2\dot{\phi}} 
    =& \frac{\lambda_2 \lambda_3 H^3}{4 k_1^3 k_2^3 k_3^3} \left[  \big( \gamma_E - 1 + \log(-k_T \eta_0) \big)\sum_{i=1}^3 k_i^3 - k_T  \sum_{i<j} k_i k_j + 4 k_1 k_2 k_3 \right]\,.
\end{aligned}
\end{equation}

\begin{figure}[htbp]
    \centering
    \includegraphics[width=0.9\textwidth]{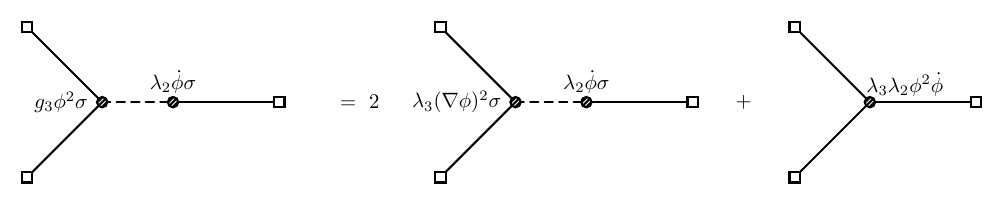}
    \caption{The diagrammatic relation for the 3-pt correlator. The non-derivative exchange diagram is equivalent to twice the derivative exchange diagram plus a contact diagram, contributed by the EoM term $\phi^2(\Box-m^2)\sigma$.}
    \label{fig:3pt_relation}
\end{figure}

Combining these results, we arrive at the final result for the bispectrum with the coupling $g_3 \phi^2\sigma$ and $\lambda_2 \dot{\phi}\sigma$:
\begin{equation}\label{eq:3ptrelation-eq032}
    \langle \phi\phi\phi \rangle'_{g_3\phi^2\sigma\times\lambda_2\dot{\phi}\sigma} = 2 \langle \phi\phi\phi \rangle'_{\lambda_3(\nabla\phi)^2\sigma\times\lambda_2\dot{\phi}\sigma} + \langle \phi\phi\phi \rangle'_{\lambda_2\lambda_3\phi^2\dot{\phi}} \,,
\end{equation}
where we have dropped the boundary terms that vanish at late times. The diagrammatic representation of this relation is illustrated in \cref{fig:3pt_relation}.

It is important to note that the late-time divergence (manifested as $\log \eta_0$) is entirely captured by the contact diagram $\langle \phi\phi\phi \rangle'_{\lambda_2\lambda_3\phi^2\dot{\phi}}$.\footnote{These secular growing terms are infrared (IR) effects that should be resummed non-perturbatively, as shown in \cite{Cespedes:2023aal,Cespedes:2025ple}. In this work, we focus on the leading IR contributions from massive exchanges in the perturbative regime.} This structural feature---where $(\Box-m^2)\sigma$ operator collapses the exchange diagram and isolates the late-time behaviour---is not unique to the 3-pt function. As we will see in the next subsection, the same mechanism still applies for the 4-pt correlator.

\subsubsection{\texorpdfstring{de Sitter Trispectrum $\langle \phi\phi\phi\phi \rangle'$ from Exchange Diagram}{de Sitter Trispectrum phi phi phi phi from Exchange Diagram}}
\label{sec:4pt_exchange}

Finally, we study the 4-pt exchange diagram constructed by connecting two cubic vertices with an internal $\sigma$-line. We consider the case where both vertices are of the non-derivative interaction $\frac{1}{2}g_3 \phi^2\sigma$. Applying the IBP relation \cref{eq:IBP_cubic_general-eq025} to both vertices generates several classes of terms.

First, let us address the boundary contributions. While most boundary terms generated by the double IBP procedure either vanish strictly or decay at late times, two specific contributions remain non-zero. One arises from the boundary-boundary sector ($\langle \phi\phi\phi\phi \rangle'_{\mathcal{B}_1\times \mathcal{B}_2}$), and the other from the sector involving a bulk interaction at one vertex and a boundary term at the other ($\langle \phi\phi\phi\phi \rangle'_{g_3\phi^2\sigma\times \mathcal{B}_2}$). Explicit evaluation gives:
\begin{equation}
\label{eq:B1-eq033}
\begin{aligned}
    \langle \phi\phi\phi\phi \rangle'_{\mathcal{B}_1\times \mathcal{B}_2}
    &\propto \int_{\partial\Sigma} \phi^2 (\partial_\eta \sigma) \times \int_{\partial\Sigma} (\partial_\eta \phi^2)\sigma
     = - \frac{\lambda_3^2 H^6 }{8 k_1^3 k_2^3 k_3^3 k_4^3} \sum_{i=1}^4 k_i^3 \,,\\
    \langle \phi\phi\phi\phi \rangle'_{g_3\phi^2\sigma\times \mathcal{B}_2}
    &\propto \int_{\Sigma} \phi^2 (m^2\sigma) \times \int_{\partial\Sigma} (\partial_\eta \phi^2)\sigma
     = - \frac{\lambda_3^2 H^6 }{8 k_1^3 k_2^3 k_3^3 k_4^3} \sum_{i=1}^4 k_i^3\,.
\end{aligned}
\end{equation}
Remarkably, these two boundary contributions cancel each other precisely. Since they appear with opposite signs in the IBP expansion, their identical explicit forms lead to a vanishing net contribution: $\langle \phi\phi\phi\phi \rangle'_{\mathcal{B}_1\times \mathcal{B}_2}  -  \langle \phi\phi\phi\phi \rangle'_{g_3\phi^2\sigma\times \mathcal{B}_2}= 0$. This cancellation is not accidental; it follows directly from the analysis in \cref{sec:building_block}, where we established that the bulk integral of the interaction $\int_{\Sigma} \phi^2 (m^2\sigma)$ is effectively equivalent to the boundary term $\int_{\partial\Sigma} \phi^2 (\partial_\eta \sigma)$. Consequently, these boundary terms cancel, leaving no net boundary contribution to the final correlator in this case\footnote{In light mass regime, this cancellation fails, and the decaying boundary contribution $\langle \phi\phi\phi\phi \rangle'_{\mathcal{B}_2\times \mathcal{B}_2}$ can survive, see \cref{Light mass} for details.}. We note, however, that for exchange diagrams where IBP is applied to only one vertex (e.g., the 4-pt exchange with one $\lambda_3(\nabla\phi)^2\sigma$ vertex and one $g_3\phi^2\sigma$ vertex), this cancellation does not occur, leaving a non-zero boundary contribution.

We can thus focus entirely on the bulk-bulk contributions, which involve bulk integrals at both vertices. The double IBP procedure relates the original derivative interactions to a structure governed by the operator product $\Box_1 \Box_2$ acting on the bulk-to-bulk propagator. Using the identity $\Box = m^2 + (\Box - m^2)$, we can decompose this interaction kernel into three terms:
\begin{equation}\label{eq:boxdecompose-eq034}
    \Box_1 \Box_2 = m^4 + m^2(\Box_1 - m^2) + \Box_1(\Box_2 - m^2) \,.
\end{equation}
This decomposition has a clear physical interpretation when acting on the bulk-to-bulk propagator $D_{ab}^\sigma$, which satisfies $(\Box - m^2)D^\sigma_{bc}(k,\eta,\eta^{\prime}) = ib(H\eta)^4\delta(\eta-\eta^{\prime})\delta_{bc}$. The first term, $m^4$, leaves the propagator intact and corresponds to the exchange diagram with two non-derivative vertices $g_3 \phi^2\sigma$.
The other two terms involve factors of $(\Box - m^2)$, which generate Dirac-$\delta$ functions that remove the internal line, effectively merging the two vertices into a single vertex.
Specifically, the second term contains an $m^2$ factor, yielding a contact diagram with a $\phi^4$ interaction.
The third term contains a $\Box_1$ operator acting on the merged vertex, leading to a higher-derivative contact interaction $(\partial\phi^2)^2$.
By rearranging these terms, we can express the late-time divergent non-derivative exchange diagram $\langle \phi\phi\phi\phi \rangle'_{g_3\phi^2\sigma\times g_3\phi^2\sigma}$ as the late-time finite derivative exchange diagram $\langle \phi\phi\phi\phi \rangle'_{\lambda_3(\nabla\phi)^2\sigma\times\lambda_3(\nabla\phi)^2\sigma}$ plus the contact diagrams:
\begin{equation}
    \label{eq:4pt_decomposition-eq035}
    \langle \phi\phi\phi\phi \rangle'_{g_3\phi^2\sigma\times g_3\phi^2\sigma} = 4 \langle \phi\phi\phi\phi \rangle'_{\lambda_3(\nabla\phi)^2\sigma\times\lambda_3(\nabla\phi)^2\sigma}+ \langle \phi\phi\phi\phi \rangle'_{m^2\lambda_3^2\phi^4} - \langle \phi\phi\phi\phi \rangle'_{\lambda_3^2(\nabla\phi^2)^2}  \,,
\end{equation}
where the subscript $(\nabla\phi^2)^2$ denotes the contracted form $\nabla_{\mu}(\phi_1\phi_2)\nabla^{\mu}(\phi_3\phi_4)$. We have also used the coupling relation $g_3 = \lambda_3 m^2$, and the factor of $4$ comes from the IBP coefficients. 
For the tree-level massive-exchange example considered here, \cref{eq:4pt_decomposition-eq035} shows that the logarithmic late-time divergence is contributed only by the contact diagram $ \langle \phi\phi\phi\phi \rangle'_{m^2\lambda_3^2\phi^4}$.

The first term in \cref{eq:4pt_decomposition-eq035}, $\langle\phi\phi\phi\phi\rangle_{\lambda_3(\nabla\phi)^2\sigma\times\lambda_3(\nabla\phi)^2\sigma}^{\prime}$, corresponds to the exchange diagram mediated by the derivative interaction $\lambda_3(\nabla\phi)^2\sigma$, which is free of the late-time divergence. Its analytic structure has been extensively studied~\cite{Arkani-Hamed:2018kmz} and we provide the explicit expression in \cref{Bootstrap details} (see \cref{inh11-eq151}). 
The explicit expressions for the two contact contributions in \cref{eq:4pt_decomposition-eq035} are given as follows. 
\begin{equation}
\label{eq:phi4-eq036}
\begin{aligned}
    \langle \phi\phi\phi\phi \rangle'_{m^2\lambda_3^2\phi^4}
    &= -\frac{m^2 \lambda_3^2 H^4}{8 \prod k_i^3} \Biggl\{\sum_i k_i^3 \left( \frac{1}{3} + \gamma_E + \log (-k_T \eta_0) \right) \\
    &+  3k_T \sum_{i<j} k_i k_j - \frac{2}{3} k_T^3 - 3 \frac{k_1 k_2 k_3 k_4}{k_T}   \Biggr\} \,,\\
    \langle \phi\phi\phi\phi \rangle'_{\lambda_3^2(\nabla\phi^2)^2}
    &=  \frac{\lambda_3^2 H^6}{8 \prod k_i^3}\sum_i k_i^3 \,.
\end{aligned}
\end{equation}
The sum $\sum k_i^3$ runs over all four external momenta, and $k_T \equiv \sum k_i$. 

\begin{figure}[H]
    \centering
    \includegraphics[width=0.95\textwidth]{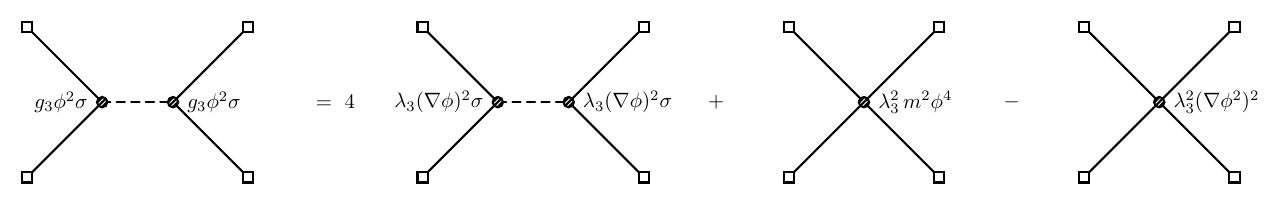}
    \caption{The diagrammatic relation for the 4-pt correlator. The late-time divergence of exchange diagram with coupling $g_3\phi^2\sigma$ has been fully isolated into the contact diagram with interaction $\lambda_3^2 m^2\phi^4$, which again comes from the contribution of EoM terms.}
    \label{fig:4pt_relation}
\end{figure}

The summation of these three terms according to the relation \cref{eq:4pt_decomposition-eq035} gives the final result for the late-time divergent four-point exchange diagram $\langle \phi\phi\phi\phi \rangle'_{g_3\phi^2\sigma\times g_3\phi^2\sigma}$. The diagrammatic representation is illustrated in \cref{fig:4pt_relation}. By adopting this IBP approach, not only the difficulty in direct time integration is circumvented but also the analytic structure becomes more transparent. 
The same relation has been identified in \cite{Cespedes:2025ple}.

\subsubsection{Inflationary Bispectrum from de Sitter Trispectrum}\label{sec:4pt_to_3pt}
It has been shown in the previous literature~\cite{Arkani-Hamed:2018kmz,Arkani-Hamed:2015bza,Qin:2023ejc} that the three-point correlators are soft limits of four-point correlators. In our context, for convenience, we will focus our discussion on the $s$-channel contributions, denoted by the subscript $\text{s-ch}$. The first example consists of the 4-pt late-time finite exchange diagram involving one derivative vertex $\lambda_3(\nabla\phi)^2\sigma$ and one non-derivative vertex $g_3\phi^2\sigma$. The relation is\footnote{Here we set $\lambda_2=g_3$ so that the relation holds including the overall coupling factors.}
\begin{equation}\label{eq:structural_relation-eq037}
    \lim_{k_4 \to 0} k_3 \partial_{k_3} \left(-2k_3^3\, 2k_4^3\,\langle \phi\phi\phi\phi \rangle'_{\lambda_3(\nabla\phi)^2 \sigma\times g_3\phi^2\sigma,\,\text{s-ch}}\right) = 2k_3^3\langle \phi\phi\phi \rangle'_{\lambda_3(\nabla\phi)^2\sigma\times \lambda_2\dot{\phi}\sigma,\,\text{s-ch}} \,.
\end{equation}
The derivation of this formula is based on the analytic form of the bulk-to-boundary propagator. We do not have to keep track of the $\eta_0$ dependence in this example since there is no late-time divergence, i.e.
\begin{equation}
\label{eqn038}
    K^\phi_{\pm}(k,\eta) = \frac{H^2}{2k^3}(1 \mp ik\eta)(1 \pm ik\eta_0)e^{\pm ik\eta \mp ik\eta_0} \sim\frac{H^2}{2k^3}(1\mp ik\eta)e^{\pm ik\eta}\,.
\end{equation}
Then we can rescale the propagator into $\hat{K}^\phi(k,\eta) \equiv 2k^3 K^\phi(k, \eta)$ such that $\hat{K}$ depends solely on $k\eta$. As a result, we may exchange $k$-derivatives with time-derivatives acting on this rescaled propagator in the following way
\begin{equation}
\label{eqn039}
    k \partial_{k} \hat{K}^\phi(k\eta) = \eta \partial_\eta \hat{K}^\phi(k \eta) \,.
\end{equation}
The action of a $\eta\partial_\eta$ operator in the Schwinger--Keldysh integral in front of a rescaled bulk-to-boundary propagator effectively converts a $\phi$ contained in the coupling into a $\dot\phi$. By swapping the time derivative with the momentum derivative, we can move the $k\partial_k$ operator outside the integral, i.e., it acts on the correlator as a whole. Therefore, if it is acted on the external leg carrying momentum $k_3$, the $\phi^2\sigma$ vertex can be related to a $\dot{\phi}\sigma$ vertex in the soft $k_4$ limit. 

To verify this relation, we present the result for the exchange diagram $\langle \phi\phi\phi\phi \rangle'_{\lambda_3(\nabla\phi)^2 \sigma\times g_3\phi^2\sigma} $. Using the IBP relation to one vertex, this correlator can be decomposed into an exchange part, a contact part, and a non-vanishing boundary term (arising from $ \langle \phi\phi\phi\phi \rangle'_{g_3\phi^2\sigma\times \mathcal{B}_2}$ in \cref{eq:B1-eq033}):
\begin{equation}
    \label{eq:4pt_mixed_result-eq040}  
    \langle \phi\phi\phi\phi \rangle'_{\lambda_3(\nabla\phi)^2 \sigma\times g_3\phi^2\sigma} = 2 \langle \phi\phi\phi\phi \rangle'_{\lambda_3(\nabla\phi)^2 \sigma \times \lambda_3(\nabla\phi)^2 \sigma} - \frac{1}{2} \langle \phi\phi\phi\phi \rangle'_{\lambda_3^2(\nabla\phi^2)^2} - \frac{\lambda_3^2 H^6(k_1^3 + k_2^3)}{16 k_1^3 k_2^3 k_3^3 k_4^3}  \,.
\end{equation}
Since all terms on the RHS of \cref{eq:4pt_mixed_result-eq040} are independent of $\eta_0$, the correlator $\langle \phi\phi\phi\phi \rangle'_{\lambda_3(\nabla\phi)^2 \sigma\times g_3\phi^2\sigma} $ is manifestly free of late-time divergence. Explicitly substituting this result and the 3-pt correlator $\langle \phi\phi\phi \rangle'_{\lambda_3(\nabla\phi)^2\sigma\times \lambda_2\dot{\phi}\sigma}$ (see \cref{eq:3ptinhsollamda3-eq130} for the explicit forms) into \cref{eq:structural_relation-eq037} confirms that the derivative relation holds. 

In contrast, subtleties arise when considering diagrams with late-time divergence, such as the contact diagram $\langle \phi\phi\phi\phi \rangle'_{\phi^4}$ and exchange diagram $\langle \phi\phi\phi\phi \rangle'_{g_3\phi^2\sigma\times g_3\phi^2\sigma}$. In these cases, the $\eta_0$-dependent factors in the bulk-to-boundary propagator cannot be neglected, since it will contribute to a finite term in the final result. Consequently, the rescaled propagator $\hat{K}^{\phi}$ depends explicitly on $k\eta_0$, breaking the equivalence between $k_3 \partial_{k_3}$ and $\eta \partial_\eta$.

Compared to the previous simple relation in \cref{eq:structural_relation-eq037}, the new relation satisfied by the correlators with late-time divergence includes explicit correction terms:
\begin{equation}
\begin{aligned}
    \label{eq:modified_relation_late_time-eq041}
    &\lim_{k_4 \to 0} k_3 \partial_{k_3} \left( -2k_3^3\, 2k_4^3\langle \phi\phi\phi\phi \rangle'_{\lambda_3^2m^2\phi^4,\,\text{s-ch}} \right) \\
    =& 2 k_3^3 \langle \phi\phi\phi \rangle'_{\lambda_2\lambda_3\phi^2\dot{\phi},\,\text{s-ch}} - \sum_{b=\pm}[k_3\partial_{k_3}C_b(k_3\eta_0)]\times I_{b,\,\lambda_2\lambda_3\phi^3,\,\text{s-ch}} \,,\\
    &\lim_{k_4 \to 0} k_3 \partial_{k_3} \left( -2k_3^3\, 2k_4^3\langle \phi\phi\phi\phi \rangle'_{g_3\phi^2\sigma\times g_3\phi^2\sigma,\,\text{s-ch}} \right) \\
    =& 2 k_3^3 \langle \phi\phi\phi \rangle'_{g_3\phi^2\sigma\times \lambda_2\dot{\phi}\sigma,\,\text{s-ch}} - \sum_{b=\pm}[k_3\partial_{k_3}C_b(k_3\eta_0)]\times I_{b,\,g_3\phi^2\sigma\times \lambda_2\dot\phi\sigma,\,\text{s-ch}} \,.
\end{aligned}
\end{equation}
Here $C_b(k_3\eta_0)=(1+ibk_3\eta_0)e^{-ibk_3\eta_0}$ encapsulates the $k_3\eta_0$-dependent factors arising from the bulk-to-boundary propagator, which can be factored out of the time integration. The term $I_b$ denotes the remaining Schwinger--Keldysh time integral, with the index $b = \pm$ labelling the branches of the in-in contour. Substituting the contact correlators \cref{eq:phi4-eq036,eq:3pt_contact_res-eq031} and the exchange correlators \cref{eq:4pt_decomposition-eq035,eq:3ptrelation-eq032} into \cref{eq:modified_relation_late_time-eq041} confirms that the modified derivative relation holds.

\subsection{Perspective from Field Redefinition}\label{field-redef}
In the previous subsection, we employed the IBP method to relate different interaction vertices. This procedure inevitably generates a lot of boundary terms. Naturally, one crucial question arises: which boundary terms are necessary, and which can be safely discarded? To address this question, one can refer to the criteria for boundary terms discussed in \cref{criteria}.\footnote{As stated in \cref{criteria}, a diagram contains boundary terms that do not vanish only if all the boundary conjugate momentum operators $\Pi_0$ are linked by external lines to corresponding field operators on the boundary, such that the original diagram cancels the variation of the correlator under field redefinitions.} Field redefinitions offer an intuitive perspective. If a specific boundary term generated by IBP corresponds to an induced correlator from the corresponding field redefinitions, then this boundary term shall contribute to the correlators of interest. Furthermore, this approach allows us to independently reproduce the relations derived from IBP.

To re-illustrate our examples from this perspective, we start with the Lagrangian generating the exchange diagram with the derivative coupling $\frac{1}{2}\lambda_3(\nabla\phi)^2\sigma$ and $\lambda_2\dot{\phi}\sigma$.  
\begin{equation}
    \mathcal{L} = -\frac{1}{2}(\nabla\phi)^2 - \frac{1}{2}(\nabla\sigma)^2 - \frac{1}{2}m^2\sigma^2 + \frac{\lambda_3}{2}(\nabla\phi)^2\sigma + \lambda_2\dot{\phi}\sigma \,.
    \label{eq:L_original-eq042}
\end{equation}
To relate the derivative interaction $(\nabla\phi)^2\sigma$ to the non-derivative coupling $\phi^2\sigma$, we need to perform two field redefinitions in sequence. The first one is 
\begin{equation}
    \phi = \tilde{\phi} + \frac{\lambda_3}{2}\tilde{\phi}\sigma \,.
    \label{eq:field_redef_1-eq043}
\end{equation}
Under this transformation, the kinetic term of $\phi$ generates a derivative interaction. Expanding the Lagrangian in terms of the new field $\tilde{\phi}$, we obtain
\begin{equation}
\label{eq:L_transformed_1-eq044}
\begin{aligned}
    \mathcal{L} =&-\frac{1}{2}(\nabla\tilde{\phi})^2 - \frac{1}{2}(\nabla\sigma)^2 - \frac{1}{2}m^2\sigma^2 - \frac{\lambda_3}{4} \nabla\sigma \nabla(\tilde{\phi}^2) + \lambda_2\dot{\tilde{\phi}}\sigma \\
    +& \frac{1}{2}\lambda_2\lambda_3 \left( \dot{\tilde{\phi}}\sigma^2 + \tilde{\phi}\dot{\sigma}\sigma \right) + \dots \,,
\end{aligned}
\end{equation}
where the ellipsis denotes terms of higher order in the coupling constants. The masses of $\phi$ and $\sigma$ remain unchanged, and the last two terms do not contribute to the correlator at the leading order. We note that the original derivative coupling $(\nabla\phi)^2\sigma$ has been replaced by a term involving $\nabla\sigma \nabla(\tilde{\phi}^2)$, which allows for further simplification.

This change of field variables leads to a transformation of the correlation functions. For the 3-pt correlator $\langle \phi \phi \sigma \rangle$, the redefinition yields
\begin{equation}
    \langle \phi(x_1) \phi(x_2) \sigma(x_3) \rangle = \langle \tilde{\phi}(x_1) \tilde{\phi}(x_2) \sigma(x_3) \rangle+ \frac{\lambda_3}{2}\langle \tilde{\phi}(x_1)\sigma(x_1) \tilde{\phi}(x_2) \sigma(x_3) \rangle + \text{perms} + O(\lambda_3^2) \,.
    \label{eq:phiphisigma_redef-eq045}
\end{equation}

Similarly, the bispectrum of $\phi$ transforms as
\begin{equation}
    \langle \phi(x_1)\phi(x_2)\phi(x_3) \rangle = \langle \tilde{\phi}(x_1)\tilde{\phi}(x_2)\tilde{\phi}(x_3) \rangle + \frac{\lambda_3}{2}\langle \tilde{\phi}(x_1)\tilde{\phi}(x_2)\tilde{\phi}(x_3)\sigma(x_3) \rangle + \text{perms} + O(\lambda_3^2) \,.
    \label{eq:bispectrum_redef_1-eq046}
\end{equation}
And the trispectrum transforms as
\begin{equation}
\label{eq:trispectrum_redef_1-eq047}
\begin{aligned}
    \langle \phi(x_1)\phi(x_2)\phi(x_3)\phi(x_4) \rangle =& \langle \tilde{\phi}(x_1)\tilde{\phi}(x_2)\tilde{\phi}(x_3)\tilde{\phi}(x_4) \rangle  + \frac{\lambda_3}{2} \langle \tilde{\phi}(x_1)\tilde{\phi}(x_2)\tilde{\phi}(x_3)\tilde{\phi}(x_4)\sigma(x_4) \rangle  \\
      +& \frac{\lambda_3^2 }{4}\langle \tilde{\phi}(x_1)\tilde{\phi}(x_2)\tilde{\phi}(x_3)\sigma(x_3)\tilde{\phi}(x_4)\sigma(x_4) \rangle + \text{perms} + O(\lambda_3^3) \,.
\end{aligned}
\end{equation}

Subsequently, to go to the non-derivative interactions $\phi^2\sigma$, we perform the second field redefinition
\begin{equation}
    \sigma = \tilde{\sigma} - \frac{\lambda_3}{4} \tilde{\phi}^2 \,.
    \label{eq:field_redef_2-eq048}
\end{equation}
Substituting this into the Lagrangian, the mass term $-\frac{1}{2}m^2\sigma^2$ generates a quartic contact interaction for $\phi$. The transformed Lagrangian reads
\begin{equation}
\label{eq:L_transformed_2-eq049}
\begin{aligned}
    \mathcal{L} =&-\frac{1}{2}(\nabla\tilde{\phi})^2 - \frac{1}{2}(\nabla\tilde{\sigma})^2 - \frac{1}{2}m^2\tilde{\sigma}^2 + \frac{\lambda_3}{4} m^2 \tilde{\sigma} \tilde{\phi}^2 + \frac{\lambda_3^2}{32} (\nabla\tilde{\phi}^2)^2 - \frac{\lambda_3^2}{32} m^2 \tilde{\phi}^4 + \lambda_2 \dot{\tilde{\phi}}\tilde{\sigma} \\
    +& \frac{1}{2}\lambda_2\lambda_3 \left( -\frac{1}{2}\dot{\tilde{\phi}}\tilde{\phi}^2 + \dot{\tilde{\phi}}\tilde{\sigma}^2 + \tilde{\phi}\dot{\tilde{\sigma}}\tilde{\sigma} \right) + \dots \,,
\end{aligned}
\end{equation}
where we have explicitly displayed all contributions up to the second order in the coupling constants. The two quartic terms, corresponding to the ones obtained via the IBP procedure in \cref{sec:4pt_exchange}, cancel the contribution from the EoM term $\phi^2(\square-m^2)\sigma$. Under this second transformation, the correlation functions transform as follows. For the 3-pt correlator $\langle \phi \phi \sigma \rangle$, the shift $\sigma \to \tilde{\sigma} - \frac{\lambda_3}{4}\tilde{\phi}^2$ introduces a new contribution involving the square of the field $\phi$:
\begin{equation}
\label{eq:phiphisigma_redef_2-eq050}
\begin{aligned}
    \langle \phi(x_1)\phi(x_2)\sigma(x_3) \rangle =& \langle \tilde{\phi}(x_1)\tilde{\phi}(x_2)\tilde{\sigma}(x_3) \rangle\ + \frac{\lambda_3}{2} \langle \tilde{\phi}(x_1)\tilde{\sigma}(x_1)\tilde{\phi}(x_2)\tilde{\sigma}(x_3) \rangle + \text{perms} \\
    -& \frac{\lambda_3}{4} \langle \tilde{\phi}(x_1)\tilde{\phi}(x_2)\tilde{\phi}(x_3)^2 \rangle + O(\lambda_3^2) \,.
\end{aligned}
\end{equation}
This relation is consistent with our previous IBP analysis. The first term on the RHS, $\langle \tilde{\phi}\tilde{\phi}\tilde{\sigma} \rangle$, represents the correlator generated by the non-derivative coupling $ \phi^2\sigma$. The second term corresponds to the boundary contribution from $(\partial_\eta \phi^2)\sigma$, which decays in the late-time limit for massive $\sigma$, yet remains non-negligible in the massless limit (see \cref{Light mass}). Crucially, the third term matches the non-vanishing boundary term $\phi^2\partial_\eta\sigma$, equal to the result of the correlator $\langle \tilde{\phi}\tilde{\phi}\tilde{\sigma} \rangle$.

\begin{figure}[tbp]
    \centering
    \includegraphics[scale=0.78, keepaspectratio]{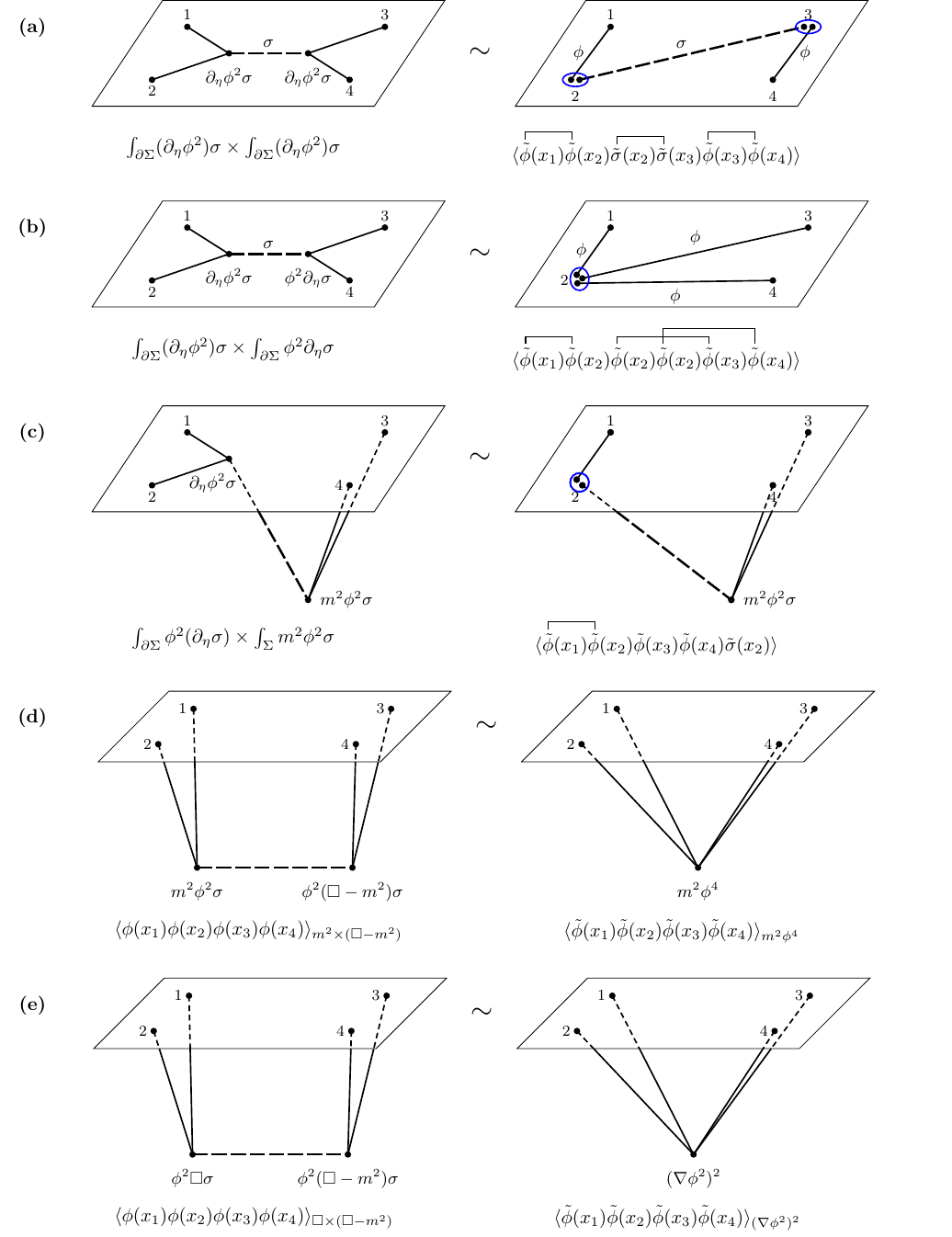}  \caption{Diagrammatic correspondence between the IBP and field redefinitions. The LHS depicts the various diagrams arising from the IBP procedure, where we use $\int_{\partial \Sigma}$ and $\int_{\Sigma}$ to denote boundary and bulk interactions respectively. The RHS illustrates the corresponding diagrams generated by the field redefinition, where we explicitly demonstrate how the field operators contract on the boundary, leaving the bulk part implicit. In the diagrams, solid and thick dashed lines represent the propagators for $\phi$ and $\sigma$ fields, respectively. The parallelograms denote the future boundary, with the region below corresponding to the bulk spacetime. The blue circles on the RHS indicate the composite operators.}
    \label{fig:full_dictionary}
\end{figure}

Then the bispectrum of $\phi$ transforms as
\begin{equation}
    \langle \phi(x_1)\phi(x_2)\phi(x_3) \rangle = \langle \tilde{\phi}(x_1)\tilde{\phi}(x_2)\tilde{\phi}(x_3) \rangle + \frac{\lambda_3}{2} \langle \tilde{\phi}(x_1)\tilde{\phi}(x_2)\tilde{\phi}(x_3)\tilde{\sigma}(x_3) \rangle + \text{perms} + O(\lambda_3^2) \,.
    \label{eq:bispectrum_redef_2-eq051}
\end{equation}
Similarly, for the bispectrum relation, the first term on the RHS corresponds to the exchange correlator generated by the non-derivative cubic interaction $\phi^2\sigma$ and the mixed coupling $\dot{\phi}\sigma$, plus the contact contribution generated by the coupling $\phi^2 \dot{\phi}$. The second term on the RHS is equivalent to the boundary-bulk contribution $\langle\phi\phi\phi\rangle'_{\mathcal{B}_2 \times \lambda_2\dot{\phi}\sigma}$ in \cref{eq:phiphiphi_decomp-eq029}, which also decays in the late-time limit.

Finally, the trispectrum transforms as
\begin{equation}
\label{eq:trispectrum_redef_2-eq052}
\begin{aligned}
    \langle \phi(x_1)\phi(x_2)\phi(x_3)\phi(x_4) \rangle =& \langle \tilde{\phi}(x_1)\tilde{\phi}(x_2)\tilde{\phi}(x_3)\tilde{\phi}(x_4) \rangle + \frac{\lambda_3}{2} \langle \tilde{\phi}(x_1)\tilde{\phi}(x_2)\tilde{\phi}(x_3)\tilde{\phi}(x_4)\tilde{\sigma}(x_4) \rangle  \\
    -& \frac{\lambda_3^2}{8} \langle \tilde{\phi}(x_1)\tilde{\phi}(x_2)\tilde{\phi}(x_3)\tilde{\phi}(x_4)^3 \rangle \\  
    +& \frac{\lambda_3^2}{4} \langle \tilde{\phi}(x_1)\tilde{\phi}(x_2)\tilde{\phi}(x_3)\tilde{\sigma}(x_3)\tilde{\phi}(x_4)\tilde{\sigma}(x_4) \rangle + \text{perms} + O(\lambda_3^3) \,.
\end{aligned}
\end{equation}
The term on the LHS represents the exchange diagram $\langle \phi\phi\phi\phi \rangle'_{\lambda_3(\nabla\phi)^2\sigma\times \lambda_3(\nabla\phi)^2\sigma}$ generated by the derivative cubic interaction $(\nabla\phi)^2\sigma$. On the RHS, the first term contains contributions from the exchange diagram $\langle \phi\phi\phi\phi \rangle'_{g_3\phi^2\sigma\times g_3\phi^2\sigma}$ generated by the non-derivative interaction $\phi^2\sigma$, together with the contact diagrams $\langle \phi\phi\phi\phi \rangle'_{m^2\lambda_3^2\phi^4}$ in \cref{eq:phi4-eq036} and $\langle \phi\phi\phi\phi \rangle'_{\lambda_3^2(\nabla\phi^2)^2}$ in \cref{eq:phi4-eq036}. The second term corresponds to the boundary-bulk contribution $\langle \phi\phi\phi\phi \rangle'_{g_3\phi^2\sigma\times \mathcal{B}_2}$ in \cref{eq:B1-eq033}, and the third term corresponds to the boundary contribution $\langle \phi\phi\phi\phi \rangle'_{\mathcal{B}_1\times \mathcal{B}_2}$ in \cref{eq:B1-eq033}. Finally, the last term corresponds to the boundary contribution $\langle \phi\phi\phi\phi \rangle'_{\mathcal{B}_2\times \mathcal{B}_2}$, which decays in the late-time limit for massive $\sigma$, but demands careful treatment in the massless limit. 

In summary, the results derived from the field redefinition are in perfect agreement with those obtained via the IBP method. To visualise the precise mapping between the IBP and field redefinition viewpoints, we present the correspondence diagrams in \cref{fig:full_dictionary}. We ignore all combinations from IBP with no contribution, the ones containing $\Box\phi$ vanish due to reduction rule (3), the other ones containing boundary terms vanish due to the violation of criterion 1. The first three pairs of diagrams can be understood as the result of applying reduction (3), while the last two pairs of diagrams come from reduction (4).     

\subsection{Recursion Relations for Higher-Derivative Interactions}\label{Recursion_Relations} 

Now we have established the relation between the exchange diagrams with  $\phi^2\sigma \times \phi^2\sigma$ couplings and $(\nabla \phi)^2\sigma \times (\nabla \phi)^2\sigma$ couplings in \cref{eq:4pt_decomposition-eq035}, while the relation between $(\nabla \phi)^2\sigma \times \phi^2\sigma$ and  $(\nabla \phi)^2\sigma \times (\nabla \phi)^2\sigma$ is given in \cref{eq:4pt_mixed_result-eq040}. 
As we have seen, these interactions with lowest derivatives may lead to late-time divergences in the final correlators, which makes the naive analysis in the dS bootstrap problematic. 
Next, we move to consider higher-derivative operators\footnote{Since off-shell propagation of $\phi$ is absent in these exchange diagrams, all dS-invariant couplings can be recast into this schematic form via IBP and the EoM for $\phi$. Boundary terms from IBP involve time derivatives on $\sigma$ and vanish for the diagrams considered here.} 
\begin{equation}
\label{eqn053}
    O^{(n)} = \lambda_3^{(n)} (\nabla_{\mu_1}\nabla_{\mu_2}\dots\nabla_{\mu_n} \phi)^2 \sigma \,, \quad \text{with} \quad \lambda_3^{(n)} \equiv \lambda_3 m^{-2n+2} \,,
\end{equation}
where for $n\geq1$ late-time divergences are in general absent. We shall build the relation among the correlators of these interactions using IBP.

While this category includes infinitely many higher-derivative couplings, the exchange diagrams generated by these vertices are not mutually independent. Instead, they are related to a simpler set of box-type vertices plus contact contributions. 
It is cleaner to first rewrite every covariant higher-derivative vertex in terms of the box-type vertices
\begin{equation}\label{eq:box_basis_direct-eq054}
    2^{-\ell}\Box^\ell(\phi^2)\,\sigma,\qquad \ell=0,1,2,\ldots ,
\end{equation}
which are also the vertices used in \cref{Bootstrap details}. For a massless external field, repeated IBP and the $\phi$ EoM give
\begin{equation}\label{eq:covariant_higher_derivative_identity-eq055}
    (\nabla_{\mu_1}\cdots\nabla_{\mu_j}\phi)
    (\nabla^{\mu_1}\cdots\nabla^{\mu_j}\phi)\,\sigma
    \simeq
    \frac{1}{2^j}\Box
    \prod_{a=1}^{j-1}\!\left(\Box-2a(a+2)H^2\right)\phi^2\,\sigma,
\end{equation}
where $\simeq$ denotes equality modulo boundary terms and terms proportional to $\Box\phi$. The curvature-dependent shifts come from commuting covariant derivatives. Equivalently, one may write
\begin{equation}\label{eq:Sm_box_basis-eq056}
    (\nabla^j\phi)^2\sigma
    \simeq
    \sum_{\ell=1}^{j} b_{j\ell}\,2^{-\ell}\Box^\ell(\phi^2)\sigma,\qquad
    \sum_{\ell=1}^{j} b_{j\ell}\,2^{-\ell}z^\ell
    =
    2^{-j}z\prod_{a=1}^{j-1}\!\left(z-2a(a+2)H^2\right).
\end{equation}
We can now derive the recursion directly for the box-type vertices. Consider an exchange diagram whose two cubic vertices are
\begin{equation}
\label{eqn057}
    2^{-r}\Box^r(\phi^2)\sigma
    \quad\hbox{and}\quad
    2^{-s}\Box^s(\phi^2)\sigma,\qquad l=r+s.
\end{equation}
If $s\geq1$, one can integrate one $\Box$ from the second vertex onto the internal $\sigma$ propagator. Using $(\Box-m^2)D^\sigma=i\delta$ gives
\begin{equation}\label{eq:box_recursion_step-eq058}
    \langle\phi^4\rangle'_{r\times s}
    =
    \frac{m^2}{2}\,
    \langle\phi^4\rangle'_{r\times(s-1)}
    +\langle\phi^4\rangle'_{\mathrm{con},l-1}
    +\langle\phi^4\rangle'_{\rm bdy}.
\end{equation}
The contact term does not depend on how the box operators were split between the two cubic vertices, because
\begin{equation}\label{eq:box_contact_independent-eq059}
    2^{-l}\int d^4x\sqrt{-g}\,
    \Box^r(\phi^2)\Box^{s-1}(\phi^2)
    \simeq
    2^{-l}\int d^4x\sqrt{-g}\,
    \phi^2\Box^{l-1}(\phi^2),
\end{equation}
again up to boundary terms and the $\phi$ EoM. Reducing the diagram with vertices $2^{-(r+1)}\Box^{r+1}(\phi^2)\sigma$ and $2^{-(s-1)}\Box^{s-1}(\phi^2)\sigma$ gives the same right-hand side. Therefore, for $l=r+s\geq2$ and $s\geq2$,
\begin{equation}\label{eq:fixed_l_independence-eq060}
    \langle\phi^4\rangle'_{r\times s}
    =
    \langle\phi^4\rangle'_{(r+1)\times(s-1)},
\end{equation}
after dropping boundary contributions, which vanish by the criteria of \cref{criteria}. Thus all box-type exchange diagrams with the same total number $l$ of boxes give the same exchange shape, which we denote by $\langle\phi^4\rangle'_{\mathrm{ex},l}$. Choosing any representative with $s\geq1$, the recursion between total order $l$ and total order $l-1$ can be written as
\begin{equation}\label{eq:box_basis_recursion_position-eq061}
    \langle\phi^4\rangle'_{\mathrm{ex},l}
    =
    \frac{m^2}{2}\,
    \langle\phi^4\rangle'_{\mathrm{ex},l-1}
    +\langle\phi^4\rangle'_{\mathrm{con},l-1},\qquad l\geq 2 .
\end{equation}
The second term is local, and its form is independent of the split $(r,s)$ by \cref{eq:box_contact_independent-eq059}. This is the position-space origin of the recursion displayed in \cref{Bootstrap details}. For the original covariant-derivative vertices, one first expands each vertex by \cref{eq:Sm_box_basis-eq056} and then applies this universal box-basis result term by term.

The low-derivative cases $l<2$ are exceptional. The $(0,0)$ relation is the decomposition in \cref{eq:4pt_decomposition-eq035}, where the EoM collapse isolates the logarithmic contact term. The mixed $(1,0)$ and $(0,1)$ relations are described by \cref{eq:4pt_mixed_result-eq040}; there the surviving boundary terms distinguish which side carries the derivative, producing the local boundary contribution displayed earlier. These boundary/anomaly effects are why the universal $l\geq2$ recursion should not be extrapolated to the low-derivative cases.

\section{Boost-Breaking Correlators with Boundary Contributions}\label{CH4}

In this section, we extend the previous analysis of dS-invariant theories to a more general EFT framework that allows for boost-breaking interactions, including different sound speeds for the fields $\phi$ and $\sigma$. 
The effects of boundary terms in single-field inflation have been discussed in the context of boostless bootstrap \cite{Pajer:2020wnj}, while our focus here is on scenarios with intermediate massive states.
The bootstrap analysis for these boost-breaking theories is introduced and extensively discussed in \cite{Pimentel:2022fsc,Jazayeri:2022kjy,Wang:2022eop}.
{Below, we use IBP and EoM relations to reduce boost-breaking exchange diagrams to a basis of bulk interactions.} 
In \cref{general-EFT}, we construct a complete basis for the shape functions generated by the $\sigma$-exchange diagram with boost-breaking bulk interactions. In \cref{BEFT}, we classify all non-vanishing boundary contributions arising from IBP among these boost-breaking couplings. This systematic procedure for classifying exchange bulk interactions and boundary terms can be readily generalised to arbitrary polynomial couplings and diagrams.

\subsection{The Bulk Interactions in a General EFT Setup}\label{general-EFT}
 The most general three-point bulk $\phi\phi\sigma$ interaction at dimension $n$ can be written in position space and, after Fourier transforming the spatial derivatives, in momentum space as
\begin{equation}
\label{eqn062}
\begin{aligned}
        \mathcal{L}^n_\mathrm{bulk}=&a^{-n}(\eta)(\partial_\eta^{\alpha_1}\left(\partial_i^2\right)^{\beta_1}\partial_{j_1}\partial_{j_2}...\partial_{j_{\gamma_3}}\partial^{k_1}\partial^{k_2}...\partial^{k_{\gamma_2}}\phi)(\partial_\eta^{\alpha_2}\left(\partial_i^2\right)^{\beta_2}\partial_{l_1}\partial_{l_2}...\partial_{l_{\gamma_1}}\partial^{j_1}\partial^{j_2}...\partial^{j_{\gamma_3}}\phi)\\
        \times&(\partial_\eta^{\alpha_3}\left(\partial_i^2\right)^{\beta_3}\partial_{k_1}\partial_{k_2}...\partial_{k_{\gamma_2}}\partial^{l_1}\partial^{l_2}...\partial^{l_{\gamma_1}}\sigma),\\
    \mathcal{L}_\mathrm{bulk}^{n'}=& a^{-n}(\eta){k_1}^{2\beta_1}{k_2}^{2\beta_2}{s}^{2\beta_3}({\bf k}_1\cdot {\bf k}_2)^{\gamma_3}({\bf k}_1\cdot {\bf s})^{\gamma_2}({\bf s}\cdot {\bf k}_2)^{\gamma_1}\\
    \times&(\partial_\eta^{\alpha_1}\phi(\eta,k_1))(\partial_\eta^{\alpha_2}\phi(\eta,k_2))(\partial_\eta^{\alpha_3}\sigma(\eta,s)).
\end{aligned}
\end{equation}
with the momentum-conserving Dirac-$\delta$ function stripped off.
We have also introduced ${\bf s}\equiv{\bf k}_1+{\bf k}_2$.
The dot products can be expanded in terms of the momentum magnitudes, allowing us to focus on individual monomials of the form
\begin{equation}\label{eq4.3-eq063}
    \mathcal{L}_\mathrm{bulk}^{n''}= a^{-n}(\eta){k_1}^{2m_1}{k_2}^{2m_2}{s}^{2m_3}(\partial_\eta^{\alpha_1}\phi(\eta,k_1))(\partial_\eta^{\alpha_2}\phi(\eta,k_2))(\partial_\eta^{\alpha_3}\sigma(\eta,s)),
\end{equation}
where $m_1,m_2,m_3$ are non-negative integers satisfying $m_{123}=\beta_{123}+\gamma_{123}$.

Next, we apply the EoM relations, restricting to on-shell $\phi$ excitations on external legs and off-shell internal $\sigma$ modes (generalisation to other configurations is straightforward):
\begin{equation}
\label{eqn064}
\begin{aligned}
    {c_\phi}^2 k^  2\phi(\eta,k)
    =-&\partial_\eta^2\phi(\eta,k)+\frac{2}{\eta}\partial_\eta\phi(\eta,k),\\
   {c_\sigma}^2 k^2\sigma(\eta,k)
   =-&\partial_\eta^2\sigma(\eta,k)+\frac{2}{\eta}\partial_\eta\sigma(\eta,k)-\frac{m^2}{H^2\eta^2}\sigma(\eta,k).
\end{aligned}
\end{equation}
For time/anti-time ordered bulk-to-bulk $\sigma$ propagators, the equation of motion can also introduce $\delta$-function terms, which generate contact diagrams.\footnote{{These contact contributions can themselves be related by IBP, and the boundary terms generated in this way give local shapes. We discuss these boundary contributions in \cref{BEFT}.}} The general contact diagrams have already been systematically classified in previous boostless bootstrap analyses \cite{Pajer:2020wxk,Jazayeri:2021fvk,Bonifacio:2021azc}.
 Repeated applications of these relations convert even powers of spatial momentum into time derivatives, yielding terms in the schematic form $\partial_\eta^{\alpha_1^{\prime}}\phi\partial_\eta^{\alpha_2^{\prime}}\phi\partial_\eta^{\alpha_3^{\prime}}\sigma$.
 
Finally, time derivatives acting on $\sigma$ can be transferred to the $\phi$ fields via IBP with respect to the conformal time $\eta$. One novelty here (compared to the covariant-derivative IBP) is the explicit time dependence in the measure factor $\sqrt{-g} = a^4(\eta)$, which generates additional terms:
\begin{equation}
\label{eqn065}
    \begin{split}
        \int \mathrm{d}\eta (-H\eta)^{\alpha_{123}'-4}\partial_\eta^{\alpha_1'}\phi\partial_\eta^{\alpha_2'}\phi\partial_\eta^{\alpha_3'}\sigma=&(\alpha_{123}'-4)\int \mathrm{d}\eta H(-H\eta)^{\alpha_{123}'-5}\partial_\eta^{\alpha_1'}\phi\partial_\eta^{\alpha_2'}\phi\partial_\eta^{\alpha_3'-1}\sigma\\-&\int \mathrm{d}\eta (-H\eta)^{\alpha_{123}'-4}\partial_\eta^{\alpha_1'+1}\phi\partial_\eta^{\alpha_2'}\phi\partial_\eta^{\alpha_3'-1}\sigma\\-&\int \mathrm{d}\eta (-H\eta)^{\alpha_{123}'-4}\partial_\eta^{\alpha_1'}\phi\partial_\eta^{\alpha_2'+1}\phi\partial_\eta^{\alpha_3'-1}\sigma+\text{bdy terms},
    \end{split}
\end{equation}
Iterating this process removes all time derivatives on $\sigma$.

{
After this reduction, the counting only depends on the resulting time derivative permutations on the external $\phi$ legs. We therefore count the number of the independent exchange shapes\footnote{The basis of exchange shapes cannot be reduced more because they obey the recursion relations below, which change the squeezed/OPE behaviour of the massive seed.} by enumerating all possible interactions with time derivatives on $\phi$. 

}

\paragraph{Trispectrum shapes.}
Consequently, shape functions generated by $\partial^{\alpha_1}_{\eta}\phi\partial^{\alpha_2}_{\eta}\phi\sigma \times \partial^{\alpha_3}_{\eta}\phi\partial^{\alpha_4}_{\eta}\phi\sigma$ are related to the reduced time-derivative vertices. For $\alpha_1,\alpha_2,\alpha_3,\alpha_4\geq1$, they obey the recursion relations
\begin{equation}\label{recursion2-eq066}
\begin{split}
&k_1^3k_2^3k_3^3k_4^3\braket{\phi\phi\phi\phi}^{\prime}_{(\alpha_1,\alpha_2,\alpha_3,\alpha_4)}=\prod_{i=1}^4\prod_{n_i=1}^{\alpha_i} (k_i\partial_{k_i}-n_i)\left(k_1^3k_2^3k_3^3k_4^3\braket{\phi\phi\phi\phi}^{\prime}_{(1,1,1,1)}\right),
\end{split}
\end{equation}
where the base shape $\braket{\phi\phi\phi\phi}^{\prime}_{(1,1,1,1)}$ is free of late-time divergence. For one $\phi\phi\sigma$-vertex with $m$ time derivatives, the two $\phi$ legs are identical, so the number of derivative distributions is
\begin{equation}
\label{eqn067}
    a_m=\left\lfloor\frac m2\right\rfloor+1 .
\end{equation}
The four-point exchange count is then the unordered-pair sum
\begin{equation}
\label{eqn068}
    N_{4,{\rm ex}}(l)=
    \sum_{\substack{m+n=l\\m<n}}a_m a_n
    +\sum_{\substack{2m=l}}\frac{a_m(a_m+1)}{2}.
\end{equation}
This is an unordered count because, in a fixed $s$-channel, the two cubic vertices are both of the $\phi\phi\sigma$-type and the diagram is symmetric under $(12)\leftrightarrow(34)$ after the $s,t,u$ channel sum. 
The counting up to 6 total derivatives is shown in \cref{table-1}.
\begin{table}[tbp] 
    \centering
    \begin{tabular}{|c|c|}
       \hline Total number &  Number of  \\
       of derivatives & exchange shapes\\ \hline
         0 & 1 \\ \hline
         1 & 1\\ \hline
         2 & 2+1=3 \\ \hline
         3 & 2+2=4\\ \hline
         4 & 3+2+3=8\\ \hline
         5 & 3+3+4=10\\ \hline
         6 & 4+3+6+3=16\\ \hline
    \end{tabular}
    \caption{{The number of independent exchange trispectrum shapes counted up to 6 derivatives. The entries are written additively by the unordered time-derivative permutations on the external $\phi$ legs of the two $\phi\phi\sigma$-vertices.}}
    \label{table-1}
\end{table}

\paragraph{Bispectrum shapes.}
{A similar classification applies to bispectra from $\phi\phi\sigma \times \phi\sigma$ couplings. In a fixed channel, let $p$ be the number of time derivatives on the two $\phi$ legs in the $\phi\phi\sigma$-vertex and $q$ the number of time derivatives on the $\phi\sigma$-vertex. Since the two vertices are different, $(p,q)$ is ordered.\footnote{{Moving a time derivative from the $\phi\phi\sigma$ side to the $\phi\sigma$ side changes the boundary differential operator acting on the massive seed, and hence changes the exchanged-channel squeezed/OPE behaviour. These ordered exchange representatives should therefore be counted separately.}} The exchange bispectrum count is therefore
\begin{equation}
\label{eqn069}
    N_{3,{\rm ex}}(l)=\sum_{p+q=l}\left(\left\lfloor\frac p2\right\rfloor+1\right),
\end{equation}
where the factor in parentheses counts the symmetric distribution of the $p$ derivatives on the two identical $\phi$ legs. Up to $l=6$, the number of independent shapes is listed in \cref{table-2}.

}

\begin{table}[tbp] 
    \centering
    \begin{tabular}{|c|c|}
       \hline Total number &  Number of  \\
       of derivatives & exchange shapes\\ \hline
         0 & 1 \\ \hline
         1 & 1+1=2 \\ \hline
         2 & 1+1+2=4 \\ \hline
         3 & 1+1+2+2=6 \\ \hline
         4 & 1+1+2+2+3=9\\ \hline
         5 & 1+1+2+2+3+3=12\\ \hline
         6 & 1+1+2+2+3+3+4=16\\ \hline
    \end{tabular}
    \caption{{Bispectrum exchange count up to 6 derivatives. The entries count ordered time-derivative permutations on the external $\phi$ legs of the  $\phi\phi\sigma$- and $\phi\sigma$-vertices. 
    }}
    \label{table-2}
\end{table}

These reductions in massive-exchange shape functions can simplify the observational tests of cosmological collider signals. In the current data analysis \cite{Cabass:2024wob,Sohn:2024xzd,Suman:2025vuf,Suman:2025tpv,Philcox:2026njr}, the focus is on a few collider templates from lowest-order EFT operators. 
If we introduce higher-derivative interactions, more templates need to be included.
The above analysis demonstrates that up to a certain total number of derivatives, the number of independent shapes is limited, while the rest can be constructed as linear combinations from this basis via IBP and EoM.

\subsection{Classifying General Boundary Contributions}\label{BEFT} 

In this subsection, we systematically investigate the general boundary terms generated via IBP from boost-breaking $  \phi\phi\sigma  $-type and $\phi\sigma$-type bulk interactions in exchange diagrams.
As we can safely neglect boundary terms with spatial derivatives, the most general $\phi\phi\sigma$-type boundary terms take the following form
\begin{equation}
\label{eqn070}
    \mathcal L_{\text{bdy}}^{\alpha_{123}}=a^{-\alpha_{123}}(\eta_0)\partial_{\eta_0}^{\alpha_1}\phi\partial_{\eta_0}^{\alpha_2}\phi\partial_{\eta_0}^{\alpha_3}\sigma.
    \end{equation}
 Furthermore, by applying the equations of motion for both $\phi$ and $\sigma$, we can reduce the number of time derivatives acting on each field.
 After performing these reductions, only six independent boundary terms remain
\begin{equation}
\label{eqn071}
    \phi^2\sigma~,~~~a^{-1}(\eta_0)\phi^\prime\phi\sigma~,~~~a^{-1}(\eta_0)\phi^2\sigma^{\prime}~,~~~a^{-2}(\eta_0)(\phi^\prime)^2\sigma~,~~~a^{-2}(\eta_0)\phi^{\prime}\phi\sigma^{\prime}~,~~~a^{-3}(\eta_0)(\phi^\prime)^2\sigma^{\prime}~.
\end{equation}
Similarly, there are only four independent $\phi\sigma$-like boundary terms. They are
\begin{equation}
\label{eqn072}
    \phi\sigma~,~~~~~~    a^{-1}(\eta_0)\phi^\prime\sigma~,~~~~~~ a^{-1}(\eta_0)\phi\sigma^{\prime}  ~,~~~~~~   a^{-2}(\eta_0)\phi^{\prime}\sigma^{\prime}~.
\end{equation}

Before proceeding with the classification, let us illustrate how the criteria of \cref{criteria} apply beyond the simple $\Pi_0 f$ vertices used in \cref{CH3}. The field-redefinition interpretation is most direct for boundary terms with exactly one conjugate momentum, but the underlying boundary contraction is more general: an equal-time commutator between a boundary field and a conjugate momentum can cut a line and leave the remaining operators on the future boundary. This allows boundary terms with zero or with more than one time derivative to contribute in special configurations.

First consider a boundary term with two time derivatives,
$\lambda a^{-2}(\eta_0)\phi'\phi\sigma'$. At first sight, one may relate it to field redefinitions in two ways. One is
$\phi=\tilde\phi+\lambda a^{-1}(\eta)\phi\sigma'$, which yields
$\Delta\mathcal L=-\lambda\partial_\mu(a^{-1}(\eta)(\partial^\mu\phi)\phi\sigma')
+\lambda a^{-1}(\eta)\square\phi\,\phi\sigma'$.
Alternatively, we may use
$\sigma=\tilde\sigma+\lambda a^{-1}(\eta)\phi\phi'$, which yields
$\Delta\mathcal L=-\lambda\partial_\mu(a^{-1}(\eta)\phi'\phi\partial^\mu\sigma)
+\lambda a^{-1}(\eta)\phi'\phi(\square-m^2)\sigma$.
The corresponding three-point function makes the two possible boundary contractions explicit:
\begin{equation}\label{counterexample-eq073}
\begin{split}
     &\sum_{b=\pm}\frac{ib\lambda}{(-H\eta_0)}\partial_{\eta_0}K^{\phi}_b(k_1,\eta_0)K^{\phi}_b(k_2,\eta_0)\partial_{\eta_0}K^{\sigma}_b(k_3,\eta_0)\\
     =&\frac{i\lambda}{(-H\eta_0)}\left(\partial_{\eta_0}\left(K^{\phi}_+(k_1,\eta_0)-K^{\phi}_-(k_1,\eta_0)\right)K^\phi(k_2,\eta_0)\partial_{\eta_0}K^{\sigma}_+(k_3,\eta_0)\right.\\
     +&\left.\partial_{\eta_0}K^{\phi}_-(k_1,\eta_0)K^\phi(k_2,\eta_0)\partial_{\eta_0}\left(K^{\sigma}_+(k_3,\eta_0)-K^{\sigma}_-(k_3,\eta_0)\right)\right)\\
     =&\lambda H\eta_0\left(K^\phi(k_2,\eta_0)\partial_{\eta_0}K^{\sigma}_+(k_3,\eta_0)+\partial_{\eta_0}K^{\phi}_-(k_1,\eta_0)K^\phi(k_2,\eta_0)\right).
\end{split}
\end{equation}
The two terms correspond to cutting either the $\phi'$ line or the $\sigma'$ line. They match the two field redefinitions above, where each of them accounts for one boundary contraction. The full boundary Feynman rule should contain both contractions.\footnote{One needs to be careful with derivative-dependent field redefinitions. In the equivalence-theorem perspective they must satisfy the Faddeev condition, namely that the derivative-dependent change of variables defines an equivalent configuration-space path integral without an extra measure factor; otherwise the induced Jacobian must be kept \cite{Chisholm:1961tha,Kamefuchi:1961sb,Divakaran:1963yxz,Kallosh:1972ap,Bergere:1975tr}. In the Schwinger--Keldysh path integral, such redefinitions can also transform the final-time sewing condition between the $+$ and $-$ branches. A clean treatment of both issues is obtained by lifting the transformation to a canonical transformation in phase space \cite{Braglia:2024zsl}.} For the present classification, however, the boundary Feynman rule already gives the relevant contractions: each external line connected to a conjugate momentum operator is cut separately, and the remaining operators are pulled up to the future boundary.


Next consider a boundary term with no time derivative. By itself such a vertex cancels in the branch sum. For example,
\begin{equation}
\label{eqn074}
\left.\int d^3x\, a^3(\eta)\,\phi^2(\eta,{\bf x}) \, \sigma (\eta,{\bf x})\right|_{\eta=\eta_0}
\end{equation}
does not contain a conjugate-momentum operator that can generate an equal-time commutator. Its direct contribution is therefore killed by the $+$/$-$ sum, in agreement with Criterion 1.

Such a vertex can nevertheless contribute when another boundary operator supplies the required conjugate-momentum contraction. Consider the exchange diagram with one boundary vertex $a^{-2}(\eta_0)\phi'\phi\sigma'$ and another $\phi^2\sigma$ vertex pulled to the boundary. Using \cref{properties-eq009}, the four-point correlator is
\begin{equation}\label{eq4.12-eq075}
\begin{split}
&\langle \phi(\vec{k}_1) \phi(\vec{k}_2) \phi(\vec{k}_3) \phi(\vec{k}_4) \rangle_{s{\rm -channel}}\\
     =&\lim_{ 
         \eta_1,\eta_2\rightarrow\eta_0
     }\sum_{b=\pm,c=\pm}\frac{-bc}{H^4\eta_1\eta_2^3}K^{\phi}_{b}(k_1,\eta_1)\partial_{\eta_1}K^{\phi}_{b}(k_2,\eta_1)
    \partial_{\eta_1}D^{\sigma}_{bc}(s,\eta_1,\eta_2)K^{\phi}_{c}(k_3,\eta_2)K^{\phi}_{c}(k_4,\eta_2)\\
    =&\sum_{b=\pm}\frac{-ibH^2}{4k_3^3k_4^3\eta_0^2}K^{\phi}_{b}(k_1,\eta_0)\partial_{\eta_0}K^{\phi}_{b}(k_2,\eta_0)
    =\frac{H^6}{8 k_1^{3}k_3^3k_4^3}.
\end{split}
\end{equation}
Here the $\phi'$ in the first boundary vertex contracts with a boundary $\phi$ operator and cuts one line. After this contraction, the remaining $\phi^2\sigma$ factor is left on the boundary and no branch-sum cancellation occurs. This is precisely the generalised boundary contraction described by Criterion 1. The post-reduction diagram must then satisfy the criteria of \cref{criteria}.

These examples show that the classification should be performed after the boundary contractions, rather than by counting time derivatives in the original boundary vertex alone. We can now classify boundary contributions by first applying all possible boundary contractions and then testing the reduced diagrams against Criteria 1, $1'$ and 2. 

\paragraph{Exchange trispectrum.} The surviving boundary-boundary combinations are as follows:
\begin{itemize}
    \item Non-decaying contributions: only the following two combinations remain, namely 
    \begin{equation}
\label{eqn076}
    \phi'\phi\sigma \times  \phi^2\sigma' ~~~~~~ {\rm and }~~~~~~  \phi^2\sigma  \times \phi'\phi\sigma'  .
    \end{equation}
    The former is demonstrated in \cref{eq:B1-eq033}, and the latter is in \cref{eq4.12-eq075}. Both of them lead to the local shape of the trispectrum.
    \item Moderately decaying contributions:
    \begin{equation}
    \label{eqn077}
    \begin{aligned}
        \phi^{\prime}\phi\sigma\times\phi^{\prime}\phi\sigma~~~~~ &\Rightarrow  ~~~~~~ \langle \phi^2 \rangle \langle \sigma^2 \rangle \langle \phi^2 \rangle \sim \eta_0^{3-2\nu} \\
        \phi^{\prime}\phi\sigma\times\phi^{\prime}\phi\sigma^{\prime} ~~~~&\Rightarrow~~~~~~ a^{-1}(\eta_0)\braket{\phi^2}\braket{\sigma^\prime\sigma}\braket{\phi^2}\sim (3-2\nu)\eta_0^{3-2\nu} \\
        \phi^{\prime}\phi\sigma^{\prime}\times\phi^{\prime}\phi\sigma^{\prime}~~~~&\Rightarrow~~~~~~ a^{-2}(\eta_0)\braket{\phi^2}\braket{\sigma^{\prime2}}\braket{\phi^2}\sim (3-2\nu)^2\eta_0^{3-2\nu} .
    \end{aligned}
    \end{equation}
    The first contribution is shown in the first row of \cref{fig:full_dictionary}. While being neglected in \cref{CH3}, it becomes relevant when $  \sigma  $ is nearly massless. 
\end{itemize}
All other combinations are excluded by the criteria in \cref{criteria}. 

We now turn to boundary-bulk mixing terms. All combinations involving $  \phi^2\sigma  $ or $  \phi^2\sigma'  $ as the boundary vertex vanish strictly (violating Criterion 1). Those with $  \phi'^2\sigma  $ or $  \phi'^2\sigma'  $ decay at the late-time boundary due to remaining $  \phi'  $ in the boundary operators (violating the extra criterion).
The remaining cases are:
\begin{itemize}
    \item $\phi^{\prime}\phi\sigma|_\text{boundary} \times \mathcal L_\text{bulk}$: equivalent to the redefined correlator $\braket{\phi^2}\braket{\phi^2\sigma}|_{\mathcal L_{bulk}}$.
    For the non-derivative bulk vertex $  \phi^2\sigma|_{\rm bulk}  $, the imaginary part behaves as
    \begin{equation}
\label{eqn078}
        \begin{split}
        \Im \Bigl( i \sigma^*(\eta_0) \int \frac{d\eta}{\eta^4} K^\phi(\eta) K^\phi(\eta) \sigma(\eta) \Bigr)
        \sim &\Im \Bigl( i \eta_0^{3/2 - \nu} \int_{-\infty}^{\eta_0} d\eta \, \eta^{-5/2 + \nu} (1 + O(\eta/\eta_0)) \Bigr) \\
        \sim &\text{const} + O(\eta_0).
        \end{split}
    \end{equation}
This is the contribution computed in \cref{eq:B1-eq033} for dS-invariant interactions. 
Bulk vertices with derivatives introduce higher positive powers of $  \eta  $ in the integrand ($  \eta^{-5/2 + \nu + \alpha}  $), which leads to decaying contributions at the late-time boundary.
\item $  \phi'\phi\sigma'|_\text{boundary} \times \mathcal{L}_{\rm bulk}  $: equivalent to $  a^{-1}(\eta_0) \langle \phi^2 \rangle \langle \phi^2 \sigma' \rangle|_{\mathcal{L}_{\rm bulk}}  $.
As shown in \cref{sec:building_block}, the non-derivative bulk vertex $  \phi^2\sigma|_{\rm bulk}  $ is related by IBP to the boundary vertex $  \phi^2 \partial_\eta \sigma  $. The corresponding integral thus reduces to
\begin{equation}
\label{eq:boundary-bulk-zero-temp-eq079}
    \sum_{b=\pm} \frac{i b}{(-H \eta_0)} K^\phi_b(\eta_0) K^\phi_b(\eta_0) \partial_{\eta_0} \sigma^*(\eta_0) \partial_{\eta_0} \sigma(\eta_0) = 0.
\end{equation}
Derivative bulk vertices again produce higher positive powers of $  \eta_0  $, and thus decay on the late-time boundary.
\end{itemize}




\paragraph{Exchange bispectrum.} 
The surviving boundary-boundary combinations are as follows: 
\begin{itemize}
    \item Non-decaying contributions:  
    \begin{equation}
\label{eq:bispec-nondecy-temp-eq080}
    \phi^2\sigma\times\phi^\prime\sigma^\prime~,~~~~ \phi^\prime\phi\sigma\times\phi\sigma^\prime~,~~~~
    \phi^2\sigma^\prime\times\phi^{\prime}\sigma~,~~~~
    \phi^\prime\phi\sigma^\prime\times\phi\sigma~.
    \end{equation}
    \item Moderately decaying contributions: 
    \begin{equation}
    \label{eqn081}
    \begin{aligned}
        \phi^{\prime}\phi\sigma\times\phi^\prime\sigma ~~~~&\Rightarrow ~~~~ \braket{\phi^2}\braket{\sigma^2}\sim\eta_0^{3-2\nu} \\
        \phi^{\prime}\phi\sigma^\prime\times\phi^\prime\sigma^\prime ~~~~&\Rightarrow ~~~~ a^{-2}(\eta_0)\braket{\phi^2}\braket{\sigma^{\prime2}}\sim(3-2\nu)^2\eta_0^{3-2\nu}.
    \end{aligned}
    \end{equation}
\end{itemize}
Regarding boundary-bulk mixing terms, there are two types: $\phi\phi\sigma|_{boundary} \times \phi\sigma|_{bulk}$ and 
$\phi\phi\sigma|_{bulk} \times \phi\sigma|_{boundary}$. Similarly to the discussion of the exchange four-point function, the only non-vanishing contribution in the first type is $\phi^\prime\phi\sigma|_{boundary} \times \phi\sigma|_{bulk}$, which corresponds to $\braket{\phi^2}\braket{\phi\sigma}|_{\phi\sigma|_{bulk}}\sim\frac{1}{k_1^3k_2^3}\frac{1}{m^2}$. While the only non-vanishing contribution in the second class is $\phi^2\sigma|_{bulk} \times \phi^{\prime}\sigma|_{boundary}$, which corresponds to $\braket{\phi^2\sigma}|_{\phi^2\sigma|_{bulk}}\sim\frac{1}{k_1^3k_2^3}\frac{1}{m^2}$.



\section{Conclusion and Discussion}\label{Conclusion}


In this paper, we investigate the systematic classification of boundary contributions to cosmological correlators and explicitly identify the redundancies of bulk interactions that can be removed via IBP or field redefinitions.

First, we start with a general setup of the Schwinger--Keldysh path integral representation of cosmological correlators and explore the connections between boundary terms and field redefinitions. To achieve this goal, we derive four reduction rules on Feynman diagrams by applying the Schwinger--Dyson equation. As a byproduct, we also propose a set of criteria to tell when boundary terms can give non-vanishing contributions to late-time correlators.

Next, we move to consider correlators from massive-exchange diagrams in both dS-invariant and boost-breaking theories.
These three-point and four-point functions with intermediate particles are important observables in cosmological collider physics, and have been systematically studied in the bootstrap analysis.
Here we focus on various possibilities of boundary contributions to the massive-exchange correlators:
\begin{itemize}
    \item 
    In dS-invariant theories, for non-derivative couplings, we identify the nontrivial late-time divergent contributions exposed by the IBP/EoM reduction, tracing them to the corresponding contact diagrams, and distinguishing these effects from the accompanying boundary contributions.
    These behaviors were overlooked in the previous analysis of the dS bootstrap. For higher-derivative interactions, we derive a recursive relation to connect the resulting exchange correlators via IBP.
    \item In boost-breaking scenarios, we first reduce the various bulk interactions to a smaller basis using IBP and EoM. This basis provides independent collider shape functions of bispectra and trispectra in the EFT framework, and can help us simplify the templates used in the observational tests of PNG in the near future. Then we present a classification of the non-vanishing boundary contributions to the exchange correlators.
\end{itemize}

Finally, we outline several promising avenues for future investigation.
First, a natural extension parallel to the classification of boundary terms is a systematic study of general field redefinitions --- including non-local ones --- in multi-field theories. Of particular interest is whether boundary terms involving arbitrary time derivatives can be equivalently generated by (possibly non-local) field redefinitions. Addressing this question requires careful treatment of the path integral over conjugate momenta and the associated measure/Jacobian terms.


Furthermore, considering the rich structure of boundary contributions to cosmological correlators, it would be interesting to construct a boundary effective field theory (BEFT) of inflation to describe physical processes that take place around the reheating surface.
For instance, in the scenario with instant reheating, 
the reheating process is confined to a thin temporal layer near the end of inflation, which can be approximated as the late-time boundary of dS. The particle production and decay processes then may be encoded by operators of the BEFT, bypassing the need for a full bulk treatment of the dynamical background.


Finally, for infrared divergences in dS spacetime, it would be useful to understand the possible connection between conformal anomalies on the boundary and field redefinitions. 
These divergences in dS lead to violations of boundary conformal Ward identities. 
As shown in the current work, field redefinitions offer a promising tool for making this correspondence between late-time divergences and conformal anomalies more manifest. 
Interestingly, the renormalisation group flow can itself be interpreted as a continuous family of field redefinitions~\cite{Latorre:2000qc,Anselmi:2012aq}. One may wonder whether a de Sitter analogue of holographic renormalisation exists. 
We leave these intriguing questions for future research.


\section*{Acknowledgment}

We thank Xingang Chen, Zhehan Qin, Xi Tong and Yuhang Zhu for helpful discussions. This work is supported in part by the National Key R$\&$D Program of China (2021YFC2203100), and the RGC Research Fellow Grant RFS2425-6S02 from the Research Grants Council of Hong Kong. DGW is supported by GRF Grant 16306425 from the Research Grants Council of Hong Kong. 

\appendix

\section{Path Integral Issues under Field Redefinition}\label{path integral issues}
In this appendix, we employ the invariance of the path integral under local invertible field redefinitions to derive the Schwinger--Dyson equations. For the purpose of deriving the identities used in \cref{CH2}, we treat redefinitions that are pointwise invertible on the final field configuration. If a redefinition contains time derivatives, the final-time sewing condition is transformed because the boundary field mixes with the canonical momentum; such cases should be treated in phase space or by the canonical-transformation construction discussed in \cref{BEFT}.

Consider a generic field redefinition of the form $\Phi=\tilde{\Phi}+f[\tilde\Phi]$, where $f$ is a local functional of $\tilde{\Phi}$. The Schwinger--Keldysh path integral expression for a correlator remains unchanged under this transformation:
\begin{equation}\label{eq:PI-invariance-eq082}
    \begin{split}
         &\braket{\Omega|\mathcal{O}[\Phi]|\Omega}|_S\\
         =&\int_{\Omega[\Phi^+],\Omega[\Phi^-]}^{\Phi^+(\eta_0)=\Phi^-(\eta_0)}\left(\prod_{b=\pm}\mathcal{D}\Phi^b\right)
         \mathcal{O}[\Phi^+]e^{\sum_{b=\pm}ib S[\Phi^b]}\\=&\int_{\Omega[\Phi^++f[\Phi^+]],\Omega[\Phi^-+f[\Phi^-]]}^{\Phi^+(\eta_0)=\Phi^-(\eta_0)} \left(\prod_{b=\pm}\mathcal{D}\Phi^b\left|\frac{\partial(\Phi^b+f[\Phi^b])}{\partial\Phi^b}\right|\right)\mathcal{O}[\Phi^++f[\Phi^+]]e^{\sum_{b=\pm}ibS[\Phi^b+f[\Phi^b]]}
         ,
    \end{split}     
    \end{equation}
Here, the subscript $|_S$ indicates evaluation under the original action $S[\Phi]$. After relabeling the integration variable $\tilde{\Phi} \to \Phi$, the action and the functional $\mathcal{O}$ acquire the form they would have under the infinitesimal variation $\Delta\Phi = f[\Phi]$. Moreover, both the path integral measure and the initial conditions defining the in vacuum state $|\Omega\rangle$ (denoted as $\Omega[\Phi]$) are also modified. We now argue that, for local invertible field redefinitions of scalar fields, these effects either give local counterterms or are fixed by the same $i\epsilon$ prescription, and therefore do not modify the tree-level reduction identities.\par

\subsection{Invariance of the Measure and Initial Conditions}\label{equivalence}
For simplicity, we only consider a single real scalar field; the generalisation to multiple fields is straightforward.\par

\paragraph{The measure} The Jacobian determinant associated with the change of variables reads
\begin{equation}\label{Jacobian-eq083}
    \begin{split}
        \left|\frac{\partial(\Phi(x)+f[\Phi(x)])}{\partial\Phi(y)}\right|=&\left|(1+f^{\prime}[\Phi(x)])\delta^{(4)}(x-y)\right|\\
        =&\exp\left[\mathrm{Tr}\log\left((1+f^{\prime}[\Phi(x)])\delta^{(4)}(x-y)\right)\right]\\
        =&\exp\left[\delta^{(4)}(0)\int d^4x\log \left(1+f^{\prime}[\Phi(x)]\right)\right],    
    \end{split}
\end{equation}
This Jacobian contributes an imaginary term to the effective Lagrangian. When $f$ involves derivatives of the field (e.g., $f[\tilde{\Phi}] = \alpha (\partial \tilde{\Phi})^2$), the Jacobian acquires terms involving derivatives of the delta function, such as
\begin{equation}
\label{eqn084}
    \exp\left[\int d^4x\log \left[1+2\alpha \partial_\mu\Phi(x)\cdot\partial^\mu\delta^{(4)}(x-x)\right]\right].
\end{equation}
For local perturbative scalar redefinitions, such contributions are UV counterterms\footnote{In contrast to scalar fields, local chiral rotations on spinor fields generate a quantum anomaly via the path integral Jacobian (Fujikawa method), leading to an anomalous Ward–Takahashi identity for the axial current. The classical chiral symmetry is thereby violated at the quantum level (e.g., Adler–Bell–Jackiw anomaly), while the vector current remains conserved.}. This is a well-known result in quantum field theory, often referred to in the context of the equivalence theorem for on-shell amplitudes \cite{Chisholm:1961tha,Kamefuchi:1961sb,Divakaran:1963yxz,Kallosh:1972ap} (see also discussions in effective field theory \cite{Passarino:2016saj,Manohar:2018aog,Criado:2018sdb}). In dimensional regularisation, scaleless counterterms vanish, and in other schemes they can be absorbed into local counterterms.\\

An alternative and more explicit demonstration uses the Faddeev-Popov representation of the Jacobian (see~\cite{tHooft:1973wag} for details):
\begin{equation}
\label{eqn085}
    \left|\frac{\partial(\Phi(x)+f[\Phi(x)])}{\partial\Phi(y)}\right|=\int\mathcal{D}c\mathcal{D}\bar{c} \exp\left[-i\int d^4x d^4 y\left(\bar{c}(x)\frac{\partial(\Phi(x)+f[\Phi(x)])}{\partial\Phi(y)}c(y)\right)\right],
\end{equation}
where $c$ is a ghost field. The ghost propagator is trivial ($\sim 1$), and the interaction vertices are local polynomials in momenta when $f$ is a local function. Since ghost fields cannot appear on external legs, they only contribute at loop level\footnote{Consistent with the fact that the Jacobian only contributes to an imaginary part to the Lagrangian, thus it has no impact on the saddle point approximation, which only contains tree level effects.}. In dimensional regularisation, these ghost loops vanish due to symmetric integration over internal momenta, yielding no contribution to physical correlation functions. 

\begin{figure}
    \centering
    \includegraphics[width=0.9\linewidth]{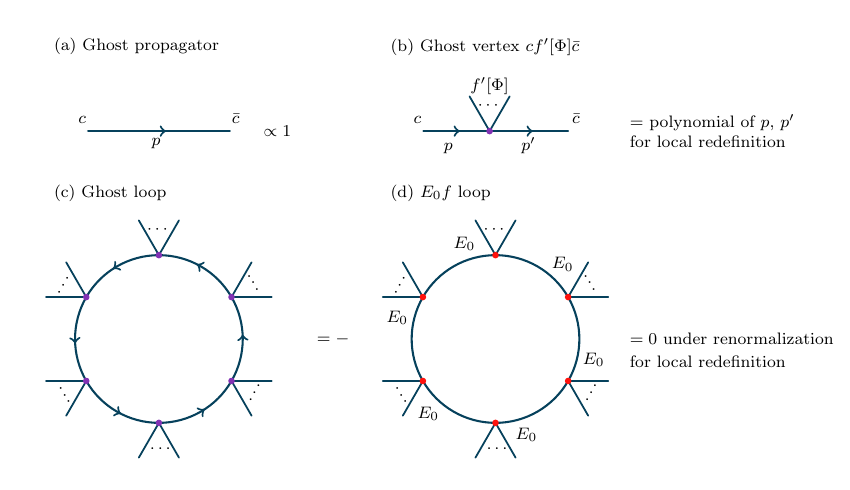}
    \caption{Diagrammatic rules for ghost fields. Arrowed lines denote ghost propagators, while purple dots denote ghost vertices. The external legs with ellipses indicate the field propagators carried by $f'[\Phi]$. For local field redefinitions, ghost loops vanish under renormalization. More generally, they are cancelled by the corresponding loops generated by $E_0 f$ vertices, independently of whether the field redefinition is local.}
    \label{ghost}
\end{figure}

The above statement remains valid even in FRW cosmologies without time-translation invariance: local redefinitions still produce polynomial momentum dependence in the ghost sector, and the spatial-momentum integrals vanish in dimensional regularisation. 
For nonlocal redefinitions (e.g., those involving the inverse Laplacian), genuine quantum anomalies may arise, but it can be shown that the ghost loops always cancel with loops from $E_0f$ vertices (see discussions in \cite{Cohen:2024fak}, and also see \cref{ghost} for illustration of ghost Feynman diagrams). We defer the investigation of such cases to future work.\\

\paragraph{Initial states.}
Another important subtlety arises from the fact that the quantum state $|\Omega\rangle$ is defined via initial conditions at past infinity. The corresponding weight functional is $\lim_{\eta\rightarrow-\infty}\Omega[\Phi_+]\equiv\lim_{\eta\rightarrow-\infty}\braket{\Phi_+(\eta)|\Omega}$, with the $-$ branch obtained by complex conjugation. Here, $|\Phi(\eta)\rangle$ denotes the field eigenstate satisfying $\hat{\Phi}(\eta, \vec{x}) |\Phi(\eta)\rangle = \Phi(\eta, \vec{x}) |\Phi(\eta)\rangle$ for all $\vec{x}$. 
Under field redefinition, the initial condition with respect to the redefined field may be different from the original field.

Under a field redefinition $\Phi = \tilde{\Phi} + f[\tilde{\Phi}]$, the initial conditions on the new field $\tilde{\Phi}$ generally differ from those on the original $\Phi$. In inflationary cosmology, the standard choice is the Bunch--Davies (BD) vacuum, whose weight functional takes a Gaussian form \cite{Chen:2017ryl,Bunch:1978yq}:
\begin{equation}
\label{eqn086}
    \langle \Phi | \Omega \rangle = \mathcal{N} \exp\!\left[ -\frac{1}{2} \int d^3x\, d^3y\, \mathcal{E}(\eta \to -\infty, \vec{x}, \vec{y})\, \Phi(\vec{x}) \Phi(\vec{y}) \right],
\end{equation}
which, equivalently, can be written as
\begin{equation}
\label{eqn087}
    \langle \Phi | \Omega \rangle = \mathcal{N} \exp\!\left[ -\frac{\epsilon}{2} \int d\eta \int d^3x\, d^3y\, \mathcal{E}(\eta, \vec{x}, \vec{y})\, \Phi(\vec{x}) \Phi(\vec{y})\, e^{\epsilon \eta} \right],
\end{equation}
where the kernel is
\begin{equation}
\label{eqn088}
    \mathcal E(\eta,\vec x,\vec y)=a^2(\eta)\int\frac{d^3 k}{(2\pi)^3}e^{i\vec k\cdot(\vec x-\vec y)}k.
\end{equation}
This contribution can be absorbed into the action by adding the term
\begin{equation}
\label{eqn089}
    S[\Phi]\rightarrow  S[\Phi]+\frac{i\epsilon}{2}\int d\eta \hspace{2pt}a^2(\eta)\int\frac{d^3 k}{(2\pi)^3}k\Phi(\vec k)\Phi(-\vec k),
\end{equation}
which is equivalent to a small Wick rotation in conformal time: $\eta \to \eta(1 - i\epsilon)$ for the $+$ branch and $\eta \to \eta(1 + i\epsilon)$ for the $-$ branch in the Schwinger--Keldysh contour. Crucially, this $i\epsilon$-prescription (and thus the BD initial condition) remains invariant under local field redefinitions, since it acts on the time contour rather than on the field variables themselves.



In cases with spontaneous symmetry breaking (SSB), the Gaussian form must be shifted: $\Phi \to \Phi - \bar{\Phi}$, where $\bar{\Phi}$ is the vacuum expectation value (vev). A general local invertible field redefinition $f[\Phi]$ can induce an additional shift in the vev, potentially generating tadpole diagrams. To maintain symmetries (or vanishing vev for fluctuations), one can expand around the true vev as $\Phi(x) = \bar{\Phi} + \Delta\Phi(x)$ and perform a covariant redefinition on the fluctuation field:
\begin{equation}
\label{eqn090}
    \Delta\Phi = \Delta\tilde{\Phi}+f[\Delta\tilde{\Phi} + \bar{\Phi}] - f[\bar{\Phi}].
\end{equation}
This ensures that the redefined fluctuation field $\Delta\tilde{\Phi}$ has zero vev. Such a procedure corresponds to a symmetry-preserving (covariant) field redefinition in the broken phase.

In perturbative calculations, this requirement can be enforced by imposing tadpole cancellation as a renormalisation condition, thereby guaranteeing that physical observables remain unaffected by the choice of field basis.

\subsection{Deriving Schwinger--Dyson Equations}\label{SD-eqn}
To derive diagrammatic relations from the invariance under field redefinitions, we employ the generating functional approach in the Schwinger--Keldysh (in-in) formalism. A general equal-time correlator can be expressed as
\begin{equation}
\label{eqn091}
    \begin{split}
        &\braket{\Omega|\mathcal{O}[\Phi]|\Omega}\\
        =&\mathcal{O}\left[\frac{\delta}{i\delta J^+}\right]Z[J^\pm]|_{J^\pm=0}\\     
        =&\mathcal{O}\left[\frac{\delta}{i\delta J^+}+f\left[\frac{\delta}{i\delta J^+}\right]\right]\exp\left(\sum_{b=\pm}ib\left(S\left[\frac{\delta}{ib\delta J^b}+f\left[\frac{\delta}{ib\delta J^b}\right]\right]-S\left[\frac{\delta}{ib\delta J^b}\right]\right) \right)
        Z[J^\pm]|_{J^\pm=0},
    \end{split}
\end{equation}
where the second line follows from the invariance of the path integral under the local field redefinition $\Phi^b \to \Phi^b + f[\Phi^b]$.

Expanding the exponential to $n$-th order in the variation $\Delta \Phi = f[\Phi]$ and setting sources to zero yields the higher-order Schwinger--Dyson equations:
    \begin{equation}\label{SD-eq092}
    \begin{split}
        &\left( \Delta^n\mathcal{O}\left[\frac{\delta}{i\delta J^+}\right]+\Delta^{n-1}\mathcal{O}\left[\frac{\delta}{i\delta J^+}\right]\sum_{b=\pm}ib\Delta S\left[\frac{\delta}{ib\delta J^b}\right]+\Delta^{n-2}\mathcal{O}\left[\frac{\delta}{i\delta J^+}\right]\right.\\
        \times&\left.\left(\sum_{b=\pm}ib\Delta^{2}S\left[\frac{\delta}{ib\delta J^b}\right]+\frac{1}{2}\left(\sum_{b=\pm}ib\Delta S\left[\frac{\delta}{ib\delta J^b}\right]\right)^2\right)+\ldots\right)Z[J^\pm]|_{J^\pm=0}=0.  
    \end{split}
    \end{equation}
\\
At leading order, we obtain a simple form:
\begin{equation}\label{1st-eq093}
\begin{split}
    \left(\Delta \mathcal{O}\left[\frac{\delta}{i\delta J^+}\right]+\mathcal{O}\left[\frac{\delta}{i\delta J^+}\right]\sum_{b=\pm}ib\Delta S\left[\frac{\delta}{ib\delta J^b}\right]\right)Z[J^\pm]|_{J^\pm=0}=0,
    \end{split}
\end{equation}
The first term represents the variation of the correlator itself, while the second contains the variation of the action induced by the field redefinition. After IBP, this action variation separates into a boundary contribution and a bulk EoM contribution. To see this separation explicitly, we first look at the first variation of the free action $S_0$:
\begin{equation} \label{IBP0-eq094}
\begin{split}
    \Delta S_0=&\int d^4x\hspace{2pt}a^4\left(\nabla_\mu\left(\frac{\partial\mathcal{L}_0}{\partial (\nabla_\mu\Phi)}\Delta \Phi\right)+\left(\frac{\partial\mathcal L_0 }{\partial\Phi}-\nabla_\mu \left(\frac{\partial \mathcal L_0}{\partial(\nabla_\mu \Phi)}\right)\right)\Delta\Phi\right)\\
    =&\int d^3x\hspace{2pt}a^4(\eta_0)\Pi_{0}[\Phi]f[\Phi]+\int d^4 x \hspace{2pt}a^4(\eta)E_{0}[\Phi]f[\Phi] ,
\end{split}
\end{equation}
where $\mathcal{L}_0$ is the free Lagrangian, $\Pi_{0}[\Phi]=a^{-2}\partial_\eta\Phi$ is the conjugate momentum of the free theory of $\Phi$, and $E_{0}[\Phi]=(\square-m^2)\Phi$. After IBP, the boundary term is evaluated on the final time slice, while the bulk EoM term vanishes on-shell at leading order. 
This boundary contribution can also be understood via canonical quantization in the in-in formalism:
\begin{equation}
\label{eqn095}
    \left\langle\left[\mathcal{O}[\hat\Phi],i\int d^3x \hspace{2pt}\Delta H_0[\hat \Pi_0,\hat \Phi]\right]\right\rangle=\left\langle\left[\mathcal{O}[\hat\Phi],-i\int d^3x \hspace{2pt}a^4(\eta_0) \hat\Pi_{0}f[\hat\Phi]\right]\right\rangle=\braket{\Delta \mathcal{O}[\hat\Phi]},
\end{equation}
 arising directly from the equal-time canonical commutation relation
\begin{equation}
\label{eqn096}
    [\hat\Phi(\vec x),\hat\Pi_{0}(\vec y)]=ia^{-4}(\eta_0)\delta^{(3)}(\vec x-\vec y) .
\end{equation}

When interactions are included, the EoM term $E[\Phi] = (\square - m^2 + \cdots) \Phi$ contributes at higher orders, and $\Delta S_{\rm int}$ yields analogous IBP terms. To disentangle boundary and bulk contributions, we decompose the field redefinition into retarded and advanced parts:
\begin{equation}
\label{eqn097}
    f_r[\Phi] = f[\Phi] \theta(\eta - \eta_0), \quad f_a[\Phi] = f[\Phi] \theta(\eta_0 - \eta),
\end{equation}
with corresponding variations $\Delta_r$ and $\Delta_a$. Then,
\begin{equation}
\label{eqn098}
    \begin{split}
        \Delta_r S=&\int d^4 x \hspace{2pt}a^4\nabla_\mu\left(\frac{\partial\mathcal{L}}{\partial (\nabla_\mu\Phi)}f[\Phi]\right)=\int d^3x \hspace{2pt}a^4(\eta_0) \Pi f[\Phi],\\
        \Delta_a S=&\int d^4 x \hspace{2pt}a^4  \hspace{2pt} \left(\frac{\partial\mathcal L }{\partial\Phi}-\nabla_\mu \left(\frac{\partial \mathcal L}{\partial(\nabla_\mu \Phi)}\right)\right)f[\Phi]= \int d^4 x \hspace{2pt}a^4(\eta)   E[\Phi]f[\Phi],\\
        \Delta_r \mathcal{O}=&\sum_{i}\frac{\partial \mathcal{O}}{\partial\Phi(\vec x_i)}f[\Phi(\vec x_i)]=\delta \mathcal{O},\\
        \Delta_a \mathcal{O}=&0,
    \end{split}
\end{equation}
Splitting \cref{1st-eq093} accordingly yields the leading-order relations (equivalent to the reduction rules in \cref{reduction} after perturbative expansion in $S_{\rm int}$):
\begin{equation}
\label{eqn099}
\begin{split}
     &\Delta_r:\left(\Delta \mathcal{O}\left[\frac{\delta}{i\delta J^+}\right]+\mathcal{O}\left[\frac{\delta}{i\delta J^+}\right]\sum_{b=\pm}ib\int d^3x \hspace{2pt}a^4(\eta_0) \Pi\left[\frac{\delta}{ib\delta J^b}\right]f\left[\frac{\delta}{ib\delta J^b}\right]\right)Z[J^\pm]|_{J^\pm=0}=0,\\
    &\Delta_a:\mathcal{O}\left[\frac{\delta}{i\delta J^+}\right]\sum_{b=\pm}ib\int d^4 x \hspace{2pt}a^4(\eta)E\left[\frac{\delta}{ib\delta J^b}\right]f\left[\frac{\delta}{ib\delta J^b}\right]
    Z[J^\pm]|_{J^\pm=0}=0.
\end{split}
\end{equation}

For the second-order variation, similar decompositions give
\begin{equation}\label{2rd-eq100}
    \begin{split}
        \Delta_r^2 S=&2\int d^3x \hspace{2pt}a^4(\eta_0) \left(\frac{\partial^2\mathcal L}{\partial\Phi\partial(\partial_\eta\Phi)} f^2[\Phi]+\frac{\partial^2\mathcal L}{\partial(\partial_\eta\Phi)^2} f[\Phi]f^{\prime}[\Phi]\right)
        =-2\Delta_r\Delta_a S,\\
        \Delta_a^2 S=&\int d^4 x \hspace{2pt}a^4(\eta)\left(\frac{\partial^2\mathcal L}{\partial\Phi^2} f^2[\Phi]+2\frac{\partial^2\mathcal L}{\partial\Phi\partial(\partial_\eta\Phi)} f[\Phi]f^{\prime}[\Phi]+\frac{\partial^2\mathcal L}{\partial(\partial_\eta\Phi)^2} f^{\prime2}[\Phi]\right)=\Delta^2S,\\
        \Delta_r^2 \mathcal{O}=&\sum_{i,j}\frac{\partial^2 \mathcal{O}}{\partial\Phi(\vec x_i)\partial\Phi(\vec x_j)}f[\Phi(\vec x_i)]f[\Phi(\vec x_j)]=\Delta^2 \mathcal{O},\\
        \Delta_r\Delta_a \mathcal{O}=&\Delta_a^2 \mathcal{O}=0,
    \end{split}
\end{equation}
The resulting second-order Schwinger--Dyson equation splits into three parts
\begin{equation}\label{2.12-eq101}
\begin{aligned}
         \Delta_r^2&:\left(\Delta^2 \mathcal{O}\left[\frac{\delta}{i\delta J^+}\right]+2\Delta \mathcal{O}\left[\frac{\delta}{i\delta J^+}\right]\sum_{b=\pm}ib\int d^3x \hspace{2pt}a^4(\eta_0) \Pi\left[\frac{\delta}{ib\delta J^b}\right]f\left[\frac{\delta}{ib\delta J^b}\right]\right.\\
    +&\left.\mathcal{O}\left[\frac{\delta}{i\delta J^+}\right]\left(\sum_{b=\pm}ib\Delta_r^2S\left[\frac{\delta}{ib\delta J^b}\right]+\left(\sum_{b=\pm}ib\int d^3x \hspace{2pt}a^4(\eta_0)\Pi\left[\frac{\delta}{ib\delta J^b}\right]f\left[\frac{\delta}{ib\delta J^b}\right]\right)^2\right)\right)\\
    \times& Z[J^\pm]|_{J^\pm=0}=0, \\
    \Delta_r\Delta_a&:\left(\Delta \mathcal{O}\left[\frac{\delta}{i\delta J^+}\right]\sum_{b=\pm}ib\int d^4 x \hspace{2pt}a^4(\eta)E\left[\frac{\delta}{ib\delta J^b}\right]f\left[\frac{\delta}{ib\delta J^b}\right]\right.\\
    +&\left.\mathcal{O}\left[\frac{\delta}{i\delta J^+}\right]\left(\sum_{b=\pm}ib\Delta_r\Delta_aS\left[\frac{\delta}{ib\delta J^b}\right]+\left(\sum_{b=\pm}ib\int d^3x \hspace{2pt}a^4(\eta_0) \Pi\left[\frac{\delta}{ib\delta J^b}\right]f\left[\frac{\delta}{ib\delta J^b}\right]\right)\right.\right. \\
    \times&\left.\left.\left(\sum_{b=\pm}ib\int d^4 x \hspace{2pt}a^4(\eta)E\left[\frac{\delta}{ib\delta J^b}\right]f\left[\frac{\delta}{ib\delta J^b}\right]\right)\right)\right) \\
    \times& Z[J^\pm]|_{J^\pm=0}=0, \\
    \Delta_a^2&:\mathcal{O}\left[\frac{\delta}{i\delta J^+}\right]\left(\sum_{b=\pm}ib\Delta_a^2 S\left[\frac{\delta}{ib\delta J^b}\right]+\left(\sum_{b=\pm}ib\int d^4 x \hspace{2pt}a^4(\eta) E\left[\frac{\delta}{ib\delta J^b}\right]f\left[\frac{\delta}{ib\delta J^b}\right]\right)^2\right) \\
    \times& Z[J^\pm]|_{J^\pm=0}=0.
\end{aligned}
\end{equation}

Although \cref{2.12-eq101} appears rather involved at first glance, we can exploit the structure already established at leading order in \cref{2.2-eq013} (and the associated diagrammatic reduction rules illustrated in \cref{reduction}) to simplify it significantly. First, replace $\mathcal{O}$ in the leading-order relation \cref{2.2-eq013} by its first variation $\Delta \mathcal{O}$. The same set of diagrammatic reduction rules then applies to the modified Feynman diagrams involving $\Delta \mathcal{O}$ as the ``observable.'' Furthermore, the reduction rules remain valid even when the diagram structure is altered by inserting additional $\Delta_r S$ or $\Delta_a S$ vertices that are disconnected from the cuts (i.e., the lines severed during the reductions). If two $E_0f$ vertices form a bubble loop, it was shown in \cite{Cohen:2024fak} that this loop cancels with the corresponding ghost loop, whether or not the ghost loop itself vanishes. By subtracting the contributions of these modified (disconnected-vertex) diagrams from the full second-order expression in \cref{2.12-eq101}, the extra terms cancel, leaving a unified form for the remaining connected piece:
\begin{equation}\label{2.18-eq102}
\begin{split}
       &\mathcal{O}\left[\frac{\delta}{i\delta J^+}\right]\left(\sum_{b=\pm}ib\Delta^2_{xy}S\left[\frac{\delta}{ib\delta J^b}\right]+\left(\sum_{b=\pm}ib\Delta_xS\left[\frac{\delta}{ib\delta J^b}\right]\right)\left(\sum_{b=\pm}ib\Delta_yS\left[\frac{\delta}{ib\delta J^b}\right]\middle)\right|_{c}\right)\\&Z[J^\pm]|_{J^\pm=0}=0.
\end{split}
\end{equation}
where $x,y \in \{r,a\}$ and the subscript $c$ restricts to diagrams where $\Delta_x S_0$ (via $\Pi_0$ or $E_0$) links to another $\Delta_y S$. Thus, by replacing $S_{\rm int}$ with $\Delta_xS$ in the second and fourth reduction rules, we directly obtain \cref{2.18-eq102}. Using $\Delta_r^2S=-2\Delta_r\Delta_aS$, equivalently $\frac12\Delta_r^2S+\Delta_r\Delta_aS=0$, the first two parts cancel with each other, leaving no boundary contributions from the second-order variation; the remaining $\Delta_a^2$ part yields \cref{2.3-eq021}.

We claim that higher-order variations similarly produce Schwinger--Dyson-like equations encoding the four diagrammatic reduction rules of \cref{reduction}, which in turn correspond to fundamental principles in canonical quantization:
\begin{itemize}
    \item Reduction rule (1): Canonical commutation relations 
    \item Reduction rule (2): Heisenberg equation (Unitary evolution)
    \item Reduction rule (3)$\&$(4): Green function identities: $(\square-m^2)\Delta_{bc}(x,y)=ib\delta^{(4)}(x-y)\delta_{bc}$
\end{itemize}
In particular, in the canonical (in-in) picture, Reduction rule (2) corresponds to the leading perturbative expansion of the Heisenberg equation:
\begin{equation}
\label{eqn103}
\begin{split}
     \Delta_a \int dt H_{int}[\hat\Phi]\simeq&\Delta_a \int dtH_{int}[\Phi]\\
     \simeq&-\Delta_a(S[\Phi]-S_0[\hat\Phi])\\
     =&-\int d^3x (\Pi[\Phi]f[\Phi]-\Pi_0[\hat\Phi]f[\hat \Phi])\\
     \simeq&-i\int d^3x\left[\Pi_0[\hat\Phi]f[\hat\Phi],\int dt H_{int}\left[\hat \Phi,\Pi_0[\hat\Phi]\right]\right].
\end{split}
\end{equation}

\section{Details of Bootstrap}\label{Bootstrap details} 
In this appendix, we employ the bootstrap method to calculate the cosmological correlators discussed in the main text \cref{IBP}, including the correlator $\langle \phi\phi\sigma \rangle'_{g_3}$, the bispectrum $\langle \phi\phi\phi \rangle'_{g_3\times\lambda_2}$, and the trispectrum $\langle \phi\phi\phi\phi \rangle'_{g_3\times g_3}$. These rigorous computations serve as a cross-check for the results derived via the IBP relations.

For the 3-pt correlator, we introduce the dimensionless variables:
\begin{equation}
\label{eqn104}
    u \equiv \frac{k_3}{k_{12}} \,, \quad x \equiv -k_{3} \eta \,, \quad x_0 \equiv -k_3 \eta_0 \,,
\end{equation}
where we use the shorthand $k_{ij} = k_i + k_j$.
For the 4-pt case, we define the variables:
\begin{equation}
\label{eqn105}
    u \equiv \frac{s}{k_{12}} \,, \quad v \equiv \frac{s}{k_{34}} \,, \quad x \equiv -s \eta, \quad x_0 \equiv -s \eta_0 \,.
\end{equation}
We use hatted quantities for the dimensionless seed integrals after stripping off the overall momentum factors and coupling constants. 

\subsection{\texorpdfstring{3-pt Correlator $\langle \phi\phi\sigma \rangle'$}{3-pt Correlator phi phi sigma}}
The 3-pt correlator $\langle \phi \phi \sigma\rangle'$ generated by the interaction $g_3 \phi^2\sigma$ is given by
\begin{equation}
\label{eqn106}
    \langle \phi\phi\sigma \rangle'_{g_3} = g_3 \sum_{b=\pm} (ib) \int_{-\infty}^{\eta_0} \frac{d\eta}{(-H\eta)^4} K^\phi_b(k_1, \eta) K^\phi_b(k_2, \eta) K^\sigma_b(k_3, \eta) \,,
\end{equation}
For convenience, we introduce the dimensionless propagator of $\sigma$:
\begin{equation}
\label{eqn107}
    K^\sigma_b(k; \eta) \equiv \frac{H^2}{k^3} \hat{K}_{b}^\sigma(k\eta, k\eta_0) \,.
\end{equation}
Substituting these into the integral, we obtain:
\begin{equation}
\label{eq:integralpps-eq108}
\begin{aligned}
    &\langle \phi\phi\sigma \rangle'_{g_3} \\
    =& \frac{g_3 H^2}{4 k_1^3 k_2^3} \sum_{b=\pm} (ib)\, \int_{-\infty}^{\eta_0} \frac{d\eta}{\eta^4} \prod_{i=1}^2(1+ib k_i \eta_0)(1-ibk_i\eta)e^{ibk_{12}(\eta-\eta_0)} \frac{\hat{K}_b^\sigma(k_3\eta, k_3\eta_0)}{k_3^3}\\
    =&\frac{g_3 H^2}{4 k_1^3 k_2^3} O_{12}\hat{\mathcal{I}}^{\phi^2\sigma}(u,x_0) \,.
\end{aligned}
\end{equation}
The seed integral $\hat{\mathcal{I}}^{\phi^2\sigma}(u,x_0)$ is
\begin{equation}
\label{eqn109}
\begin{split}
        \hat{\mathcal{I}}^{\phi^2\sigma}(u,x_0)=&\sum_{b=\pm}(ib)\int^{\infty}_{x_0}\frac{\mathrm{d}x}{x^4}\left(1-ib\frac{x_0}{u}\right)\left(1+ib\frac{x}{u}\right)e^{-ib\frac{x-x_0}{u}}\hat{K}_b^\sigma(x,x_0)\\
    =&\sum_{b=\pm}(ib)\,\mathcal C_b(u,x_0)\int^{\infty}_{x_0}\frac{\mathrm{d}x}{x^4}\left(1+ib\frac{x}{u}\right)e^{-ib\frac{x}{u}}\hat{K}_b^\sigma(x,x_0)\\
    \equiv&\int^{\infty}_{x_0}\frac{\mathrm{d}x}{x^4}G^{\phi^2\sigma}[u,x,x_0],
\end{split}
\end{equation}
where the factor $C_b(u,x_0)$ and the operator $O_{12}$ are defined as
\begin{equation}
\label{eqn110}
\begin{aligned}
   \mathcal{C}_b(u,x_0)=&\left(1-ib\frac{x_0}{u}\right)e^{ib\frac{x_0}{u}},\\
    O_{12}\equiv&1-\frac{k_1k_2}{k_{12}}\partial_{k_{12}}=1+\frac{k_1k_2}{k_{12}^2}{u\partial_{u}}.
\end{aligned}
\end{equation}
Now we introduce the bootstrap operator $\tilde\Delta_u \equiv (u^2 - u^4)\partial_u^2 + (4u - 4u^3)\partial_u$ for $\Delta_{\phi}=0$, which satisfies the commutation relation:
\begin{equation}
\label{eqn111}
    \tilde\Delta_u \mathcal{C}_b(u, x_0)=\mathcal{C}_b(u, x_0)\tilde\Delta_u+\mathcal O(x_0^2),
\end{equation}
where the $\mathcal O(x_0^2)$ part can be safely dropped when acting on the seed integral. We can derive the bootstrap equation by applying the equality
\begin{equation} \label{bootstrap-equality-eq112}
    (x^2\partial_{x}^2-2x\partial_x+x^2)\left(1+ib\frac{x}{u}\right)e^{-ib\frac{x}{u}}=\tilde\Delta_u\left(1+ib\frac{x}{u}\right)e^{-ib\frac{x}{u}},
\end{equation}
and using IBP:
\begin{equation}\label{eq:master_eq_3pt-eq113}
\begin{split}
    & \left(\tilde\Delta_u +\frac{m^2}{H^2}\right)\hat{\mathcal{I}}^{\phi^2\sigma}(u, x_0)\\
     =&\sum_{b=\pm}(ib)C_b(u,x_0)\int^{\infty}_{x_0}\frac{\mathrm{d}x}{x^4}\left(x^2\partial_{x}^2-2x\partial_x+x^2+\frac{m^2}{H^2}\right)\left[\left(1+ib\frac{x}{u}\right)e^{-ib\frac{x}{u}}\right]\hat{K}_b^\sigma(x,x_0)\\
     =&\sum_{b=\pm}(ib)C_b(u,x_0)\Biggl[\frac{\hat{K}_b^\sigma(x_0,x_0)\partial_{x_0}\left[\left(1+ib\frac{x_0}{u}\right)e^{-ib\frac{x_0}{u}}\right]-\partial_{1}\hat{K}_b^\sigma(x_0,x_0)\left(1+ib\frac{x_0}{u}\right)e^{-ib\frac{x_0}{u}}}{x_0^2}\Biggr.\\
    +& \Biggl.\int^{\infty}_{x_0}\frac{\mathrm{d}x}{x^4}\left(1+ib\frac{x}{u}\right)e^{-ib\frac{x}{u}}\left(x^2\partial_{x}^2-2x\partial_x+x^2+\frac{m^2}{H^2}\right)\hat{K}_b^\sigma(x,x_0)\Biggr]\\
    =&\sum_{b=\pm}(ib)C_b(u,x_0)\frac{e^{-ib\frac{x_0}{u}}[x_0 \hat{K}_b^\sigma(x_0,x_0)-u^2(1+ib\frac{x_0}{u})\partial_1 \hat{K}_b^\sigma(x_0,x_0)]}{u^2x_0^2}\\
    =&1-\frac{2}{u^3}\hat K^\sigma(x_0,x_0),
\end{split}  
\end{equation}
The RHS of \cref{eq:master_eq_3pt-eq113} serves as the source term for the bootstrap equation. We refer to such source terms as boundary conformal anomalies, namely anomalous source terms in momentum-space conformal Ward identities induced by the late-time cutoff \cite{Bzowski:2015pba,Bzowski:2022rlz,Wang:2022eop,Bzowski:2023nef}. For the heavy mass regime, only the term proportional to the Wronskian $\partial_1 \hat K^\sigma_b(x_0,x_0)$ survives. The inhomogeneous solution for the integral $\hat{\mathcal{I}}(u, x_0)$ simplifies to a constant:
\begin{equation}
\label{eq:inhom-phi2sigma-temp-eq114}
    \hat{\mathcal{I}}^{\phi^2\sigma}_{\text{inh}}(u, x_0) = \frac{H^2}{m^2} \, ,
\end{equation}
while the homogeneous solution decays when $x_0 \to 0$ (because contact diagram cannot carry oscillatory signal). Therefore, we obtain the result for the correlator $\langle \phi\phi\sigma \rangle'_{g_3}$:
\begin{equation}
\label{eqn115}
    \langle \phi\phi\sigma \rangle'_{g_3} = \frac{g_3 H^4}{4 m^2 k_1^3 k_2^3} \,,
\end{equation}
which is consistent with the direct boundary evaluation discussed in \cref{sec:building_block}. Although this seed function has no late-time divergence, it is still different from the corresponding triple-K integral $z_0^\Delta I_{0,0,\Delta}$ \footnote{In this combination of conformal weights, the triple-K integral is badly divergent, i.e. with cutoff $z_0$, it behaves as $I_{0,0,\Delta}^{z_0}\sim \frac{z_0^{-\Delta}}{\Delta}$. When $\Delta\rightarrow0$, the divergence becomes moderate, $I_{0,0,\Delta}^{z_0}\sim \frac{z_0^{-\Delta}}{\Delta}\sim\frac{1}{\Delta}-\log z_0$.} \cite{Bzowski:2013sza,Coriano:2013jba,Bzowski:2015pba}, since it does not satisfy the standard conformal Ward identities, which come from boundary conformal symmetry.
But we find the modified bootstrap equation (with anomalous source) can be written into
\begin{equation}\label{modified-bootstrap-equation-eq116}
\begin{split}
    &\left(\tilde\Delta_u+\frac{m^2}{H^2}\right)\hat{\mathcal{I}}^{\phi^2\sigma}(u,x_0)\\
    =& (x_0^2\partial_{x_0}^2-2x_0\partial_{x_0})\hat{\mathcal{I}}^{\phi^2\sigma}(u,x_0)-\int^{\infty}_{x_0}\frac{dx}{x^4}(x_0^2\partial_{x_0}^2-2x_0\partial_{x_0}) G^{\phi^2\sigma}[u,x,x_0],
\end{split}
\end{equation}
which can be understood as a direct consequence of dS isometry. Naively, from dS isometry to conformal Ward identities, one replaces $\eta_0\partial_{\eta_0}$ acting on a boundary operator $\Phi_i$ by $\Delta_i$, because $\Phi_i(\eta_0)\sim\eta_0^{\Delta_i}$. However, in late-time divergent cases, the integration upper bound and the next-to-leading terms in the OPE also play important roles. We should therefore recover these non-trivial contributions by reintroducing $\eta_0\partial_{\eta_0}$ operators, while subtracting the trivial contributions $\propto \Delta$. 
The bootstrap equation comes from a non-linear combination of conformal Ward identities. Since in this case there is only one field with $\Delta\neq0$, each $\Delta$ appearing in the bootstrap equation indicates an action of $\eta_0\partial_{\eta_0}$. Thus $m^2/H^2=3\Delta-\Delta^2\sim2\eta_0\partial_{\eta_0}-\eta_0^2\partial_{\eta_0}^2\sim2x_0\partial_{x_0}-x_0^2\partial_{x_0}^2$. By reintroducing these operators and subtracting the trivial contributions (the second terms on the RHS), we obtain \cref{modified-bootstrap-equation-eq116}\footnote{Here the situation is different from the case considered in \cite{Wang:2022eop}, where the integral has a late-time divergence; instead, we have a constant term plus a decaying term. Therefore the criteria proposed there can no longer determine whether the dilation conformal Ward identity is broken in this case. However, \cref{modified-bootstrap-equation-eq116} provides an effective way to obtain the modified bootstrap equation whether or not the anomalous source term is late-time divergent.}.

\subsection{\texorpdfstring{Bispectrum $\langle\phi\phi\phi\rangle'$}{Bispectrum phi phi phi}}

In this subsection, we extend the bootstrap analysis to the 3-pt exchange diagram of the inflaton. As detailed in the main text, the exchange process is mediated by the massive scalar $\sigma$ via the mixed coupling $\lambda_2 \dot{\phi}\sigma$. We consider two types of cubic interactions for $\sigma$: the non-derivative interaction $g_3 \phi^2 \sigma$ and the derivative interaction $\lambda_3 (\partial_\mu\phi)^2 \sigma$.

We first compute the 3-pt exchange diagram generated by the cubic interaction $g_3 \phi^2 \sigma$, denoted as $\langle\phi\phi\phi\rangle'_{g_3\times\lambda_2}$. To incorporate the mixed coupling $\lambda_2 \dot{\phi}\sigma$, it is convenient to introduce the mixed propagator $\mathcal{K}_b(k, \eta_1, \eta_0)$:
\begin{equation}
\label{eqn117}
\begin{aligned}
    \mathcal{K}_b(k, \eta_1, \eta_0) =& \lambda_2 \sum_{c=\pm} (ic) \int_{-\infty}^{\eta_0} \frac{d\eta_2}{(-H\eta_2)^3} D^\sigma_{bc}(k,\eta_1, \eta_2) \partial_{\eta_2} K^\phi_c(k, \eta_2)\\
    =& \frac{\lambda_2 H}{k^3} \hat{\mathcal{K}}_b(k\eta, k\eta_0)\,.
\end{aligned}
\end{equation}
where $\hat{\mathcal{K}}_b$ is the dimensionless mixed propagator by rescaling. With the help of the mixed propagator, the 3-pt correlator can be written as
\begin{equation}
\label{eqn118}
\begin{aligned}
    \langle\phi\phi\phi\rangle'_{g_3\times\lambda_2} =& g_3  \sum_{b=\pm} (ib) \int_{-\infty}^{\eta_0} \frac{d\eta}{(-H\eta)^4} K^\phi_b(k_1, \eta) K^\phi_b(k_2, \eta) \mathcal{K}_b(k_3, \eta, \eta_0) \\
    =&\frac{g_3 \lambda_2 H}{4 k_1^3 k_2^3} O_{12} \hat{\mathcal{I}}^{\phi^3}(u, x_0).
\end{aligned}
\end{equation}
The seed integral $\mathcal{I}^{\phi^3}(u, x_0)$ is defined as
\begin{equation}
\label{eqn119}
    \hat{\mathcal{I}}^{\phi^3}(u, x_0) = \sum _{b=\pm}(ib)\mathcal C_b(u,x_0)\int^{\infty}_{x_0}\frac{dx}{x^4}\left(1+ib\frac{x}{u}\right)e^{-ib\frac{x}{u}}\hat{\mathcal K}_b(x,x_0)\equiv \int^{\infty}_{x_0}\frac{dx}{x^4} G^{\phi^3}[u,x,x_0],
\end{equation}
where $\hat{\mathcal K}_b(x,x_0)$ reads
\begin{equation}
\label{eqn120}
    \hat{\mathcal K}_b(x,x_0)=-\sum_{c=\pm}(ic)\int_{x_0}^{\infty}\frac{dy}{y^2}\hat{D}^{\sigma}_{bc}(x,y)(1-icx_0)e^{-ic(y-x_0)},
\end{equation}
and the dimensionless propagator satisfies
\begin{equation}
\label{eqn121}
    \left(x^2\partial_{x}^2-2x\partial_x+x^2+\frac{m^2}{H^2}\right)\hat{D}^{\sigma}_{bc}(x,y)=ibx^4\delta_{bc}\delta(x-y).
\end{equation}
We obtain a bootstrap equation similar to \cref{eq:master_eq_3pt-eq113}:
\begin{equation}\label{eq:master_eq_phiphiphi-eq122}
\begin{aligned}
    &\left(\tilde\Delta_u+\frac{m^2}{H^2}\right) \hat{\mathcal{I}}^{\phi^3}(u, x_0)\\
    =& \sum_{b=\pm}(ib)  \mathcal{C}_b(u,x_0)\left[\frac{e^{-ib\frac{x_0}{u}}[x_0 \hat{\mathcal K}_b(x_0,x_0)-u^2(1+ib\frac{x_0}{u})\partial_1 \hat{\mathcal K}_b(x_0,x_0)]}{u^2x_0^2}\right.\\
    +&\left.\int_{x_0}^{\infty}\frac{\mathrm{d}x}{x^4}\left(1+ib\frac{x}{u}\right)e^{-ib\frac{x}{u}}x^2(1-ibx_0)e^{-ib(x-x_0)}\right].
\end{aligned}
\end{equation}
Since the mixed propagator satisfies
\begin{equation}
\label{eqn123}
    \hat{\mathcal K}_b(x_0,x_0)\sim\mathcal O(x_0^{2}),\, \partial_1\left(\hat{\mathcal K}_+(x_0,x_0)-\hat{\mathcal K}_-(x_0,x_0)\right)=0,
\end{equation}
the boundary contribution in the first line vanishes in the limit $x_0 \to 0$. In contrast, the bulk contribution in the second line survives and manifests the late-time divergence. This bulk contribution is precisely the 3-pt contact diagram with the interaction $\phi^2\dot{\phi}$, so the bootstrap equation simplifies to
\begin{equation}
    \left(\tilde\Delta_u+\frac{m^2}{H^2}\right) \hat{\mathcal{I}}^{\phi^3}(u, x_0) = \gamma_E - 1 + \log x_0 + \log\left(1+\frac{1}{u}\right) - \frac{1}{u} \,.
    \label{eq:diff_eq_simplified-eq124}
\end{equation}
 The inhomogeneous solution is
\begin{equation}
    \begin{split}\label{eq:3ptinh-eq125}
    \hat{\mathcal{I}}^{\phi^3}_{inh}(u, x_0)=&\frac{H^2}{m^2}\left[\gamma_E-1+\log\left(x_0\left(1+\frac{1}{u}\right)\right)-\frac{1}{u}\right.\\
    +&\left.\frac{3H^2}{m^2}-\frac{2H^2}{m^2-2H^2}\frac{1}{u}+2\sum_{n=1}^\infty a_n u^n\right].
    \end{split}
\end{equation}
where the first two coefficients of the series are
\begin{equation}
\label{eqn126}
    a_1=-\left(1-\frac{2}{\frac{1}{4}-\nu^2}\right)\frac{1}{\frac{25}{4}-\nu^2}, \quad a_2=\frac{1}{\frac{49}{4}-\nu^2},
\end{equation}
and the recurrence relation is
\begin{equation}
\label{eqn127}
    a_n=\frac{(-1)^n+(n+1)(n-2)a_{n-2}}{(n+\frac{3}{2})^2-\nu^2}.
\end{equation}
The homogeneous solution is given by
\begin{equation}\label{eq:3pthomsol-eq128} 
	\hat{\mathcal{I}}^{\phi^3}_{h}(u) = \sum_{\pm} \mathcal{N}_\pm \, \, u^{-\frac{3}{2} \pm \nu}\, {}_2F_1\left(-\frac{3}{4} \pm \frac{\nu}{2},\, \frac{3}{4} \pm \frac{\nu}{2};\, 1 \pm \nu;\, u^2\right) \,.
\end{equation}
By avoiding folded singularity at $u\rightarrow 1$ and matching the total energy singularity at $u\rightarrow-1$, we can fix the coefficients as 
\begin{equation} \label{eq:N_pm-eq129}
	\mathcal{N}_\pm = (-1)^{-\frac{3}{4} \pm \frac{\nu}{2}}\frac{(1+i)\pi^{3/2}}{16 \cos(\pi\nu)} e^{\mp i\frac{\pi\nu}{2}} 2^{\mp \nu} (3 \pm 2\nu) \left( 1 \pm \frac{1}{\sin(\pi\nu)} \right) \frac{\Gamma(\frac{1}{2} \pm \nu)}{\Gamma(1 \pm \nu)} \,.
\end{equation}
For the derivative interaction $\lambda_3 (\partial_\mu\phi)^2 \sigma$, the corresponding exchange diagram $\langle\phi\phi\phi\rangle'_{\lambda_3\times\lambda_2}$ is free of late-time divergence. This diagram has been well-studied in~\cite{Pimentel:2022fsc}, via the weight-shifting operator acting on a conformal seed integral. The inhomogeneous solution is 
\begin{equation}\label{eq:3ptinhsollamda3-eq130}
    \hat{\mathcal{I}}^{\phi^3}_{inh}(u) = \frac{3 H^2}{2m^2} - \frac{H^2}{m^2-2H^2}\frac{1}{u} +  \sum_{n=1}^{\infty} a_n u^n \,,
\end{equation}
where the coefficients $a_n$ are the same as those in \cref{eq:3ptinh-eq125}. The homogeneous solutions are also identical.

Therefore, by comparing the explicit analytic results derived above for the two exchange diagrams, one can check that their difference corresponds precisely to the contribution from the contact diagram with the coupling $\phi^2\dot{\phi}$. This explicitly verifies the relation derived in the main text via the IBP method, given in \cref{eq:3ptrelation-eq032}.

\subsection{\texorpdfstring{Trispectrum $\langle\phi\phi\phi\phi\rangle'$}{Trispectrum phi phi phi phi}}
In this subsection, we compare the explicit results from IBP with those derived by the bootstrap method (see Appendix D in \cite{Arkani-Hamed:2018kmz}). We first list the results here; the contact sources read:
\begin{equation}\label{contact-bootstrap-eq131}
    \begin{split}
        C_n=&s^3 O_{12}O_{34}\hat{C}_n,\\
        \hat{C}_0=&\frac{1}{3}\left[\left(\frac{1}{u^3}+\frac{1}{v^3}\right)\log\left(\frac{uv}{u+v}\right)+\left(\frac{1}{u}+\frac{1}{v}\right)\frac{1}{uv}\right],\\
        \hat{C}_1=&\frac{u^2+v^2+uv-1}{uv(u+v)}=\tilde\Delta_u\hat{C}_{0}+\frac{1}{u^3},\\
        \hat{C}_n=&\tilde\Delta_u\hat{C}_{n-1}, \text{ for }n\geq2.
    \end{split}
\end{equation}
and exchange solutions are:
\begin{equation}
\label{eqn132}
\begin{split}
    \left(\tilde\Delta_u+\frac{m^2}{H^2}\right)\hat F_n=& (-1)^n \hat C_n,\\
    \hat F_{n}=&\frac{m^2}{H^2}\hat F_{n-1}+(-1)^n\hat C_{n-1}-\frac{H^2}{m^2}\frac{1}{u^3}\delta_{n1}.
    \end{split}
\end{equation}
In their derivation, there are no conformal anomalous source terms. In the actual 4-pt exchange diagrams, however, extra source terms may emerge. We now use a derivation similar to the $\braket{\phi^2\sigma}^\prime$ and $\braket{\phi^3}^\prime$ cases. We first consider the 4-pt $s$-channel exchange diagram generated by two cubic interactions $g_3 \phi^2 \sigma$, denoted as $\langle\phi\phi\phi\phi\rangle'_{g_3\times g_3}$. In this case, the exchange process involves a massive propagator connecting two pairs of external legs, $(k_1, k_2)$ and $(k_3, k_4)$. As in the 3-pt case, we define the composite propagator $\mathcal K_b(k_3,k_4,\eta,\eta_0)$:
\begin{equation}
\label{eqn133}
\begin{split}
        \mathcal K_b(k_3,k_4,\eta_1,\eta_0)=&g_3\sum_{c=\pm}(ic)\int^{\eta_0}_{-\infty}\frac{d\eta_2}{(-H\eta_2)^4}D^{\sigma}_{bc}(s,\eta_1,\eta_2)K_c^{\phi}(k_3,\eta_2)K^{\phi}_c(k_4,\eta_2)\\
        =&\frac{g_3H^2}{4k_3^3k_4^3}O_{34}\hat K_b(s\eta_1,s\eta_0,k_{34}\eta_0).
\end{split}
\end{equation}
The 4-pt correlator can be written as:
\begin{equation}
\label{eqn134}
   \begin{aligned}
    \langle\phi\phi\phi\phi\rangle'_{g_3\times g_3} =& g_3 \sum_{b=\pm}(ib) \int_{-\infty}^{\eta_0} \frac{d\eta}{(-H\eta)^4} K^\phi_b(k_1; \eta) K^\phi_b(k_2; \eta) \mathcal K_b(k_3,k_4,\eta,\eta_0)\\=&\frac{g_3^2 H^2 s^3}{16 k_1^3 k_2^3 k_3^3 k_4^3}O_{12}O_{34}\hat{\mathcal{I}}^{\phi^4}_{(0,0)}(u,v,x_0).
\end{aligned} 
\end{equation}
The seed integral $\mathcal I^{\phi^4}_{(0,0)}(u,v,x_0)$ is defined as
\begin{equation}
\label{eqn135}
\begin{split}
    \hat{\mathcal{I}}^{\phi^4}_{(0,0)}(u,v,x_0) =& \sum_{b=\pm}(ib)\mathcal C_b(u,x_0)\int_{x_0}^{\infty}\frac{dx}{x^4} \left(1+ib\frac{x}{u}\right)e^{-ib\frac{x}{u}} \hat{\mathcal K}_b(x,x_0,v)\\
    \equiv& \int_{x_0}^{\infty}\frac{dx}{x^4} G^{\phi^4}[u,v,x,x_0],
\end{split}
\end{equation}
where $\hat{\mathcal K}_b(x,x_0,v)$ reads
\begin{equation}
\label{eqn136}
    \hat{\mathcal K}_b(x,x_0,v)=\sum_{c=\pm}(ic)\mathcal C_c(v,x_0)\int_{x_0}^{\infty}\frac{dy}{y^4}\hat{D}^\sigma_{bc}(x,y)(1+ic\frac yv)e^{-ic\frac yv}, 
\end{equation}
Similar to the $\langle\phi\phi\phi\rangle'$ case, we obtain the bootstrap equation satisfied by $ \hat{\mathcal{I}}^{\phi^4}(u,v,x_0)$:
\begin{equation}
\begin{aligned}
    &\left(\tilde\Delta_u+\frac{m^2}{H^2}\right) \hat{\mathcal{I}}^{\phi^4}_{(0,0)}(u,v,x_0)\\
    =& \sum_{b=\pm}  (ib)\mathcal{C}_{b}(u, x_0)\left[\frac{e^{-ib\frac{x_0}{u}}[x_0 \hat{\mathcal K}_b(x_0,x_0,v)-u^2(1+ib\frac{x_0}{u})\partial_1 \hat{\mathcal K}_b(x_0,x_0,v)]}{u^2x_0^2}\right.\\
    -&\left.\mathcal{C}_{b}(v, x_0)\int_{x_0}^{\infty}\frac{\mathrm{d}x}{x^4}\left(1+ib\frac{x}{u}\right)e^{-ib\frac{x}{u}}\left(1+ib\frac{x}{v}\right)e^{-ib\frac xv}\right].
    \label{eq:four_point_bootstrap_eq-eq137}
\end{aligned}
\end{equation}
Similar to the contact diagram $\braket{\phi^2\sigma}'$, this bootstrap equation can also be written as:
\begin{equation}\label{bootsrapexchange4pt-eq138}
\begin{split}
     &\left(\tilde\Delta_u+\frac{m^2}{H^2}\right)\hat{\mathcal{I}}^{\phi^4}_{(0,0)}(u,v,x_0)\\
     =&(x_0^2\partial_{x_0}^2-2x_0\partial_{x_0})\hat{\mathcal{I}}^{\phi^4}_{(0,0)}(u,v,x_0)-\int^{\infty}_{x_0}\frac{dx}{x^4}(x_0^2\partial_{x_0}^2-2x_0\partial_{x_0}) G^{\phi^4}[u,v,x,x_0]\\
     +&2\hat C_0-\frac23\left(\frac1{u^3}+\frac1{v^3}\right)(\gamma_E-1+\log x_0),
\end{split}
\end{equation}
except the normal contact source $\hat C_0$ (with a local correction), other anomalous source terms arise from the action of $x_0^2\partial_{x_0}^2-2x_0\partial_{x_0}$ on the upper bound of time integration.\\
Since the composite propagator satisfies
\begin{equation}
\label{eqn139}
    \hat{\mathcal K}_b(x_0,x_0,v)=\hat{\mathcal{I}}^{\phi^2\sigma}(v,x_0)=\frac{H^2}{m^2}, \, \partial_1(\hat{\mathcal K}_+(x_0,x_0,v)-\hat{\mathcal K}_-(x_0,x_0,v))=0,
\end{equation}
the boundary term (the first line) contributes a local term $2\frac{H^2}{m^2}\frac{1}{u^3}$. The bootstrap equation thus simplifies to 
\begin{equation}\label{eq:bootsrapequationphi4-eq140}
\begin{aligned}
\left(\tilde\Delta_u+\frac{m^2}{H^2}\right) \hat{\mathcal
{I}}^{\phi^4}_{(0,0)}(u,v,x_0)
=-\frac{2H^2}{m^2}\frac{1}{u^3}+2\hat{C}_0-\frac23\left(\frac1{u^3}+\frac1{v^3}\right)(\gamma_E-1+\log x_0).
\end{aligned}
\end{equation}
The inhomogeneous solution is 
\begin{equation}\label{inh00-eq141}
\begin{split}
   \hat{\mathcal{I}}^{\phi^4}_{(0,0),<}(u,v,x_0) =& \frac{H^2}{m^2}\left(2\hat C_0-\frac23\left(\frac1{u^3}+\frac1{v^3}\right)(\gamma_E-1+\log x_0)+2\sum_{p,q=0}^{\infty} d_{pq} u^{2p-3} \left(\frac{u}{v}\right)^ q \right)\\  
   =&2\hat F_0-2\frac{H^4}{m^4}\frac{1}{u^3}-\frac{2}{3}\frac{H^2}{m^2}\left(\frac1{u^3}+\frac1{v^3}\right)(\gamma_E-1+\log x_0).
\end{split}   
\end{equation}
The coefficients $d_{pq}$ appearing in the series are
\begin{equation}
\label{eqn142}
    d_{pq} = 
    \begin{cases} 
        -\dfrac{q-1}{2p+q-3} \dfrac{m^2/H^2}{q^2-3q+m^2/H^2} c_{p-1, q}, & p \neq 0, q \neq 1, \\
        c_{0, q-2}, & p=0, q \notin \{0, 1\}, \\
        0, & \text{otherwise},
    \end{cases}
\end{equation}
with 
\begin{equation}
\label{eqn143}
    c_{pq} = \frac{(-1)^q (q+1)_{2p}}{2^{2p+2} \left(\frac{q}{2} + \frac{1}{4} -\frac{\nu}{2}\right)_{p+1}  \left(\frac{q}{2} + \frac{1}{4} +\frac{\nu}{2}\right)_{p+1}}.  
\end{equation}
We also need to work out the homogeneous solution. The two eigenfunctions of the equation are given by
\begin{equation}
\label{eqn144}
    \mathcal{F}_{\pm}(u)=\frac{1}{\nu^2}\left(-\frac{u}{2\nu}\right)^{-\frac32\pm\nu}\text{}_{2}F_1\left(-\frac{3}{4} \pm \frac{\nu}{2},\, \frac{3}{4} \pm \frac{\nu}{2};\, 1 \pm \nu;\, u^2\right),
\end{equation}
since the integral is symmetric under $u\leftrightarrow v$, and the inhomogeneous solution satisfy
\begin{equation}
\label{eqn145}
\begin{split}
         \hat{\mathcal{I}}^{\phi^4}_{(0,0),<}(u,v,x_0)-\hat{\mathcal{I}}^{\phi^4}_{(0,0),>}(u,v,x_0)=&\hat{\mathcal{I}}^{\phi^4}_{(0,0),<}(u,v,x_0)-\hat{\mathcal{I}}^{\phi^4}_{(0,0),<}(v,u,x_0)\\
     =&\frac{H^2}{m^2}\frac{2\pi}{\cos(\pi \nu)}(\mathcal{F}_{+}(u)\mathcal{F}_{-}(v)-\mathcal{F}_{-}(u)\mathcal{F}_{+}(v) ),
\end{split}
\end{equation}
We can write the full solution as
\begin{equation}
\label{eqn146}
\begin{split}
    \hat{\mathcal{I}}^{\phi^4}_{(0,0)}(u,v,x_0)=& \hat{\mathcal{I}}_{(0,0),<}^{\phi^4}(u, v, x_0) - \frac{H^2}{m^2}\frac{\pi}{\cos(\pi \nu)}\left[(\mathcal{F}_{+}(u)\mathcal{F}_{-}(v)-\mathcal{F}_{-}(u)\mathcal{F}_{+}(v) )\right.\\
 +&   \left.\beta_+\mathcal{F}_{+}(u)\mathcal{F}_{+}(v)+\beta_-\mathcal{F}_{-}(u)\mathcal{F}_{-}(v)+\beta_0(\mathcal{F}_{+}(u)\mathcal{F}_{-}(v)+\mathcal{F}_{-}(u)\mathcal{F}_{+}(v))\right].
 \end{split}
\end{equation}
By avoiding folded singularity at both $u\rightarrow 1$ and $v\rightarrow 1$, and matching the partial energy singularity $u,v\rightarrow -1$, we obtain
\begin{equation}
\label{eqn147}
\begin{split}
     \tilde\alpha_{\pm}\equiv& \frac{1}{2\nu^2}\left(-\frac{1}{2\nu}\right)^{-\frac32\pm\nu}\frac{\Gamma(1 \pm \nu)}{\Gamma(-\frac{3}{4} \pm \frac{\nu}{2})\Gamma(\frac{3}{4} \pm \frac{\nu}{2})},\\  
     \beta_{\pm}=&-(\beta_0\pm1)\tilde\alpha_{\mp}/\tilde\alpha_{\pm},\\
     \beta_0=&\frac{1}{\sin(\pi\nu)}.
\end{split}
\end{equation}
Redefine 
 \begin{equation}
\label{eqn148}
\hat{\mathcal{F}}_{\pm}(u) = \pm \frac{\pi  }{2\tilde{\alpha}_{\pm}} \mathcal{F}_{\pm}(u),
\end{equation}
We get a simple form 
\begin{equation}
\label{eqn149}
    \hat{\mathcal{I}}^{\phi^4}_{(0,0)}(u,v,x_0)=\hat{\mathcal{I}}_{(0,0),<}^{\phi^4}(u, v, x_0)+\sum_{b=\pm}\frac{H^2\tilde\alpha_+\tilde\alpha_-(\beta_0+b)}{\pi m^2\cos(\pi\nu)}(\hat{\mathcal{F}}_{+}(v)+\hat{\mathcal{F}}_{-}(v))\hat{\mathcal{F}}_{b}(u).
\end{equation}
We now turn to derivative couplings. For the 4-pt exchange diagram generated by the derivative interaction $\lambda_3 (\partial_\mu \phi)^2 \sigma$, the corresponding seed function is free of the late-time divergences and was studied in~\cite{Arkani-Hamed:2018kmz}. The bootstrap equation with respect to $\hat{\mathcal{I}}^{\phi^4}_{(1,1)}(u, v) $ is given by
\begin{equation}
    (\tilde\Delta_u+\frac{m^2}{H^2})\hat{\mathcal{I}}^{\phi^4}_{(1,1)}(u, v)=\frac{(1-u^2)(1-v^2)(u^2+3uv+v^2)}{uv(u+v)^3}=\frac{1}{2}\hat{C}_2.
    \label{eq:eqm-eq150}
\end{equation}
The inhomogeneous solution can be expressed as
\begin{equation}\label{inh11-eq151}
   \hat{\mathcal{I}}^{\phi^4}_{(1,1),<}(u, v) =  \frac{1}{2}\hat{C}_1 + \frac{m^2}{2H^2} \sum_{p,q=0}^{\infty} d_{pq} u^{2p-3} \left(\frac{u}{v}\right)^q=\frac{1}{2}\hat F_2 \,,
\end{equation}
For the 4-pt exchange diagram generated by mixing $\lambda_3 (\partial_\mu \phi)^2 \sigma$ and $g_3\phi^2\sigma$, the bootstrap equations with respect to $\hat{\mathcal{I}}^{\phi^4}_{(1,0)}(u, v)$ and $\hat{\mathcal{I}}^{\phi^4}_{(0,1)}(u, v)$ are given by
\begin{equation}
\label{eqn152}
\begin{split}
        &(\tilde\Delta_u+\frac{m^2}{H^2})\hat{\mathcal{I}}^{\phi^4}_{(1,0)}(u, v)=-\hat{C}_1+\frac{1}{u^3},\\
&(\tilde\Delta_u+\frac{m^2}{H^2})\hat{\mathcal{I}}^{\phi^4}_{(0,1)}(u, v)=-\hat{C}_1+\frac{1}{v^3}.
\end{split}
\end{equation}
The inhomogeneous solutions can be expressed as
\begin{equation}\label{inh10-eq153}
\begin{split}
       & \hat{\mathcal{I}}^{\phi^4}_{(1,0),<}(u, v) =\sum_{p,q=0}^{\infty} d_{pq} u^{2p-3} \left(\frac{u}{v}\right)^q+\frac{H^2}{m^2}\frac{1}{u^3}=\hat F_1+\frac{H^2}{m^2}\frac{1}{u^3} ,\\
    & \hat{\mathcal{I}}^{\phi^4}_{(0,1),<}(u, v) =\sum_{p,q=0}^{\infty} d_{pq} u^{2p-3} \left(\frac{u}{v}\right)^q+\frac{H^2}{m^2}\frac{1}{v^3}=\hat F_1+\frac{H^2}{m^2}\frac{1}{v^3}.
\end{split}
\end{equation}
By comparing these inhomogeneous solutions \cref{inh00-eq141}, \cref{inh11-eq151} and \cref{inh10-eq153}, we can verify the IBP relations \cref{eq:4pt_decomposition-eq035} and \cref{eq:4pt_mixed_result-eq040}. By comparing \cref{inh00-eq141}, \cref{inh10-eq153} with \cref{eq:3ptinh-eq125}, \cref{eq:3ptinhsollamda3-eq130}, we can verify the derivative relation between dS trispectrum and inflationary bispectrum \cref{eq:modified_relation_late_time-eq041} and \cref{eq:structural_relation-eq037}. \\
For higher-derivative exchange diagrams, the recursion is most directly stated in the box basis used in \cref{eq:box_basis_direct-eq054}. We therefore compute the contact sources for the box-basis contact vertex $2^{-l}\phi^2\square^l(\phi^2)$; covariant-derivative vertices such as $\phi^2(\nabla^j\phi)^2$ should first be projected onto this basis using \cref{eq:Sm_box_basis-eq056}, rather than identified with it term by term.
The $l=0$ and $l=1$ case are shown by direct computation:
\begin{equation}
\label{eqn154}
\begin{split}
       \mathcal I_0=&s^3O_{12}O_{34}\hat{\mathcal{I}}_{0}=2s^3O_{12}O_{34}\left(\hat C_0-\frac13\left(\frac1{u^3}+\frac1{v^3}\right)(\gamma_E-1+\log x_0)\right),\\
       \mathcal I_1=&s^3O_{12}O_{34}\hat{\mathcal{I}}_{1}=-s^3O_{12}O_{34}\hat C_1.
\end{split}
\end{equation}
For $l\geq2$, the original integral can be rewritten as
\begin{equation}
\label{eqn155}
\begin{split}
      \mathcal I_{l}=&s^3O_{12}O_{34}\hat{\mathcal{I}}_{l},\\
      =&2^{-l}O_{12}O_{34}\sum_{b=\pm}ib\int \frac{d\eta}{\eta^4}\left[(\eta^2\partial_{\eta}^2-2\eta\partial_\eta+s^2\eta^2)^l(1-ibk_{12}\eta)e^{ibk_{12}\eta}\right](1-ibk_{34}\eta)e^{ibk_{34}\eta}\\  =& (-2)^{-l}s^3O_{12}O_{34}\sum_{b=\pm}ib\int \frac{dx}{x^4}\left[(x^2\partial_{x}^2-2x\partial_x+x^2)^l\left(1+ib\frac{x}{u}\right)e^{-ib\frac{x}{u}}\right]\left(1+ib\frac{x}{v}\right)e^{-ib\frac{x}{v}},
\end{split}
\end{equation}
while from \cref{contact-bootstrap-eq131} and the $l=1$ case, $C_l$ can be written as 
\begin{equation}
\label{eqn156}
    C_l=\frac{1}{2}s^3O_{12}O_{34}\sum_{b=\pm}ib\int \frac{dx}{x^4}\tilde\Delta_u^l\left(1+ib\frac{x}{u}\right)e^{-ib\frac{x}{u}}\left(1+ib\frac{x}{v}\right)e^{-ib\frac{x}{v}}.
\end{equation}
Because of \cref{bootstrap-equality-eq112} and the commutativity between $x^2\partial_{x}^2-2x\partial_x+x^2$ and $\tilde\Delta_u$, the box-basis source is $-(-2)^{1-l}\hat C_l$. This matches the contact diagrams generated by the IBP/EoM reduction in \cref{eq:box_contact_independent-eq059}. Thus, for $l=r+s\geq2$, there are no boundary conformal anomalies in the box-basis exchange seed, and every split $(r,s)$ corresponds to the same exchange solution $\hat{\mathcal I}^{\phi^4}_{l}=2^{1-l}\hat F_l$.
We summarise the results in \cref{table-6}.
\begin{table}[htbp]
 \centering
    \begin{tabular}{|c|c|c|}
       \hline Number of derivatives & seed functions & sources\\ \hline
         $(0,0)$ & $\begin{matrix}
             2\hat F_0-2\frac{H^4}{m^4}\frac{1}{u^3}\\-\frac23\frac{H^2}{m^2}\left(\frac1{u^3}+\frac1{v^3}\right)(\gamma_E-1+\log x_0)
         \end{matrix}$ & $\begin{matrix}
             2\hat{C}_0-\frac{2H^2}{m^2}\frac{1}{u^3}\\-\frac23\left(\frac1{u^3}+\frac1{v^3}\right)(\gamma_E-1+\log x_0)
         \end{matrix}$ \\ \hline $(1,0)$ & $\hat F_1+\frac{H^2}{m^2}\frac{1}{u^3}$ & $-\hat C_1+\frac{1}{u^3}$\\ \hline
         $(0,1)$ & $\hat F_1+\frac{H^2}{m^2}\frac{1}{v^3}$ & $-\hat C_1+\frac{1}{v^3}$\\ \hline
         $l=r+s\geq 2$ & $2^{1-l} \hat F_l$ & $-(-2)^{1-l}  \hat C_l$\\ \hline
    \end{tabular}
    \caption{Seed functions and sources with conformal anomaly.}
    \label{table-6}
\end{table}

\section{Light Mass Regime}\label{Light mass}
In this appendix we turn to the light mass regime $0<\nu<\frac32$, with particular interest in the near massless behaviour of bootstrap equations and solutions. For this purpose, we adopt $\epsilon=\frac32-\nu$ as a regulator, and expand the equations and solution in orders of $\epsilon$. We take the 3-pt contact diagram $\braket{\phi^2\sigma}^\prime_{g_3}$ and the 4-pt exchange diagram $\braket{\phi^4}^\prime_{g_3\times g_3}$ as examples, and use different symbols $\tilde{\mathcal I}$ to denote the seed functions in this regime.

Near the massless limit, the bootstrap equation \cref{eq:master_eq_3pt-eq113} can be expanded up to $\mathcal O(\epsilon^1)$:
\begin{equation} \label{light-contact-eq157}
    \left(\tilde\Delta_u+3\epsilon\right)\tilde{\mathcal I}^{\phi^2\sigma}(u,x_0)\simeq1-\frac{x_0^{2\epsilon}}{u^3}\left(1-\frac23 \epsilon\right)\simeq1-\frac{1}{u^3}-2\frac{\epsilon\log x_0}{u^3}+\frac{2}{3}\frac{\epsilon}{u^3},
\end{equation}
where we demand $\tilde{\mathcal I}^{\phi^2\sigma}(u,x_0)\sim \mathcal O(\epsilon^0)$ to avoid mismatching $\mathcal O(\epsilon^1)$ terms, because we neglect $-\epsilon^2\tilde{\mathcal I}^{\phi^2\sigma}(u,x_0)$ on the LHS. The original inhomogeneous solution, which is of order $\mathcal O(\epsilon^{-1})$, does not satisfy this condition. We therefore seek a new inhomogeneous solution without a divergence as $\epsilon\rightarrow 0$.\footnote{Even if we match up to $\mathcal O(\epsilon^1)$ in \cref{light-contact-eq157}, there are still ambiguous terms $\epsilon\left(\alpha+\beta\frac{1}{u^3}\right)$ whose coefficients cannot be fixed unless we expand the bootstrap equation to the next order, $\mathcal O(\epsilon^2)$. These terms are beyond the order needed for the matching performed here and amount to higher-order scheme dependence in the light-mass expansion.}
To achieve this goal, we solve the differential equation
\begin{equation}
\label{eqn158}
    \tilde\Delta_u f(u)=1-\frac{1}{u^3},
\end{equation}
One inhomogeneous solution is
\begin{equation}
\label{eqn159}
f_1(u)=\frac{1}{3u^3}\left(u(1+u)+(1+u^3)\log\left(\frac{u}{u+1}\right)\right),
\end{equation}
Then we solve 
\begin{equation}
\label{eqn160}
    \tilde\Delta_u f(u)=f_1(u),
\end{equation}
and denote one inhomogeneous solution by $f_2(u)$. Its explicit form contains dilogarithm functions, but will not be needed below.
With $f_1(u)$ and $f_2(u)$, we can obtain a suitable inhomogeneous solution of \cref{light-contact-eq157}:
\begin{equation}
\label{eqn161}
    \tilde{\mathcal I}^{\phi^2\sigma}_{inh}(u,x_0)=f_1(u)-3\epsilon f_2(u)+\frac{1}{3}\left(-2\frac{\log x_0}{u^3}+\frac{2}{3}\frac{1}{u^3}\right),
\end{equation}
We expect that, when $\epsilon\rightarrow0$, the full solution reduces to the massless 3-pt contact diagram:
\begin{equation}
\label{eqn162}
    \tilde{\mathcal I}^{\phi^3}(u,x_0)=\frac19\left[\frac{3\gamma_E-4}{u^3}+\frac{3}{u^2}+\frac{3}{u}+3\gamma_E-4+3\left(1+\frac{1}{u^3}\right)\log\left(\left(1+\frac{1}{u}\right)x_0\right)\right],
\end{equation}
so we need to fix the coefficients of homogeneous solutions to match the $\mathcal O(\epsilon^{0})$ terms. 
The homogeneous solutions become
\begin{equation}
    \begin{split}\label{eqn163}
    &c_1u^{-3+\epsilon} {}_2F_1\left(-\frac{3}{2} + \frac{\epsilon}{2},\, \frac{\epsilon}{2} ;\, \epsilon - \frac12;\, u^2\right) + c_2u^{-\epsilon} {}_2F_1\left(-\frac\epsilon 2,\, \frac{3}{2} - \frac\epsilon 2;\, \frac52 -\epsilon ;\, u^2\right)\\
    =&c_1\frac{1}{u^3}+c_2+\mathcal O(\epsilon),
    \end{split}
\end{equation}
so we have 
\begin{equation}
\label{eqn164}
    c_1=\frac{\gamma_E-2+\log x_0}{3},\,c_2=\frac{3\gamma_E-4+9\log x_0}{9}.
\end{equation}
We can further expand the homogeneous solutions up to $\mathcal O(\epsilon^2)$, 
\begin{equation}
\label{eqn165}
\begin{split}
     u^{-3+\epsilon}{}_2F_1\left(-\frac{3}{2} + \frac{\epsilon}{2},\, \frac{\epsilon}{2} ;\, \epsilon - \frac12;\, u^2\right)=\frac{1}{u^3}+\epsilon  f_{11}(u)+\epsilon^2 f_{12}(u),
\end{split}
\end{equation}
where $f_{11}(u)$ and $f_{12}(u)$ satisfy:
\begin{equation}
\label{eqn166}
    \tilde\Delta_u  f_{11}(u)=-3\frac{1}{u^3}, \, \tilde\Delta_u  f_{12}(u)=-3f_{11}(u),
\end{equation}
One can choose 
\begin{equation}
\label{eqn167}
    f_{11}(u)=-\frac{1}{2u^3}\left(-2u^2+u^3 \log\left(\frac{1+u}{1-u}\right)+\log\left(\frac{1-u^2}{u^2}\right)\right),
\end{equation}
while $f_{12}$ again has a complicated form containing dilogarithm functions. These functions can also be obtained from series expansion of hypergeometric functions. Similarly, 
\begin{equation}
\label{eqn168}
\begin{split}
     u^{-\epsilon} {}_2F_1\left(-\frac\epsilon 2,\, \frac{3}{2} - \frac\epsilon 2;\, \frac52 -\epsilon ;\, u^2\right)=1+\epsilon  f_{21}(u)+\epsilon^2 f_{22}(u),
\end{split}
\end{equation}
where 
\begin{equation}
\label{eqn169}
    f_{21}(u)=-\frac{1}{2u^3}\left(2u-u^3\log\left(\frac{1-u^2}{u^2}\right)-\log\left(\frac{1+u}{1-u}\right)\right)=f_{11}(u)-3f_1(u),
\end{equation}
and 
\begin{equation}
\label{eqn170}
    f_{22}(u)=f_{12}(u)-3f_2(u),
\end{equation}
Finally, the full solution up to $\mathcal O(\epsilon^1)$ is
\begin{equation}\label{contact-light-eq171}
\begin{split}
     \tilde{\mathcal I}^{\phi^2\sigma}(u,x_0)=&\frac19\left[\frac{3\gamma_E-4}{u^3}+\frac{3}{u^2}+\frac{3}{u}+3\gamma_E-4+3\left(1+\frac{1}{u^3}\right)\log\left(\left(1+\frac{1}{u}\right)x_0\right)\right]\\
     +&\epsilon\left(-3f_2(u)-\frac{3\gamma_E-4+9\log x_0}{9u^3}\left(u-u^2+(u^3-1)\log\left(\frac{u}{1-u}\right)\right)\right.\\
     +&\left.\frac{3\log x_0+1}{9u^3}\left(-2u^2+u^3 \log\left(\frac{1+u}{1-u}\right)+\log\left(\frac{1-u^2}{u^2}\right)\right)+\alpha+\beta\frac{1}{u^3}\right).
\end{split}
\end{equation}
Now one may wonder why the solution has entirely different behaviour near the massless limit. We can see the answer if we adopt the inhomogeneous solution as the heavy mass regime:
\begin{equation}
\label{eqn172}
    \frac{H^2}{m^2}\left(1-\frac{2\hat K^{\sigma}(x_0,x_0)}{u^3}\right)\simeq\frac{1}{3\epsilon}\left(1-\frac1{u^3}\right)+\frac{1}{3}\left(-2\frac{\log x_0}{u^3}+\frac{1}{3}\left(1+\frac{1}{u^3}\right)\right), 
\end{equation}
which extracts the boundary contributions (with respect to the derivative coupling $(\partial\phi)^2\sigma$). By matching with \cref{contact-light-eq171}, we find that the coefficients of the homogeneous solutions become
\begin{equation}
\label{eqn173}
    \tilde c_1= \frac{2\hat K^{\sigma}(x_0,x_0)(1+\epsilon(\gamma_E-1+\log x_0))}{3\epsilon},\, \tilde c_2=\frac{2\hat K^{\sigma}(x_0,x_0)(-1+\epsilon(\gamma_E-\frac73+3\log x_0))}{3\epsilon}, 
\end{equation}
which decay as $\epsilon$ increases. Thus, as the mass increases, the homogeneous solutions decay to zero, and the inhomogeneous solution becomes $\frac{H^2}{m^2}$, returning to the heavy-mass result. However, if $\epsilon\sim1/100$, i.e. $\frac{m}{H}\sim1/10$, $\epsilon\log x_0$ can reach $\mathcal O(1)$ while $x_0^\epsilon\sim \mathcal O(1)$. We then expect a sizeable correction to the non-Gaussianity of a massless inflaton. This correction can originate from both the slow-roll mass of the inflaton and the exchange of a nearly massless mediator.\par

Now we turn to the exchange diagram and keep only the $\epsilon$-independent part. \Cref{bootsrapexchange4pt-eq138} becomes
\begin{equation}
    \begin{split}\label{eqn174}
    &\left(\tilde\Delta_u+\frac{m^2}{H^2}\right)\tilde{\mathcal I}^{\phi^4}_{(0,0)}(u,v,x_0)\\
    =&2\hat C_0-\frac23\left(\frac1{u^3}+\frac1{v^3}\right)(\gamma_E-1+\log x_0)-\frac{2H^2}{m^2}\frac{1}{u^3}+\frac{1}{u^3}\left(\frac{2H^2}{m^2}-2\hat{\mathcal I}^{\phi^3}(v,x_0)\right),
    \end{split}
\end{equation}
Except for the last anomalous source term, the others give the same result as in the large-mass regime, i.e. $\hat{\mathcal I}^{\phi^4}_{(0,0)}(u,v,x_0)$, which near the massless limit becomes:
\begin{equation}
\label{eqn175}
\begin{split}
    &\left(\frac{2}{3\epsilon}+\frac29\right)\left(-\frac{1}{3}\left(\frac1{u^3}+\frac1{v^3}\right)(\gamma_E-1+\log x_0)-\frac{1}{3\epsilon}\left(\frac1{u^3}+\frac1{v^3}\right)\right.\\
    -&\left.\frac{1}{3}\left(\frac{f_{21}(v)}{u^3}+\frac{f_{21}(u)}{v^3}\right)+\frac{1}{3\epsilon}\frac{1}{u^3v^3}+\frac{8-6\gamma_E}{9}\frac{1}{u^3v^3}+\frac{1}{3}\left(\frac{f_{11}(v)}{u^3}+\frac{f_{11}(u)}{v^3}\right)\right)\\
    =&\frac{2}{9\epsilon^2}\left(\frac{1}{u^3v^3}-\frac1{u^3}-\frac1{v^3}\right)+\frac{2}{9\epsilon}\left(-\left(\frac1{u^3}+\frac1{v^3}\right)(\gamma_E-1+\log x_0)+(3-2\gamma_E)\frac{1}{u^3v^3}\right.\\
    -&\left.\frac13\left(\frac1{u^3}+\frac1{v^3}\right)-\left(\frac{f_{21}(v)}{u^3}+\frac{f_{21}(u)}{v^3}\right)+\left(\frac{f_{11}(v)}{u^3}+\frac{f_{11}(u)}{v^3}\right)\right).
\end{split}
\end{equation}
We solve the remaining part $\delta{\mathcal I}^{\phi^4}_{(0,0)}$:
\begin{equation}
\label{eqn176}
    \left(\tilde\Delta_u+\frac{m^2}{H^2}\right)\delta{\mathcal I}^{\phi^4}_{(0,0)}(u,v,x_0)=\frac{1}{u^3}\left(\frac{2H^2}{m^2}-2\hat{\mathcal I}^{\phi^3}(v,x_0)\right),
\end{equation}
Demanding that the solution be symmetric under $u\leftrightarrow v$, we find an inhomogeneous solution
\begin{equation}\label{C20-eq177}
\begin{split}
       &\delta{\mathcal I}^{\phi^4}_{(0,0)}(u,v,x_0)\\
       =&-\frac{4H^4}{m^4}\frac{\hat K^{\sigma}(x_0,x_0)}{u^3v^3}-\frac{2H^2}{m^2}\left(\frac{\tilde I^{\phi^2\sigma}(v,x_0)}{u^3}+\frac{\tilde I^{\phi^2\sigma}(u,x_0)}{v^3}\right)+\frac{2H^4}{m^4}\left(\frac1{u^3}+\frac1{v^3}\right).
\end{split}
\end{equation}
This solution decays as $\epsilon$ increases, and one can check that these terms are precisely the boundary contributions (with respect to $\tilde{\mathcal I}^{\phi^4}_{(1,1)}(u,v,x)$) in the light-mass regime: the first term is decaying and corresponds to $\braket{\phi^4}^{\prime}_{\mathcal B_2\times \mathcal B_2}$, while the second and third terms correspond to $\braket{\phi^4}^{\prime}_{g_3\times \mathcal B_2}$ and $\braket{\phi^4}^{\prime}_{\mathcal B_1\times \mathcal B_2}$, respectively. However, they are not cancelled in the light-mass regime.
Near the massless limit, it becomes:
\begin{equation}
\label{eqn178}
    \begin{split}
    -&\frac{2}{9\epsilon^2}\left(\frac{1}{u^3v^3}-\frac1{u^3}-\frac1{v^3}\right)+\frac{2}{27\epsilon}\left(\frac{1}{u^3v^3}-6\frac{\log x_0}{u^3v^3}+\frac1{u^3}+\frac1{v^3}\right.\\
    -&\left.9\left(\frac{\tilde I^{\phi^2\sigma}(v,x_0)}{u^3}+\frac{\tilde I^{\phi^2\sigma}(u,x_0)}{v^3}\right)\right),
    \end{split}
\end{equation}
Adding the two parts, we find that the $\mathcal O(\epsilon^{-2})$ terms are cancelled, and the $\mathcal O(\epsilon^{-1})$ terms become
\begin{equation}
\label{eqn179}
    \frac{2}{27\epsilon}\left(\frac{2}{u^3v^3}-\frac{1}{u^3}-\frac{1}{v^3}\right),
\end{equation}
which are cancelled by a homogeneous solution without folded singularity and symmetric under $u\leftrightarrow v$:
\begin{equation}\label{C23-eq180}
    \begin{split}
    -&\frac{4\hat K^{\sigma}(x_0,x_0)}{27\epsilon}\left(\epsilon\beta_+\mathcal{F}_{+}(u)\mathcal{F}_{+}(v)+(2+\epsilon\beta_-)\mathcal{F}_{-}(u)\mathcal{F}_{-}(v)\right.\\
    +&\left.(-1+\epsilon\beta_0)(\mathcal{F}_{+}(u)\mathcal{F}_{-}(v)+\mathcal{F}_{-}(u)\mathcal{F}_{+}(v))\right),
    \end{split}
\end{equation}
As the mass increases, this homogeneous solution decays to zero. The coefficients $\beta_+$, $\beta_-$ and $\beta_0$ can be fixed by matching the $\mathcal O(\epsilon^0)$ terms with the massless exchange result~\cite{Bzowski:2022rlz,Wang:2022eop,Bzowski:2023nef}. In particular, one can check that \cref{C20-eq177,C23-eq180} recover the partial energy singularities associated with late-time divergence in the massless limit,
\begin{equation}
\label{eqn181}
\begin{split}
       -&\frac{2}{9}\left(1+\frac{1}{u^3}\right)\left(1+\frac{1}{v^3}\right)\log\left(\left(1+\frac{1}{u}\right)x_0\right)\log\left(\left(1+\frac{1}{v}\right)x_0\right)\\+&\frac{1}{18}\left(\left(1+\frac{1}{u^3}\right)\log^2\left(\left(1+\frac{1}{u}\right)x_0\right)+\left(1+\frac{1}{v^3}\right)\log^2\left(\left(1+\frac{1}{v}\right)x_0\right)\right)+\mathcal O(\log x_0).
\end{split}
\end{equation}

\bibliographystyle{utphys}
\bibliography{ref}

\end{document}